\documentclass[12pt]{article}
\pdfoutput=1
\usepackage{amsmath,amsfonts,amssymb,amsthm,amssymb,bbm,bm}
\usepackage{pdfsync}
\usepackage{graphicx,import}
\usepackage[utf8]{inputenc}
\usepackage{empheq}
\usepackage[all]{xy}
\usepackage{stmaryrd}
\usepackage{rotating}
\usepackage[dvipsnames]{xcolor}
\usepackage{slashed}
\usepackage{cancel}
\usepackage{array, makecell} %
\usepackage{comment}
\usepackage{tikz-cd}
\usepackage{hhline}
\usepackage{cancel}
\usepackage{dsfont}
\usepackage{stackengine}
\usepackage[breakable]{tcolorbox}
\usepackage{mathrsfs}
\usepackage{booktabs}
\usepackage{float}
\usepackage{changepage}

\allowdisplaybreaks

\usepackage[pagebackref]{hyperref}
\renewcommand*{\backref}[1]{}
\renewcommand*{\backrefalt}[4]{\small{
    \ifcase #1%
          \or (Cited on page~#2.)%
          \else (Cited on pages~#2.)%
    \fi%
    }}
\hypersetup{colorlinks,linkcolor={RedViolet},citecolor={Blue},urlcolor={Blue}}

\usepackage{orcidlink}

\DeclareMathAlphabet\mathrsfso{U}{rsfso}{m}{n}

\usepackage[scr=boondoxo,scrscaled=1.1]{mathalpha}

\makeatletter

\DeclareFontFamily{U}{matha}{\hyphenchar\font45}
\DeclareFontShape{U}{matha}{m}{n}{
      <5> <6> <7> <8> <9> <10> gen * matha
      <10.95> matha10 <12> <14.4> <17.28> <20.74> <24.88> matha12
      }{}
\DeclareSymbolFont{matha}{U}{matha}{m}{n}
\DeclareMathSymbol{\oright}       {2}{matha}{"69}

\usepackage[top=2cm, bottom=2.5cm, left=2.5cm, right=2.5cm]{geometry}

\newcommand{\doublehat}[1]{%
\begingroup%
  \let\macc@kerna\z@%
  \let\macc@kernb\z@%
  \let\macc@nucleus\@empty%
  \hat{\raisebox{.55ex}{\vphantom{\ensuremath{#1}}}\smash{\hat{#1}}}%
\endgroup%
}

\renewcommand{\ni}{\noindent}

\newcommand{\bit}{\begin{itemize}}
\newcommand{\eit}{\end{itemize}}
\newcommand{\bd}{\begin{description}}
\newcommand{\ed}{\end{description}}

\newcommand{\bc}{\begin{center}}
\newcommand{\ec}{\end{center}}

\newcommand{\C}{{\mathbb C}}
\newcommand{\N}{{\mathbb N}}
\newcommand{\R}{{\mathbb R}}
\newcommand{\cP}{{\cal P}}
\newcommand{\cD}{\mathscr{D}}
\newcommand{\E}{\mathsf{E}}

\newcommand{\bmsw}{\mathfrak{bmsw}}
\newcommand{\gbms}{\mathfrak{gbms}}

\newcommand{\cF}{{\mathcal F}}
\newcommand{\cM}{{\mathcal M}}

\newcommand{\cA}{{\mathcal A}}

\def\be#1\ee{\begin{align}#1\end{align}}
\newcommand{\bea}{\begin{eqnarray}}
\newcommand{\eea}{\end{eqnarray}}
\newcommand{\bs}{\begin{subequations}}
\newcommand{\es}{\end{subequations}}
\newcommand{\nn}{\nonumber}

\usepackage[numbers,sort&compress]{natbib}

\newcommand{\rd}{\mathrm{d}}
\newcommand{\pa}{\partial}
\newcommand{\Dcal}{\mathcal{D}}

\newcommand{\pui}[1][1]{\pa_u^{-#1}}

\newcommand{\dt}{\delta_\tau}
\newcommand{\dtp}{\delta_{\tau'}}
\newcommand{\dtpp}{\delta_{\tau''}}

\newcommand{\tdt}{\tilde{\delta}_\tau}
\newcommand{\tdtp}{\tilde{\delta}_{\tau'}}

\newcommand{\hdt}{\hat{\delta}_\tau}
\newcommand{\hdtp}{\hat{\delta}_{\tau'}}

\newcommand{\dT}{\delta_T}
\newcommand{\dTp}{\delta_{T'}}

\newcommand{\hdT}{\hat{\delta}_T}
\newcommand{\hdTp}{\hat{\delta}_{T'}}

\renewcommand{\d}[2]{\delta^{\scriptscriptstyle{[#1]}}_{#2}}
\newcommand{\td}[2]{\tilde{\delta}^{\scriptscriptstyle{[#1]}}_{#2}}

\newcommand{\Hd}[2]{\delta^{\scriptscriptstyle{[#1]}\mathsf{H}}_{#2}}
\newcommand{\Htd}[2]{\tilde{\delta}^{\scriptscriptstyle{[#1]}\mathsf{H}}_{#2}}

\newcommand{\Sd}[2]{\delta^{\scriptscriptstyle{[#1]}\mathsf{S}}_{#2}}
\newcommand{\Std}[2]{\tilde{\delta}^{\scriptscriptstyle{[#1]}\mathsf{S}}_{#2}}

\newcommand{\LL}{\mathscr{L}}
\newcommand{\sE}{\mathsf{E}}
\newcommand{\sI}{\mathsf{I}}

\newcommand{\nkor}{\rho}

\newcommand{\overbar}[1]{\mkern 3mu\overline{\mkern-3.5mu#1\mkern-2mu}\mkern 2mu}

\newcommand{\bC}{\overbar{C}}
\newcommand{\bN}{\overbar{N}}
\newcommand{\bz}{\bar{z}}
\renewcommand{\bm}{\overbar{m}}
\newcommand{\bY}{\overbar{Y}}
\newcommand{\bD}{\overbar{D}}
\newcommand{\bP}{\overbar{P}}
\newcommand{\Vs}{\overline{\mathsf{V}}(S)}

\newcommand{\bgN}{\bar{g}_\textsc{n}}

\newcommand{\hs}{\hat{s}}
\newcommand{\shs}{\hat{s}^*}
\newcommand{\hS}{\hat{S}}
\newcommand{\shS}{\hat{S}^*}

\newcommand{\overhat}[1]{\mkern 5mu\widehat{\mkern-4.5mu#1\mkern-1mu}\mkern 2mu}

\newcommand{\hh}{\overhat{\cH}}
\newcommand{\hq}{\widehat{\cq}}
\newcommand{\ft}{\hat{\tau}}
\newcommand{\hG}{\widehat{G}}

\makeatletter
\newcommand{\hhat}[1]{%
\begingroup%
  \let\macc@kerna\z@%
  \let\macc@kernb\z@%
  \let\macc@nucleus\@empty%
  \widehat{\mathchoice%
    {\raisebox{.2ex}{\vphantom{\ensuremath{\displaystyle #1}}}}%
    {\raisebox{.4ex}{\vphantom{\ensuremath{\textstyle #1}}}}%
    {\raisebox{.16ex}{\vphantom{\ensuremath{\scriptstyle #1}}}}%
    {\raisebox{.14ex}{\vphantom{\ensuremath{\scriptscriptstyle #1}}}}%
    \smash{\widehat{#1}}}%
\endgroup%
}
\makeatother

\makeatletter
\DeclareRobustCommand\widecheck[1]{{\mathpalette\@widecheck{#1}}}
\def\@widecheck#1#2{%
    \setbox\z@\hbox{\m@th$#1#2$}%
    \setbox\tw@\hbox{\m@th$#1%
       \widehat{%
          \vrule\@width\z@\@height\ht\z@
          \vrule\@height\z@\@width\wd\z@}$}%
    \dp\tw@-\ht\z@
    \@tempdima\ht\z@ \advance\@tempdima2\ht\tw@ \divide\@tempdima\thr@@
    \setbox\tw@\hbox{%
       \raise\@tempdima\hbox{\scalebox{1}[-1]{\lower\@tempdima\box
\tw@}}}%
    {\ooalign{\box\tw@ \cr \box\z@}}}
\makeatother

\newcommand{\hhG}{\widecheck{G}}

\newcommand{\overtilde}[1]{\mkern 5mu\widetilde{\mkern-4.5mu#1\mkern-1mu}\mkern 2mu}

\newcommand{\tQ}{\widetilde{Q}}
\newcommand{\tq}{\mkern0.7mu\widetilde{\cq}}
\renewcommand{\th}{\overtilde{\cH}}

\newcommand{\tQcal}{\widetilde{\Qcal}}

\newcommand{\gN}{g_\textsc{n}}
\newcommand{\GN}{G_{\mkern-2mu N}}
\newcommand{\X}{\mathfrak{X}}
\newcommand{\g}{\mathfrak{g}}

\newcommand{\M}{\mathcal{M}}

\newcommand{\A}{\mathcal{T}}
\newcommand{\Ah}{\widehat{\mathcal{T}}}
\newcommand{\Ao}{\mathcal{T}^+}

\newcommand{\Cc}[1]{\sigma_{#1}}

\newcommand{\G}[1]{\mathcal{G}\big(#1\big)}

\newcommand{\PS}{\mathcal{P}}

\newcommand{\cS}{\mathcal{S}}

\newcommand{\V}{\mathsf{V}}
\newcommand{\tV}{\widetilde{\mathsf{V}}}

\newcommand{\W}{\mathsf{W}}
\newcommand{\Wcal}{\mathcal{W}}

\newcommand{\sfT}{{\mathsf{T}}}
\newcommand{\sfTo}{\mathsf{T}^+}
\newcommand{\sfTh}{\widehat{\mathsf{T}}}
\newcommand{\TN}{\mathscr{N}}

\newcommand{\Ccel}[1]{\mathcal{C}^\textsf{cel}_{#1}(S)}
\newcommand{\Ccar}[1]{\mathcal{C}^\textsf{car}_{(#1)}(\scri)}
\newcommand{\dCar}{\delta}

\renewcommand{\l}{\ell}

\newcommand{\ad}{{\textrm{ad}}_\sigma}
\newcommand{\adC}{\textrm{ad}_{\scriptscriptstyle{C}}}
 \newcommand{\mTG}{\mathcal{T}^G}

\renewcommand{\t}[1]{\tau_{#1}}
\newcommand{\T}[1]{T_{#1}}
\newcommand{\tp}[1]{\tau'_{#1}}
\newcommand{\Tp}[1]{T'_{#1}}
\newcommand{\tpp}[1]{\tau''_{#1}}
\newcommand{\Tpp}[1]{T''_{#1}}

\newcommand{\scri}{\mathrsfso{I}}

\newcommand{\Qt}{Q_\tau}
\newcommand{\Ht}{H_\tau}
\newcommand{\cH}{\mathrsfso{H}}
\newcommand{\Qtp}{Q_{\tau'}}

\newcommand{\Qcal}{\mathcal{Q}}
\newcommand{\cq}{\mathscr{Q}}
\newcommand{\q}[2]{\hq_{#1}^{\scriptscriptstyle{\!(#2)}}}
\newcommand{\cQ}[2]{Q^u_{#1}\left[#2\right]}

\newcommand{\soft}[1]{#1^{\mathsf{S}}}
\newcommand{\hard}[1]{#1^{\mathsf{H}}}

\newcommand{\lbr}{\llbracket}
\newcommand{\rbr}{\rrbracket}
\newcommand{\poisson}[1]{\big\{#1\big\}}
\newcommand{\paren}[1]{\,\Lbag #1\Rbag\,}

\newcommand{\cyc}{\overset{\circlearrowleft}{=}}

\newcommand{\Ham}{\eta}

\newcommand{\hHam}{\eta_{\scriptscriptstyle{G}}}

\newcommand{\bfm}[1]{\boldsymbol{#1}}
\newcommand{\bft}{\bfm{\tau}}
\newcommand{\bfCc}{\bfm{\sigma}}

\usepackage{tikz}
\newcommand*\circled[1]{\tikz[baseline=(char.base)]{
            \node[shape=circle,draw,inner sep=2pt] (char) {#1};}}
       
\definecolor{myorange}{RGB}{223, 109, 20}

\definecolor{argile}{RGB}{239, 239, 239}
\definecolor{beige}{RGB}{254, 253, 240}

\begin{document}

\title{\Large{\bf Asymptotic Higher Spin Symmetries II:\\
Noether Realization in Gravity}}

\author{Nicolas Cresto\,\orcidlink{0009-0006-7263-8777}$^{1,2}$\thanks{ncresto@perimeterinstitute.ca}\,,
Laurent Freidel\,\orcidlink{0000-0001-6964-0100}$^1$\thanks{lfreidel@perimeterinstitute.ca} 
}
\date{\small{\textit{
$^1$Perimeter Institute for Theoretical Physics,\\ 31 Caroline Street North, Waterloo, Ontario, N2L 2Y5, Canada\\ \smallskip
$^2$Department of Physics \& Astronomy, University of Waterloo,\\Waterloo, Ontario, N2L 3G1, Canada
}}}

\maketitle
\begin{abstract}
In this paper we construct a non-perturbative action of the higher spin symmetry algebra on the gravitational phase space. 
We introduce a symmetry algebroid $\mathcal{T}$ which allows us to include radiation in an algebraic framework.
We show that $\mathcal{T}$ admits a non-linear realization on the asymptotic phase space generated by a Noether charge defined non-perturbatively for all spins.
Besides, this Noether charge is conserved in the absence of radiation. 
Moreover, at non radiative cuts, the algebroid can be restricted to the wedge symmetry algebra studied in \cite{Cresto:2024fhd}.  
The key ingredient for our construction is to consider field and time dependent symmetry parameters constrained to evolve according to equations of motion dual to (a truncation of) the asymptotic Einstein's equations.
This result then guarantees that the underlying symmetry algebra is also represented canonically.\\

\ni \textbf{Keywords:}
Gravity, General Relativity, Asymptotic Symmetries, Asymptotically Flat Spacetimes, Celestial Holography, Noether Realization, Canonical Representation, Wedge Algebra, $w_{1+\infty}$ Algebra, Twistor Theory, Good Cut Equation.
\end{abstract}

\newpage
\tableofcontents
\newpage

\section{Introduction}

For more than a century, Emmy Noether has guided us thanks to her theorems relating symmetries and conserved quantities \cite{noether1971invariant}.
Since the seminal work of Bondi, van der Burg, Metzner and Sachs \cite{Bondi:1962px, Sachs:1962wk}, we know that Asymptotically Flat Spacetimes (AFS) do not reduce to Minkowski space at infinity. 
In particular, the symmetry group preserving the asymptotic structure was originally  recognised to be the BMS group. Recently it was shown that it is necessary to consider extensions such as the eBMS \cite{Barnich:2010eb, Barnich:2011mi, Barnich:2016lyg}, GBMS \cite{Campiglia:2014yka, Compere:2018ylh, Campiglia:2020qvc, Grant:2021sxk} and BMSW \cite{Freidel:2021fxf, Freidel:2021qpz} groups and view BMS as a residual unbroken symmetry group \cite{Freidel:2024jyf}.  
 These symmetry groups are infinite-dimensional, and the associated Noether charges depend on the point of the celestial sphere.
This led to the notion of charge \textit{aspect} which represents, through Noether's theorem, the charge density on the celestial sphere. 
It is now understood that asymptotic symmetries can be directly related to the broader concept of \emph{corner symmetries}, which appear as global symmetries associated with gauge symmetries \cite{Donnelly:2016auv, Freidel:2020xyx, Ciambelli:2021nmv}. 
These corner symmetries are associated with codimension $2$ surfaces, on which causal diamonds can be attached and which represent entangling surfaces.

More recently, in the asymptotically flat context, the much bigger algebra $Lw_{1+\infty}$ (see \cite{Shen:1992dd} for a review of the mathematical and historical developments of $W_N$ algebras) found its relevance in the analysis of soft gravitons scattering \cite{Guevara:2021abz, Strominger:2021mtt, Himwich:2021dau} in the context of celestial holography.
This symmetry was also found to be exact in self-dual gravity \cite{Ball:2021tmb} and naturally explained via twistor methods in connection with Penrose's non-linear graviton construction \cite{Penrose:1976js, Adamo:2021lrv, Adamo:2021zpw}.
The projection of its twistor action on the asymptotic data on $\scri$ and the correspondence with the canonical realization of that symmetry at quadratic order \cite{Freidel:2021ytz, Geiller:2024bgf} was worked out in \cite{Donnay:2024qwq}.

Such symmetries have been referred to as higher spin symmetries because the symmetry generators—originally the Bondi mass aspect and the angular momentum aspect, generating respectively super-translations (spin 0) and sphere diffeomorphisms (spin 1 transformation)—included a spin 2 charge aspect. 
Allowing for such a spin 2 symmetry transformation then forces the inclusion of the whole infinite tower of positive spin in order to close the algebra.
This paper focuses on the canonical asymptotic representation of the higher spin symmetry algebra at $\scri$. 
We follow the work of \cite{Freidel:2021dfs, Freidel:2021ytz, Freidel:2023gue, Geiller:2024bgf}, which formalized that there exist charges aspects of spin $s$ that one can build from the asymptotic gravitational data. These results generalized to higher spin the classic construction of \cite{Strominger:2013jfa, Strominger:2014pwa, Campiglia:2014yka, Campiglia:2015yka, Kapec:2014opa, Pasterski:2015tva} for spin $0$ and $1$.
These charges are related to subleading soft theorems and they are known to represent corner data that incorporate the necessary information to reconstruct the bulk spacetime \cite{Freidel:2021ytz, Geiller:2024bgf}.
They are also related to gravitational multipoles \cite{Compere:2022zdz}.
Besides, they furnish an operational description of General Relativity, where the metric components are recast as a set of non-commutative  observables satisfying a well-defined Poisson algebra. Moreover, they are associated to a higher spin generalization of the memory effect (see \cite{braginskii1985kinematic, braginsky_gravitational-wave_1987, PhysRevLett.67.1486, PhysRevD.45.520, PhysRevD.46.4304} for the displacement memory effect and \cite{Pasterski:2015tva, Compere:2016gwf, Compere:2018ylh, Compere:2019odm, Nichols:2017rqr, Nichols:2018qac, Flanagan:2018yzh, Flanagan:2019ezo, Grant:2021hga, Grant:2023ged, Seraj:2021rxd, Seraj:2022qyt, Seraj:2022qqj, Faye:2024utu} for higher spin generalizations). 
This suggests that a proper non-perturbative understanding of the higher spin charges would then be invaluable for quantization. Finally, the soft components of these charges and their dual can be used to construct a discrete asymptotic basis of states \cite{Freidel:2022skz, Cotler:2023qwh}. 
\medskip

 Although there is evidence, for spin $2$ only,  that the higher spin charges could be understood as overleading diffeomorphism changing the boundary condition \cite{Campiglia:2016jdj, Campiglia:2016efb, Horn:2022acq}, a fully consistent and non-perturbative Noether analysis is missing. 
In other words, the charges have \textit{not} been built as Noether charges derived from a symmetry action on the gravitational phase space, but from consistency with the OPEs and soft theorems. The detailed derivation 
related the Ward identities for the spin $s$ charge aspect $\tQ_s$ with a tower of (sub)$^{s}$-leading conformally soft theorems.
See \cite{Strominger:2017zoo, Hamada:2018vrw, Raclariu:2021zjz} and references therein for the interpretation of soft theorems as conservation laws.

The goal of this paper is to remedy this situation and give a first principle phase space derivation of the tower of higher spin charges from the Noether's theorem applied to the gravitational phase space.
Understanding the tower of higher spin charges in a systematic, non-perturbative way, where one can leverage Noether's theorems, is primordial to determine the full set of symmetries of General Relativity (GR) and understand to which extent they constraint the gravitational dynamics and the quantum $\cS$-matrix \cite{Strominger:2013jfa, ashtekar_angular_1979, Prabhu:2019fsp, Prabhu:2021cgk, Strominger:2017zoo, Hamada:2018vrw, Raclariu:2021zjz}.

The goal of this project is, therefore, to show that  the higher spin symmetries can be realized canonically and non-perturbatively via Noether charges. 
One of the many challenges one faces is that the action of the symmetry generators on the gravitational phase space is \emph{non linear} beyond spin $2$ and this non-linearity grows with the spin.  
The fact that this non-linear action closes into an algebra up to quadratic order was shown in \cite{Freidel:2021ytz, Freidel:2023gue}. 
These promising results looked rather miraculous and were due to the combination of highly non-trivial hypergeometric identities. Quite remarkably,
the closure of the algebra  beyond the wedge required to include the commutator at cubic order, beyond the quadratic hard charges. This suggests that the validity of the higher spin symmetry algebra controls some of the non-linearity of Einstein's Equations (EE)---see also \cite{Hu:2022txx, Hu:2023geb}.
The validity of the algebra is so non-trivial that it suggested that a \textit{non-perturbative} description should exist.
This was the original motivation for the present work.

In this paper, we achieve three connected results: We identify a non-trivial higher spin symmetry algebroid whose structure constants depend explicitly on the shear and which is proven to satisfy the Jacobi identity. 
We show that the non linearity of higher spin symmetry on the gravitational phase space can be recast into a simple equation of motion for the transformation parameters while the action itself involves only spin $0$, spin $1$ and spin $2$ generators.
This radical simplification allows us to construct, to all orders in $\GN$, the gravitational representation of the higher spin generators. 
The fact that these generators satisfy the algebra non-perturbatively then follows from the power of Noether's theorem and the covariant phase space formalism \cite{kijowski1976canonical, Wald:1999wa, crnkovic1987covariant, ashtekar1991covariant, lee1990local, barnich1991covariant, Freidel:2020xyx, Freidel:2021cjp}.
Finally, we establish that the renormalization procedure developed in \cite{ Freidel:2021ytz, Geiller:2024bgf} for the charge simply amounts to evaluating the transformation parameters in terms of their initial conditions.

To understand the origin of the current work, let us review the procedure of \cite{Freidel:2021dfs, Freidel:2021ytz} while introducing the rest of the relevant literature along the way.
The starting point was to consider a set of spin-weighted functionals $\tQ_s(u,z,\bz)$, of spin $s\geqslant -2$ on $\scri$ that was proven to transform homogeneously under the asymptotic $\bmsw$ algebra \cite{Freidel:2021fxf, Freidel:2021qpz}, the symmetry algebra of null infinity.
They were related by a set of evolution equations that take the simple recursive form
\begin{equation}
    \pa_u\tQ_s=D\tQ_{s-1}+(s+1)C\tQ_{s-2}, \qquad s\geqslant 0,\label{intro1}
\end{equation}
with $C$ the asymptotic shear.
For $s\leqslant 3$, these are the \textit{exact} asymptotic EE \cite{newman_new_1968, Freidel:2021qpz, Freidel:2021ytz}. 
For instance, the $s=0$ equation can be recast into the Bondi mass loss formula. 
In general the higher spin charges aspects  $\tQ_s$ 
are  related to the asymptotic expansion of the Weyl tensor---$\Psi_0^{(s)}/r^{s+5}$ in the Newman-Penrose notation, with $r$ the radial direction; see \cite{newman_new_1968} for the original NP work but also \cite{Barnich:2019vzx, Freidel:2021ytz} and recently \cite{Geiller:2024bgf} for the most up-to-date account. 
For spins 4 and higher, \eqref{intro1} represents a truncation of EE on $\scri$. 
The spacetime interpretation of this truncation still remains mysterious, although it has been recently shown to arise naturally from a twistor formulation of self-dual gravity \cite{Kmec:2024nmu}.

One of the key ingredients of our construction is to introduce Carrollian smearing parameters $\tau_{s}(u,z,\bz)$ on $\scri$, of spin $-s$ with $s\geqslant -1$. 
These parameters are dual to the spin $s$ charges aspects and such that the pairing $Q_s[\tau_s]$ defines a scalar charge after integration on the sphere.  
We constraint the time evolution of $\t{s}$ to follow the dual equations of motion (EOM)
\begin{equation}
    \pa_u\t{s}= D\t{s+1}- (s+3) C \t{s+2} := (\Dcal \tau)_s, \label{intro4}
\end{equation}
for $s\geq 0$. 
These evolution equations ensure that the generating functional  
\be 
Q_\tau =\sum_{s} Q_s[\tau_s]
\ee 
is conserved in the absence of radiation. 
The RHS of the evolution equation \eqref{intro4} reveals a generalization of the covariant derivative on the sphere, which depends explicitly on the shear and plays an essential role in our construction. 
This derivative $\Dcal$ encompasses the algebraic non-linearity due to the presence of a non-vanishing shear.

We establish that the action of the symmetry algebra on the gravitational phase space, represented by the shear, is given by the following differential action
\begin{equation}\label{intro2}
    \dt C=\t0\pa_uC-D^2\t0+2DC\t1+3CD\t1-3C^2\t2.
\end{equation}
This action, supplemented by the dual equations of motion, reproduces to quadratic order in the shear the canonical action of higher spin symmetry derived in \cite{Freidel:2021ytz}.
Remarkably, we also prove that the commutator of this action closes thanks to the validity of the dual EOM. This reveals a shear-dependent bracket which is given in terms of the covariant derivative $\Dcal$, as a deformation of a shifted Schouten-Nijenhuis bracket for symmetric tensors on 2d manifolds \cite{Schouten, SNbracket},
\begin{equation}
[\tau,\tau']^C_s=\sum_{n=0}^{s+1}(n+1)\big(\t{s}(\Dcal \tau')_{s-n}-\tp{s}(\Dcal \tau)_{s-n}\big). \label{intro3}
\end{equation}
Notice two important features about the transformation \eqref{intro2}: Firstly, when written in terms of $\tau$, the symmetry is realized non-linearly due to the presence of the $C^2$ term associated with the spin $2$ symmetry.
Secondly, the symmetry transformation of the shear contains only the parameters $\t0,\t1$ and $\t2$.
However, they themselves contain arbitrary high spin symmetry parameters when expressed in terms of the celestial symmetry parameters given by the initial value  $\T{s}:=\tau_s(u=0)$.
When expressed in terms of the latter, the associated symmetry transformation $\delta_{\T{s}}C$ becomes polynomial in $C$ with a degree growing linearly with $s$ and also non-local along the null time $u$ direction. 
The soft component of the action is given by $\delta_{T_s}^{\mathsf{S}}=\frac{u^s}{s!} D^{s+2} T_s$. 
By definition, it vanishes when $T_s$ belongs to the wedge subalgebra, which forms a Lie algebra when $C=0$. 
Our analysis also reveals that at any non-radiative cut of $\scri$, we can define a deformation of the wedge algebra labelled by the value of the shear at the cut $\sigma = C|_{u=cst}$. 
This algebra, denoted $\Wcal_\sigma(S)$ was shown in \cite{Cresto:2024fhd} to be a Lie algebra.\footnote{The covariant wedge space precisely becomes the restriction of the anchor map $\delta$ to its kernel, which defines an algebra within an algebroid.}

To go beyond the wedge and allow for charges that interpolate between different non-radiative algebras $\Wcal_\sigma(S) \to \Wcal_{\sigma'}(S)$ requires the introduction of a symmetry algebroid bracket $\lbr\cdot\,, \cdot \rbr$ build from the C-bracket.
Our main result is that $\Qt$ furnishes a \emph{non-perturbative Noether representation of this algebroid} on the Ashtekar-Streubel gravitational phase space \cite{ashtekar1981symplectic}.
It also ensures that $\Qt$ forms a representation of the symmetry algebras $\Wcal_\sigma(S)$ at non-radiative cuts.
This result is \emph{non-perturbative} and guarantees that the symmetry is realized \emph{non-linearly} on the gravitational phase space to all order in $\GN$.

The algebroid framework we introduce in this paper is essential to carry over the calculations efficiently.
In addition to uncovering a clear algebraic structure that gives access to a non-perturbative description, we think that this framework naturally accommodates the idea of radiation as a transition between two non-radiative states. 
Indeed, if $\Wcal_\sigma(S)$ and $\Wcal_{\sigma'}(S)$ represent the symmetry algebras at two different non-radiative cuts of $\scri$, and that radiation was registered on $\scri$ in the interval, then $\sigma\neq\sigma'$ due to the memory component carried by this radiation.
A Lie algebroid describes how these algebras relate to one another and provides a notion of path in that space that we can physically interpret (at least classically) as a transition between the two non-radiative spacetimes.
\medskip

While this work was in completion, the Oxford group published a beautiful paper \cite{Kmec:2024nmu} which obtained independently results in agreement with ours. They showed, starting from a twistor space formulation of self-dual gravity, that the twistor symmetry charges satisfy, after integration over the fiber, the equation \eqref{intro1}. 
They also proved that the Noether's theorem ensures a realization of the twistor symmetry beyond the wedge on the self-dual phase space.
The dual equations of motion arise there from a gauge fixing condition that projects twistor functionals onto spacetime functionals. 
One apparent difference between their work and ours is that the symmetry bracket used in twistor theory is simply a Poisson bracket on the twistor fiber and is, therefore, independent of the shear. 
In our work, the symmetry bracket that stems from the spacetime analysis is shear-dependent. 
In section \ref{sec:twistor}, we show that, quite remarkably, the shear non-linearity can be reabsorbed through the introduction of an extra spin $1$ variable and after using the dual EOM.
It should be clear to the reader that our work, which focuses on the spacetime and canonical formulation, was done independently of \cite{Kmec:2024nmu}. 
It shows that spinor and canonical approaches lead to similar results from very different perspectives. 
For instance, the canonical framework allows for two series of charges: the one conserved when $\dot{\bC}=0$, which includes the Bondi mass, and the covariant charges conserved when  $\ddot{\bC}=0$. 
The latter are the ones that appear naturally from the twistor analysis. Both are represented canonically and differ only by how one treats the spin $-1$ transformation.
\medskip

The paper is organized as follows:
In section \ref{sec:preliminaries} we introduce our conventions of notation and recall the definition of Carrollian fields.
Section \ref{sec:Wedge} then summarizes the main results of \cite{Cresto:2024fhd}.
In section \ref{sec:celestialToScri}, we introduce time and generalize the objects discussed in \cite{Cresto:2024fhd} to adapt them to $\scri$.
We define the dual equations of motion in \ref{Sec:DualEOM} and show that they are consistent with the algebroid bracket $\lbr\cdot\,,\cdot\rbr$ in \ref{sec{timeev}}.
We include a reminder of the necessary algebroid framework in \ref{Sec:Liealg}.
Section \ref{secAlgebroid} deals with the main object of interest in this work, namely the $\A$-algebroid.
Then \ref{sec:InitialCondition} completes \ref{secAlgebroid} with a discussion about the sub-algebroids $\Ao$ and $\Ah$.
We describe the canonical realization of $\Ao$ and $\Ah$ in section \ref{SecNoether} and prove that the symmetry algebroids are realized as Noether charges on the gravitational phase space.
We also give a definition of the surface charges for arbitrary cuts of $\scri$ in \ref{sec:ArbitraryCut}.
We discuss the filtration and the gradation of $\A$, together with a class of solutions of the dual EOM in section \ref{Sec:SolDualEOM}.
This allows us to define the renormalized charges in \ref{sec:RenormCharge} and to compute the action of the associated Noether charge onto the shear in \ref{SecComparison}.
We show that we recover previously known results as a particular application of our formalism.
In section \ref{Sec:algebroidTs}, we make the connection with the covariant wedge algebra from \cite{Cresto:2024fhd}.
Finally, we discuss the relationship with twistor theory and Newman's good cut equation \cite{newman_heaven_1976} in section \ref{sec:twistor}.
We conclude and give an outlook in section \ref{secGR:conclusion}.
Besides the main text, we gather the technical proofs of our results in an extensive collection of appendices. 
We also include a glossary in section \ref{secGR:glossary}.

\section{Preliminaries \label{sec:preliminaries}}

In the following we work on $\scri = \R \times S$, where $S$ denoted a 2d complex manifold with a complex structure. 
We use the usual Bondi coordinates $(u,r,z,\bz)$ together with the null dyad $(m,\bm)$ on the sphere. $S$ can be the regular sphere $S_0$ or the sphere minus $n$ number of punctures denoted $S_n$.  
We denote $D= m^A D_A$ the covariant derivative along $m$ preserving the complex structure and $\bfm{\epsilon}_S$ the area form on $S$. 
We refer to \cite{Cresto:2024fhd} for extra details.

\subsection{Carrollian fields}
\label{Sec:Carroll}

On $\scri$, the symmetry generators include the supertranslations $T(z,\bar{z}) \pa_u$. Moreover, the action of sphere diffeomorphisms on Carrollian fields is given by the vector field  $\mathcal Y \equiv \frac{1}{2}\big(DY+\bD\bY\big)u \pa_u+\bY\bD+YD$, which includes a boost component  determined by the components $Y$ and $\bY$ of an actual diffeomorphism of $S$.
The combination of supertranslations and sphere diffeomorphisms form the generalized BMS algebra $\gbms$ \cite{Barnich:2011mi,  Campiglia:2016efb, 
Compere:2018ylh,Freidel:2021qpz}.
If we demand in addition that the vector field is holomorphic, i.e. $\bD Y=0$ then we get the algebra $\mathfrak{ebms}(S)$. We have that $\mathfrak{ebms}(S_0)=\mathfrak{sl}(2,\C)$ and $\mathfrak{ebms}(S_2)= \mathrm{Vir}\times \overline{\mathrm{Vir}}$.

The Carrollian field $\Phi(u,z,\bz) \in \Ccar{\dCar,s}$ of Carrollian weight $\delta$ and helicity (or spin-weight) $s$ admits an action of supertranslations $\delta_T\Phi= T\pa_u \Phi$. 
It also transforms under sphere diffeomorphisms as
\be
\delta_{\mathcal Y}  \Phi 
= \big(YD + \tfrac12 (\delta  + u\pa_u + s) DY   \big)\Phi
+ \left(\bY\bD+\tfrac12 (\delta + u\pa_u  - s)\bD\bY \right) \Phi .
\label{sDeltaDefCarroll}
\ee
We can similarly define the notion of celestial primary fields $\Ccel{(\Delta,s)}$ on the 2-sphere (or the punctured 2-sphere), where $\Delta$ is the celestial conformal dimension and $s$ still the helicity.
By definition a $\phi(z,\bz) \in \Ccel{(\Delta,s)}$ transforms under sphere diffeomorphisms $\mathcal Y\in \mathrm{Diff}(S)$ as 
\be 
\delta_{\mathcal Y}  \phi
&=\big(YD+\tfrac12(\Delta + s) DY\big)\phi+\big(\bY\bD + \tfrac12(\Delta - s)\bD\bY\big) \phi. \label{sDeltaDef}
\ee
We see that the connection between the Carrollian and celestial weights is simply that 
\be 
\hat\Delta = \delta + u\pa_u.
\ee 
$\hat\Delta$ is the operator which once diagonalized in the celestial basis, has eigenvalues $\Delta$. For instance, if $\Phi_\delta\in \Ccar{\dCar,s}$, we can construct a celestial field $\phi_{\Delta}\in \Ccel{(\Delta,s)}$ using the time Mellin transform \cite{Freidel:2021ytz,Freidel:2022skz}:
\begin{equation}
    \phi^\pm_\Delta (z,\bz) = 
    (\mp i)^{(\Delta+1-\delta)} \Gamma(\Delta+1-\delta)
    \int_{-\infty}^\infty\!\rd u\,\frac{\Phi_\delta(u,z,\bz)}{( u \pm i \epsilon)^{(\Delta-\dCar+1)}}. \label{CelCarMap}
\end{equation}
This means that a Carrollian field can be expressed as a direct sum (or direct integral) of celestial fields after diagonalization of the boost operator. This decomposition is either done in terms of the principal \cite{Pasterski:2017kqt} or discrete series of representation \cite{Freidel:2022skz, Cotler:2023qwh, Mitra:2024ugt}.  

Note that from our definition \eqref{sDeltaDefCarroll} we easily infer that $\pa_u$ is an operator of weight $(1,0)$,\footnote{Use that $ \delta_{\mathcal Y}(\pa_u\Phi)= \pa_u\big(\delta_{\mathcal Y}\Phi\big)$.} while $D$ is an operator of weight $(1,1)$.

Once a conformal structure is chosen one  introduces homogeneous coordinates $\lambda_\alpha$ on the sphere such that  $z= \lambda_1/\lambda_0$. 
One can understand the Carrollian primaries in $\Ccar{\dCar,s}$ as homogeneous sections over $\scri$ of degree $\big(-(\delta+s),-(\delta-s)\big)$. They transform as \cite{Eastwood_Tod_1982}
\begin{equation}
    \Phi(|a|^2 u, a \lambda_\alpha,\bar{a} \bar{\lambda}_{\dot{\alpha}})=a^{-(\delta+s)}\bar a^{-(\delta-s)}\Phi(u, \lambda_\alpha, \bar{\lambda}_{\dot{\alpha}}),\qquad a\in\C^*.
\end{equation}
In the following we will use that there is a natural projection 
\be 
\Ccar{\dCar,s} & \to \Ccel{(\dCar,s)} \cr 
\Phi &\mapsto \phi(z,\bz) = 
\Phi(u=0,z,\bz),
\ee 
which follows from a choice of embedding of $S$ into $\scri$.
This projection preserves the conformal dimension $\Delta=\delta $ since the boost operator action vanishes $(u \pa_u \Phi)|_{u=0} = 0$.\footnote{This projection can also be written in terms of \eqref{CelCarMap} as a limit $\lim_{\Delta \to \delta}(\phi^+_\Delta) $. }

\subsection{The $\tV$-space \label{secGR:tVspace}} 

We introduce the following space of  Carrollian  fields:
\be \label{VCcardef}
\tV_s\equiv  \Ccar{-1,-s}, \qquad 
\tV(\scri)\equiv \bigoplus_{s=-1}^\infty\tV_s.
\ee 
$\tV(\scri)$ is a graded vector space where the spin  $s$ denotes the grade. 
The gradation of $\tV(\scri)$ induces a filtration denoted 
\be
\tV^s\equiv \tV^s(\scri) := \bigoplus_{n=-1}^s \tV_n.
\ee
Finally we introduce the space $\V(S)$ in a similar fashion as $\tV(\scri)$, except that we now work with celestial weights (cf. subsec. \ref{Sec:Carroll}):
\be \label{Vsccel}
\V_s\equiv  \Ccel{(-1,-s)}, \qquad 
\V(S)\equiv \bigoplus_{s=-1}^\infty\V_s.
\ee 
In the following we denote by $T$ the elements of $\V(S)$ and by $\t{}$ the elements of $\tV(\scri)$. 
We denote by $(T)_s\equiv T_{s}$ and $(\tau)_s\equiv \t{s}$ their respective grade $s$ element.
In other words, $\tau$ corresponds to the series of spin weighted fields  $(\t{s})_{s+1\in\N}$ and $\t{s}\in \tV_s$ is the result of the projection onto the subspace of degree $s$. 
There is also a natural inclusion map $\iota: \tV_s \to \tV(\scri)$ which 
takes the element $\t{s}$ and sends it to the series $\iota(\t{s})=(0,\ldots,0,\t{s},0,\ldots)$.

\section{Time and Algebroid}

In this section we construct an algebroid bracket on $\scri$ which generalizes the covariant wedge bracket from our companion paper \cite{Cresto:2024fhd}.

\subsection{The Wedge bracket }\label{sec:Wedge}

In our companion paper we constructed a Lie algebra $\Wcal_\sigma(S)$ which generalizes the wedge algebra. 
It is labelled by an element $\sigma \in \Ccel{(1,2)}$ and this construction relies on the introduction of a bracket $[\cdot\,,\cdot]^\sigma$ acting on $\V(S)$ and defined by\footnote{This is a deformation of the $\W$-bracket, defined as $[\cdot\,,\cdot]^\W\equiv\big([\cdot\,,\cdot]^\sigma\big)\big|_{\sigma=0}$. \label{footWbracket}} 
\begin{equation}
[T,T']^\sigma_s:=\sum_{n=0}^{s+1}(n+1)\big(\T{n}D\Tp{s+1-n}-\Tp{n}D\T{s+1-n}\big)-(s+3)\sigma \big(\T0\Tp{s+2}-\Tp0\T{s+2}\big). \label{Wbracket}
\end{equation}
It was shown in \cite{Cresto:2024fhd} that this bracket satisfies the Jacobi identity when $T$ belongs to a subspace of $\V(S)$ called the covariant wedge space and denoted $\W_\sigma(S)$.
To describe the latter, it is necessary to introduce the key element $\hHam\in \V^0(S)$ given by $\hHam=(DG, 1, 0,0,\cdots)$, where $G \in \V_0$ is the Goldstone field connected to $\sigma$ by 
\be 
D^2 G =\sigma. \label{Goldstone}
\ee 
In other words, $\hHam $ is such that $({\hHam})_{-1}=DG$, 
$({\hHam})_0=1$ and $({\hHam})_s=0$ for $s>0$.
As we are about to see, it is convenient to think of $\hHam$ as a Hamiltonian generator and define its adjoint action $\ad \hHam: \V(S) \to \V(S)$ to be the linear operator $\ad \hHam(T)=[\hHam, T]^\sigma$ with grade $s$ element given by 
\be
\big(\ad \hHam(T)\big)_s=[\hHam, T]_s^\sigma = D T_{s+1} -\sigma (s+3) T_{s+2}\equiv(\Dcal T)_s.
\ee 
Furthermore, we define $(\Dcal T)_{-2}:=D\T{-1}-\sigma\T0$. 
We also showed that the $\sigma$-bracket has a convenient expression in terms of the covariant derivative $\Dcal$, namely
\begin{equation}
[T,T']^\sigma_s:=\sum_{n=0}^{s+1}(n+1)\big(\T{n}(\Dcal\Tp{})_{s-n}-\Tp{n}(\Dcal T)_{s-n}\big). \label{WbracketDcal}
\end{equation}
The covariant wedge is then defined as  
\be \label{defCovWedgeSpaceGR}
\W_\sigma(S) = \Big\{ T\in \V(S) ~\big|~  D\big(\ad^n\hHam (T)\big)_{-1} =  \sigma \big(\ad^n\hHam (T)\big)_{0},\, n \in \mathbb{N}\Big\}.
\ee
Note that the covariant wedge condition is also compactly written as $(\Dcal^{s+2}T)_{-2}=0,\,s\geq -1$.
One of the main result of \cite{Cresto:2024fhd} is that 
$\Wcal_\sigma(S)=\big(\W_\sigma(S),[\cdot\,,\cdot]^\sigma \big)$ is a Lie algebra. 
The Jacobi identity for the $\sigma$-bracket follows from the fact that $\hdT \sigma =0$, where 
\begin{align}
-\hdT \sigma =(\Dcal^2 T)_{-2}= D [ \hHam, T ]^\sigma_{-1} -\sigma [\hHam, T]_0^\sigma 
= D^2 T_0 -2 D(\sigma T_1) - \sigma DT_1 + 3\sigma^2 T_2. 
\end{align}

When $\sigma=0$ we recover the usual wedge algebra with bracket $[T,T']^0\equiv[T,T']^\W$ (the shifted SN bracket mentioned in the introduction) and the familiar wedge condition \cite{Strominger:2021mtt, Mason:2022hly} implying that $(D^{s+2}T)_{-2}= D^{s+2} T_{s}=0$ for $s\geq -1$.

Moreover, it was also shown that $\Wcal_\sigma$ is isomorphic as a Lie algebra to $\Wcal_0\equiv\Wcal$ with the isomorphism given as the path ordered exponential of an adjoint action,
\begin{align}
\mTG: \Wcal_\sigma &\to \Wcal_0 \cr
  T &\mapsto \overrightarrow{\mathrm{Pexp}}\left(\int_0^{1}\mathrm{ad}_{\lambda \sigma} \hG_\lambda\,\rd \lambda \right)(T). \label{PathOrderedExp}
\end{align} 
where $\hG_\lambda=\lambda\hG$ and $\hG\in \V^0$ is such that $ \hG_n= 0$ for $n\neq 0$ and $\hG_0 = G$.

\subsection{From celestial to sky  \label{sec:celestialToScri}}

The  Lie algebra $\Wcal_\sigma(S)=\big(\W_\sigma(S),[\cdot\,,\cdot]^\sigma \big)$ is a celestial symmetry algebra supported on the 2d surface $S$. It depends on the value of the shear $\sigma$ on $S$.  
In order to understand how this symmetry acts on the gravitational phase space one needs to include time into the picture and extend this symmetry from $S$ to $\scri$. 
One also needs to go beyond the wedge in order to allow for the shear to be time dependent. 
Quite remarkably the two issues are connected and there exists a choice of time dependence for the transformation parameters that allows to go beyond the wedge. This is what we now develop.

The first step consists in  promoting the transformation variables $T_s(z,\bz) \in \V(S)$ to time dependent variables $ \tau_s(u, z,\bz) \in \tV(\scri)$ where $\tV(\scri)$ is  introduced in \eqref{VCcardef}. This change of support is reflected in the change of notation $T\to \tau$.
As we have seen, $\tV(\scri)= \bigoplus_{s=-1}^\infty \tV_s$ is graded, and we denote by $\t{}$ the elements of $\tV(\scri)$ and by $\t{s}$ its grade $s$ elements.

Now that we introduced a time dependency in the transformation parameters, we need to allow $\sigma$ to also depend on time. 
For clarity, we shall henceforth use $C(u,z,\bz)$ to denote the time dependent version of the parameter $\sigma(z,\bz)$. 
We did not choose the letters $\sigma$ and $C$ without reason.
In section \ref{SecNoether}, $C$ will play the role of the asymptotic shear on the gravitational phase space. In particular this means that we take $C\in\Ccar{1,2}$.

Given these data, we  can naturally extend the $\sigma$-bracket to $\scri$:
given $\tau,\tau'\in\tV(\scri)$  we define the bracket $[\cdot\,,\cdot]^C$ to be the same as the bracket $[\cdot\,,\cdot]^\sigma$ after replacing $\sigma \to C$, i.e.
\begin{equation}
[\tau,\tau']^C_s:=\sum_{n=0}^{s+1}(n+1)\big(\t{n}D\tp{s+1-n}-\tp{n}D\t{s+1-n}\big)-(s+3)C\big(\t0\tp{s+2}-\tp0\t{s+2}\big). \label{CFbrackettau}
\end{equation}
For later convenience, we refer to the two pieces as\footnote{We refer to $\paren{\cdot\,,\cdot}$ as the Dali-bracket, as a reference to Dali's paintings with melting watches.}
\bs
\begin{align}
    [\tau,\tau']_s^{\tV}&:=\sum_{n=0}^{s+1}(n+1)\big(\t{n}D\tp{s+1-n}-\tp{n}D\t{s+1-n}\big) & & \hspace{-1.5cm}\in\tV_s, \label{tVBracket}\\
    \paren{\tau,\tau'}_s &:= (s+3)(\t0 \tp{s+2}-\tp0\t{s+2}) & & \hspace{-1.5cm}\in \Ccar{-2,-s-2}, \label{DaliBracket}
\end{align}
\es
so that $[\tau,\tau']^C=[\tau,\tau']^{\tV}-C\paren{\tau,\tau'}$.
Note that the term proportional to $C$ introduces a field dependency on $C$ in the structure constants of the bracket. 
This field dependency encodes the algebra deformation in the presence of shear and reflects the  non-linearity of the Einstein's theory. 
It will prove convenient to recast this non-linear dependence into a covariant derivative operator. 
Since the context is clear, we keep the symbol $\Dcal$ for the $C$-bracket adjoint action of $\Ham$, where the element $\Ham:=(0,1,0,0,\cdots)$ is such that $\eta_n=\delta_{n,0}$. 
This means that\footnote{We use $\Dcal$ and $\adC\Ham$ interchangeably, with appropriate multiplication by $\Ham_0\in\tV_0$ in order for the Carrollian boost weight to be consistent; see the corresponding discussion in \cite{Cresto:2024fhd}.}
\be\label{DcalC}
\Dcal\tau:=[\Ham,\tau]^C, \qquad
(\Dcal\tau)_s = D\tau_{s+1} - C(s+3) \t{s+2}.
\ee 
Although $\tau_s$ is defined only for $s\geq -1$, it will be convenient to use the previous definition of $\Dcal$ for $s=-2$, where by definition we denote $(\Dcal\tau)_{-2} := D\t{-1} - C\t0$.
Notice that the  value of $\Ham_{-1}$ is irrelevant for the definition of $\Dcal$, since the degree $-1$ elements are central for the $C$-bracket.\footnote{ This implies that the adjoint action of $\Ham$ and $\hHam$ are the same. The condition $D(\hHam)_{-1}=\sigma=\sigma(\hHam)_0$ was needed for the wedge but is not necessary here so we chose $\Ham_{-1}=0$ for simplicity.}

Now notice that $\sum_{n=0}^{s+2}(n+1)(s+3-n)\t{n}\tp{s+2-n}$ is symmetric under the exchange of $\tau$ and $\tau'$ (using the change of variable $n\leftrightarrow (s+2-n)$).
Isolating the term $n=s+2$ in the sum  implies the important identity:
\begin{equation}
    (s+3)\big(\t0\tp{s+2}-\tp0\t{s+2}\big)=\sum_{n=0}^{s+1}(n+1)(s+3-n)\big(\t{n}\tp{s+2-n}-\tp{n}\t{s+2-n}\big).
\end{equation}
This allows us to recast the  term $C\paren{\tau,\tau'}_s$ as a deformation of the sphere derivative $D\t{s+1-n}\to D\t{s+1-n}-(s+3-n)C\t{s+2-n}=(\Dcal \tau)_{s-n}$.
The $C$-bracket \eqref{CFbrackettau} then also takes the form
\begin{equation}
    \boxed{\,[\tau,\tau']^C_s= \sum_{n=0}^{s+1}(n+1)\big(\t{n}(\Dcal\tp{})_{s-n}-\tp{n}(\Dcal\t{})_{s-n}\big).\,} \label{CFbrackettauDcal}
\end{equation}
This shows that the (field dependency) of the $C$-bracket can remarkably entirely be recast in the deformation $D\to \Dcal$ of the holomorphic derivative.
The analysis of the Jacobi identity for the $C$-bracket is similar to the one done in \cite{Cresto:2024fhd} for the $\sigma$-bracket since the time dependence plays no role.
For completeness, we also report another proof in appendix \ref{AppJacobiDcal}, that uses the form \eqref{CFbrackettauDcal} of the bracket.
We find that
\be \label{JacCFC}
\big[\tau,[\tau',\tau'']^C\big]_s^C \cyc -\hdt C \paren{\tau',\tau''}_s,
\ee
where $\cyc$ denotes the equality after summing over cyclic permutation of $(\tau, \tau', \tau'')$.
The hatted variation is given by
\be \label{hataction}
\hdt  C = -D^2 \tau_0 +2 D(C \tau_1) + C D\tau_1 - 3C^2 \tau_2=-(\Dcal^2\tau)_{-2}.
\ee 
Note that \eqref{JacCFC} is not the most general way to write the Jacobi identity violation. 
We  have the freedom to add a term proportional to $\t0$ in the expression of $\hdt C$ since the following cyclic combination vanishes identically:
\begin{equation}
    \t0 \paren{\tau',\tau''}_s\cyc 0, \quad \forall \,s.
\end{equation}
This justifies the addition of a term $\alpha \t0 N$, $\alpha\in\R$, to the initial transformation $\hdt$. 
This additional term is however constrained to have definite Carrollian weights. In order for $\tau_0 N$ to have Carrollian weights $(1,2)$ we need $N$ to be in $\Ccar{2,2}$.
Next, remember that $\pa_u$ is an operator of weight $(1,0)$.
Therefore, the only local functional with the right weight is
\begin{equation}
    \boxed{N:=\pa_u C}.
\end{equation}
This allows to conclude that we have $\big[\tau,[\tau',\tau'']^C\big]_s^C \cyc -\dt C \paren{\tau',\tau''}_s$ with 
\be \label{dtauCalpha}
\dt C = \alpha \t0 \pa_u C + \hdt C.
\ee 
This possibility is a crucial difference between working solely on $S$ and working on $\scri$: the field variations $\hdT\sigma$ and $\dt C$ are \textit{not} the same. 
 In the following we assume that $\alpha\neq 0 $ since the case $\alpha=0$ was thoroughly investigated in \cite{Cresto:2024fhd}. 
 When $\alpha\neq 0$  we can always rescale the time coordinate $u \to \alpha u$ to ensure that $\alpha=1$.
This means that the expression for $\dt C$ is
\begin{equation}
    \boxed{\dt C = \t0  N -D^2\t0+2D(C\t1)+C D\t1-3 C^2\t2}\quad\in\,\Ccar{1,2}, \label{dtauC}
\end{equation}

\subsection{Time as a canonical transformation \label{Sec:DualEOM}}

So far we have left arbitrary the time dependence of $\tau_s$. As we will see, in order to have a well defined action of  symmetries generated by $\tau$ on the gravitational phase space, we need to assume that $\tau_s$ satisfies a differential equation in time.

The key idea comes from realising that the aforementioned element $\Ham=(0,1,0,0,\cdots)$ is the generator of a constant supertranslation.\footnote{See remark \ref{remarkH} for an alternative time translation generator that generalizes $\Ham$.}${}^,$\footnote{Notice in particular that the action of $\Ham$ on the phase space is consistent with this fact since $\delta_\Ham C=\pa_u C$.}
Therefore, we naturally give $\Ham$ the status of Hamiltonian in the space of $\tau$ and identify its adjoint action with the flow of time. 
We thus consider, for $s\geq 0$, the \textit{dual equations of motion} 
\be 
\boxed{\,\, \pa_u\tau= \Ham_0^{-1}[\Ham, \tau]^C \quad \Leftrightarrow \quad  
\pa_u\t{s}= (\Dcal\tau)_s=D\t{s+1}-(s+3) C\t{s+2}.\,\,} \label{dualEOM}
\ee 
In the following we refer to these dual equations as $\E_s(\tau)=0$, where $\E(\tau) :=\pa_u \tau -\Dcal\tau$ and denote the corresponding space of parameters by $\sfT$.
If the context is clear, we often write $\E_s$ instead of $\E_s(\tau)$.

\begin{tcolorbox}[colback=beige, colframe=argile]
\textbf{Definition [$\sfT$-space]}
\be \label{TspaceGR}
\sfT :=\Big\{\,\tau\in\tV(\scri)~\big|~\pa_u\t{s}=D\t{s+1}-(s+3) C\t{s+2},~  s\geqslant 0~ \Big\}.
\ee 
\end{tcolorbox}

\ni These evolution equations imply that elements of $\sfT$ are uniquely determined by their celestial data $T_s(z,\bz)\equiv \tau_s(u=0,z,\bz)$ which play the role of initial conditions for the $\tau_s$.

\subsection{Comments about Lie algebroids  \label{Sec:Liealg}}

For the reader's convenience, we provide a reminder about Lie algebroids \cite{algebroid, Ciambelli:2021ujl}, since  this notion will be essential in our understanding of the nature of the time dependent symmetry outside the wedge. 

\begin{tcolorbox}[colback=beige, colframe=argile] \label{ReminderAlgebroid}
\textbf{Reminder [Lie algebroid -- Definition]}\\
A Lie algebroid $\big(\mathcal{A},\lbr\cdot\,,\cdot\rbr,\hat{\nkor} \big)$ is a vector bundle $\mathcal{A}\to\M$ with a Lie bracket  $\lbr\cdot\,,\cdot\rbr$ on its space of sections $\Gamma(\mathcal{A})$ and a vector bundle morphism $\hat\nkor:\mathcal{A}\to T\M$, called the anchor.
The anchor gives rise to an anchor map $\nkor:\Gamma(\mathcal{A})\to\X(\M)$ with two fundamental properties, namely the compatibility between Lie algebras and the Leibniz rule:
\begin{subequations}
\begin{align}
    i)\hspace{-2cm}& &\nkor\big(\lbr\alpha,\alpha'\rbr\big) &=\big[\nkor(\alpha),\nkor(\alpha')\big]_{T\M}, \label{anchorRep}\\
    ii) \hspace{-2cm}& &\big\lbr\alpha,f\alpha'\big\rbr &=f\big\lbr\alpha,\alpha'\big\rbr+\big(\nkor(\alpha)\triangleright f\big)\alpha', \label{anchorLeib}
\end{align}
\end{subequations}
for $\alpha,\alpha'\in\Gamma(\mathcal{A})$ and $f\in \mathcal{C}^\infty(\M)$. The RHS of $i)$ is the Lie bracket of vector fields on $\M$ while in $ii)$, $\nkor(\alpha)\triangleright f$  is the derivative of $f$ along the vector field $\nkor(\alpha)$.
\end{tcolorbox}

A canonical way to construct algebroids arises when the action of a Lie algebra on a given manifold exists. In this case we can construct a \emph{symmetry algebroid} as follows.

\begin{tcolorbox}[colback=beige, colframe=argile] \label{ReminderAlgebroidbis}
\textbf{Reminder [Symmetry algebroid]}\\
Let us assume that we have a manifold $\M$ equipped with the action of a Lie algebra $(\g,[\cdot\,,\cdot]_\g)$. 
In practice this means that there exists an
infinitesimal left action  $\hat\nkor : \g\to\X(\M)$, such that 
 \be \label{symcomp}
 \big[\hat\nkor(a), \hat\nkor(b)\big]_{T\M} 
= -\hat\nkor\big([a,b]_\g\big), \qquad a,b 
\in \g.
 \ee 
One can build the associated Lie algebroid $\mathcal A$ as the trivial bundle  $\M\times\g\to  \M$ with anchor map $\nkor:\Gamma(\mathcal{A})\to\X(\M)$ such that $\nkor(\alpha)(x)=\hat\nkor(\alpha(x))$ for each $x\in \M$ and $\alpha\in\Gamma(\cA)$. 
We identify sections of $\mathcal{A}$ with Lie algebra functions on $\M$. This naturally extends the vector fields action to $\Gamma(\mathcal{A})$.
The Lie algebroid bracket is then given by
\begin{equation}
    \lbr\alpha,\alpha'\rbr :=[\alpha,\alpha'] +\nkor(\alpha')\triangleright\alpha-\nkor(\alpha)\triangleright\alpha', \label{assocLieoidBracket}
\end{equation}
where we have defined $[\alpha,\alpha'](x):= [\alpha(x),\alpha'(x)]_\g$
for $\alpha,\alpha'\in\Gamma(\mathcal{A})$. 
In other words, the bracket $[\cdot\,,\cdot]$ over $\Gamma(\mathcal A)$ is the  fiberwise lift of $[\cdot\,,\cdot]_\g$ when we identify the sections $\alpha,\alpha'\in\Gamma(\mathcal A)$ with $\g$-valued functions over $\M$.
\end{tcolorbox}

In the physics literature the notion of symmetry algebroid appears naturally in field theory (see e.g. \cite{Gomes:2019xto} and references therein). 
It is used when we have the action of a symmetry algebra on the space of fields $\cF$ defined as the space of sections over a spacetime manifold $\cM$.
In this case the algebroid is a bundle over field space $\cF$ and the algebroid bracket is a generalization of the Lie bracket to field dependent transformations. It is exactly the modified bracket introduced  in \cite{Barnich:2010eb}.

We are now in position to build a new bracket, denoted $\lbr\cdot\,,\cdot\rbr$, by combining $[\cdot\,,\cdot]^C$ and the action $\delta_\tau$.

\subsection{Consistent time evolution \label{sec{timeev}}}

In order to impose the dual evolution equations \eqref{dualEOM} we need to allow for a field dependency of the transformation parameters. 
Therefore one defines an algebroid  bracket $\lbr\cdot\,,\cdot\rbr$ associated with $[\cdot\,,\cdot]^C$ and which acts on $\sfT$.

\begin{tcolorbox}[colback=beige, colframe=argile]
\textbf{Definition [$\sfT$-bracket]} \label{Tbra}\\
The $\sfT$-bracket $\lbr\cdot\,,\cdot\rbr :\sfT\times\sfT\to\sfT$, between $\tau,\tau'\in\sfT$ takes the form\footnote{If there is no possible confusion, then we write $\big(\dt\tp{}\big)_s\equiv\dt\tp{s}$ for shortness.}
\begin{equation}
\lbr\tau,\tau'\rbr_s :=  [\tau,\tau']^C_s+\big(\dtp\tau\big)_{s}-\big(\dt\tau'\big)_{s},\qquad s\geqslant -1. \label{Tbracket}
\end{equation}
\end{tcolorbox}

\ni The first remarkable fact is that the bracket is well defined on $\sfT$. 
It satisfies the following closure property.

\begin{tcolorbox}[colback=beige, colframe=argile] \label{LemmaClosure}
\textbf{Lemma [$\sfT$-bracket closure]}\\
The $\sfT$-bracket closes, i.e. satisfies the dual EOM \eqref{dualEOM}.
If $\tau$ and $\tau'$ are in $\sfT$ then $\E_s\big(\lbr\tau,\tau'\rbr\big)=0,\, s\geqslant 0.$ 
Hence $\lbr\tau,\tau'\rbr \in \sfT$.
\end{tcolorbox}

\paragraph{Proof:}
For any bracket $[\cdot\,,\cdot]$ and differential operator $\mathscr{D}$, we define the Leibniz rule anomaly as
\begin{equation}
    \cA\big([\cdot\,,\cdot],\mathscr{D} \big)=\mathscr{D}[\cdot\,,\cdot]-[\mathscr{D}\cdot\,,\cdot]-[\cdot\,,\mathscr{D}\cdot].
\end{equation}
In appendix \ref{Apptppsevol}, we compute that
\begin{align}
    \cA_s\big([\tau,\tau']^C,\Dcal \big)=-(s+3)\big(\t{s+2}\,\hdtp C-\tp{s+2}\,\hdt C\big). \label{AnomalyLeibniz1}
\end{align}
This anomaly disappears for the algebroid bracket and is replaced by 
\begin{equation}
    \cA_s\big(\lbr\tau,\tau'\rbr,\Dcal \big)=-N\paren{\tau,\tau'}_s+ \delta_{\Dcal\tau}\tp{s}-\delta_{\Dcal\tau'}\t{s}. \label{AnomalyLeibniz2}
\end{equation}
Moreover, $\pa_u$ is also anomalous:
\begin{equation}
    \cA_s\big(\lbr\tau,\tau'\rbr,\pa_u \big)=-N\paren{\tau,\tau'}_s+ \delta_{\pa_u\tau}\tp{s}-\delta_{\pa_u\tau'}\t{s}. \label{AnomalyLeibniz3}
\end{equation}
Hence, taking the difference gives 
\begin{equation}
    \cA_s\big(\lbr\tau,\tau'\rbr,\pa_u-\Dcal \big)= \delta_{(\pa_u\tau-\Dcal\tau)}\tp{s} -\delta_{(\pa_u\tau'-\Dcal\tau')}\t{s},
\end{equation}
which amounts to
\begin{equation} \label{AnomalyLeibniz}
    (\pa_u-\Dcal)\lbr\tau,\tau'\rbr= \big\lbr(\pa_u-\Dcal)\tau,\tau'\big\rbr +\big\lbr\tau,(\pa_u-\Dcal)\tau'\big\rbr+\delta_{(\pa_u-\Dcal)\tau}\tau'-\delta_{(\pa_u-\Dcal)\tau'}\tau.
\end{equation}
The RHS vanishes on shell of the dual EOM \eqref{dualEOM}, i.e. if $\tau,\tau'\in\sfT$, which proves that the bracket $\lbr\cdot\,,\cdot\rbr$ closes on $\sfT$.

\subsection{A symmetry algebroid \label{secAlgebroid}}

The next step is to show that  $\dt$  is an algebroid action for the $\sfT$-bracket. 
$\dt$ is a vector field in the gravitational phase space denoted $\PS$, namely the space of functionals of $C$ and $\bC$, cf. \eqref{actiondelta}. 
Holomorphic functionals on $\PS$ are simply functionals of $C \in \Ccar{1,2}$.
When acting on 0-forms on fields space, $\dt$ is nothing else than the Lie derivative $L_{\dt}$ in $\PS$ along the vector field generated by the symmetry parameter $\tau$.

\begin{tcolorbox}[colback=beige, colframe=argile]
\textbf{Lemma [Algebroid action and dual EOM]} \label{LemmaDualEOM}\\
The action $\dt$ admits a representation of the bracket $\lbr\cdot\,,\cdot\rbr$ onto $\PS$ when $\tau\in\sfT$:
\begin{equation}
    \tau,\tau'\in \sfT\quad \Rightarrow \quad 
    \big[\dt,\dtp\big]C =-\delta_{\lbr \tau,\tau'\rbr}C. \label{RepdeltaPS}
\end{equation}
\end{tcolorbox}

\ni While we already gave a physical motivation for the form of the time evolution of $\tau$, the validity of this morphism property for the $\dt$ action is what formally justifies the introduction of the dual equations of motion \eqref{dualEOM}. 
Indeed, the proof shows that the violation of the morphism property 
is parametrized by   $\E_s$ (see \eqref{algebroidRepinterm}).

The next step requires to show that  if $\delta$ is an algebroid action,  then the $\sfT$-bracket $\lbr\cdot\,,\cdot\rbr$ is indeed a Lie (algebroid) bracket, so that the Jacobi identity holds.

\begin{tcolorbox}[colback=beige, colframe=argile]
\textbf{Lemma [Jacobi identity]} \label{PropJacId}\\
The Lie algebroid bracket $\lbr\cdot\,,\cdot\rbr$ satisfies Jacobi if and only if  $\dt$ is an algebroid action:
\begin{equation}
    \big[\dt,\dtp\big]+\delta_{\lbr \tau,\tau'\rbr}=0\qquad\Leftrightarrow \qquad \big\lbr \tau,\lbr \tau',\tau''\rbr\big\rbr_s\cyc 0.
\end{equation}
\end{tcolorbox}

Since the penultimate lemma means that $\tau \to \delta_\tau$ is a  Lie algebra anti-homomorphism when $\tau \in \A$, then piecing everything together, the following theorem represents one of the main results of this paper. 

\begin{tcolorbox}[colback=beige, colframe=argile] \label{theoremAlgebroid}
\textbf{Theorem [$\A$-algebroid]}\\
The space $\A\equiv\big(\sfT,\lbr\cdot\,,\cdot\rbr,\delta\big)$ equipped with the $\sfT$-bracket \eqref{Tbracket} and the anchor map $\delta$,
\begin{align}
    \delta : \,&\A\to \X(\PS) \nn\\
     &\tau\mapsto\dt \label{defdelta}
\end{align}
is a Lie algebroid over $\PS$.
\end{tcolorbox}

\paragraph{Remark:} \label{remark1}
The  $\sfT$-bracket construction is similar to the notion of symmetry algebroid explained in \ref{ReminderAlgebroid}, where $\delta$ plays the role of the anchor map $\rho$, $\dt$ plays the role of  the vector field $\nkor(\alpha)$ and the $C$-bracket plays the role of the Lie bracket $[\cdot\,,\cdot]$ (or equivalently $[\cdot\,,\cdot]_\g$) while $\PS$ corresponds to the base manifold $\M$.
The difference in our case is that the $C$-bracket is field dependent and is \textit{not} a Lie bracket. 
What is remarkable is that the violation of its Jacobi identity is cancelled by the $\sfT$-bracket field dependency.
This result is therefore a \textit{generalization} of the symmetry algebroid construction (see also the next remark).

\paragraph{Proof of lemma [Jacobi identity]:}
We rewrite the property \eqref{JacCFC} for convenience,
\begin{equation}
    \big[\tau,[\tau',\tau'']^C\big]^C_s\cyc
    -\dt C \paren{\tau',\tau''}_s.
\end{equation}
We then establish that we have a Leibniz anomaly involving the $C$-bracket and the variation $\dt$,
\begin{equation} \label{vanomaly}
 \cA_s\big([\tau',\tau'']^C,\dt \big):= \dt \big[\tau',\tau''\big]^C_s-\big[\dt \tau',\tau''\big]^C_s-\big[\tau',\dt \tau''\big]^C_s = -\dt C\paren{\tau',\tau''}_s.
\end{equation}
We then turn to the quantity of interest: the expansion of the double commutator of the $\sfT$-bracket reads 
\begin{align}
    \big\lbr \tau,\lbr \tau',\tau''\rbr\big\rbr & = \big[\tau,[\tau',\tau'']^C\big]^C+\big[\tau,\dtpp \tau'-\dtp \tau''\big]^C +\delta_{\lbr \tau',\tau''\rbr}\tau-\dt\lbr \tau',\tau''\rbr\cr
    &\cyc \big[\tau,[\tau',\tau'']^C\big]^C 
    - \cA\big([\tau',\tau'']^C,\dt \big)
    +\delta_{\lbr \tau',\tau''\rbr}\t{} + [\dtp, \dtpp] \tau. 
\end{align}
In the second equality we have used the cyclic permutation to reconstruct the variational Leibniz anomaly \eqref{vanomaly}.
In the context we are in, the first two terms cancel each other and we are simply left with 
\vspace{-0.1cm}
\begin{equation}
    \big\lbr \tau,\lbr \tau',\tau''\rbr\big\rbr_s \cyc \delta_{\lbr \tau',\tau''\rbr}\t{s} +[\dtp,\dtpp]\t{s}, \qquad s\geqslant -1.
\end{equation}
Therefore the RHS vanishes iff $\big[\dtpp,\dtp\big]=\delta_{\lbr \tau',\tau''\rbr}$, on all functionals of $C$. 
This concludes the proof of this lemma.

\paragraph{Remark:}
The fact that $\lbr\cdot\,,\cdot\rbr$ is a Lie algebroid bracket was not guaranteed since the $C$-bracket  fails to be a Lie bracket.
The validity of the former lemma follows from  the fact that if $[\cdot\,,\cdot]$ is a bracket with Jacobi anomaly  $\big[\cdot\,,[\cdot\,,\cdot]\big]\cyc\mathcal{B}$ 
and with Leibniz anomaly $\cA\big([\cdot\,,\cdot] ,\nkor(\cdot)\big) =\mathcal{B'}$ one can still build a Lie algebroid bracket $\lbr\cdot\,,\cdot\rbr$ via the \hyperref[ReminderAlgebroidbis]{`symmetry algebroid' procedure}, provided that both anomalies are equal, i.e. $\mathcal{B}=\mathcal{B'}$.
We are using  the notation of the reminder of section \ref{Sec:Liealg}.

\paragraph{Remark:}
As mentioned in section \ref{Sec:DualEOM}, the flow of time can be generated through the action of  $\Ham$ which corresponds to a constant supertranslation.
Indeed, we can choose the flow generated by $\Ham$ to \textit{define} the flow of time $\pa_u$, i.e imposing that $\delta_\Ham C=:\pa_u C$.\footnote{This in turn tells us that $N$ has to be $\pa_u C$ (since what we call $N$ is after all just $\delta_\Ham C$).}
Quite remarkably this means that the dual time evolution equation can then be simply written as a constraint using the algebroid bracket. 
Indeed, from what we just said around equation \eqref{dtauCalpha}, the $C$-bracket is only a Lie bracket in the associated covariant wedge where $\dt C=0$.
The introduction of time therefore comes hand to hand with the introduction of the algebroid bracket $\lbr\cdot\,,\cdot\rbr$, cf.\,\eqref{Tbracket}, for which $\dt C\neq 0$.
The adjoint action of $\Ham$ is then given by 
\begin{equation}
    \lbr\Ham,\tau\rbr=\Dcal\tau-\delta_\Ham\tau=:\Dcal\tau-\pa_u\tau.
\end{equation}
Time evolution thus amounts to imposing the \textit{`dual Hamiltonian' constraint}
\begin{equation} \label{HamiltonianConstraint}
    \boxed{\lbr\Ham,\tau\rbr\overset{!}{=}0}.
\end{equation}
If Jacobi holds, then we immediately infer that
\begin{equation}
    \big\lbr\Ham,\lbr\t{},\tp{}\rbr\big\rbr= \big\lbr\lbr\Ham,\t{}\rbr,\tp{}\big\rbr+ \big\lbr\t{},\lbr\Ham,\tp{}\rbr\big\rbr,
\end{equation}
which is another proof that the \hyperref[LemmaClosure]{$\sfT$-bracket closes on-shell}.\\

We now give the proof of the \hyperref[LemmaDualEOM]{algebroid action lemma}. This requires to investigate under which conditions $\dt$ is an algebroid action.

\begin{tcolorbox}[colback=beige, colframe=argile]
\textbf{Lemma [Pre-algebroid action]}\\
Let us assume that $\tau,\tau' \in \tV(\scri)$, hence that no evolution equations are imposed, then 
\begin{equation}
    \big[\dt,\dtp\big]C+\delta_{\lbr \tau,\tau'\rbr}C =\tp0\delta_{(\dot\tau-\Dcal\tau)}C-
    \t0\delta_{(\dot\tau'-\Dcal\tau')}C
    \label{algebroidRepinterm}.
\end{equation}
The RHS vanishes if the $\tau$'s satisfy the recursion relations 
\be \label{trecursion}
\dot\tau_s =(\Dcal\tau)_s,\qquad s\geq 0.
\ee
In particular, no condition is required on $\tau_{-1}$.
\end{tcolorbox}

\paragraph{Proof:}
Notice that $\big[\dt,\dtp\big]C$ splits into 3 contributions, namely
\bs
\label{ddC0}
\begin{align}
    \big[\dt,\dtp\big]C &=
    \dt\Big\{ N\tp0 -D^2\tp0+2DC\tp1+3C D\tp1-3 C^2\tp2\Big\}-\tau\leftrightarrow \tau' \nn\\
    &=\Big\{ N\dt\tp0 -D^2\big(\dt\tp0\big) +2DC\dt\tp1+3CD\big(\dt\tp1\big)-3 C^2\dt\tp2 \label{ddC01}\\
    &\qquad +2\tp1 D\big(\hdt C\big)+3D\tp1\big(\hdt C\big)-6C\tp2\big(\hdt C\big) \label{ddC02}\\
    &\qquad + \tp0\dt N+2\tp1 D(N\t0)+3 N\t0 D\tp1-6 CN\t0\tp2\Big\}-\tau\leftrightarrow \tau'. \label{ddC03}
\end{align}
\es
The first line is equal to $\delta_{\dt\tau'} C$, it contains all the terms of the type $\dt\tau'$; 
the second line all the terms that do not involve any news $N$ nor any variations of $\tau$ (it can be written using the hatted action \eqref{hataction}).
The third line collects all the terms that depend on $N$. 
Next, $\delta_{\lbr \tau,\tau'\rbr}C$ also splits into the 3 same contributions:
\begin{align} \label{dttC} 
    \delta_{\lbr \tau,\tau'\rbr}C =  
    \delta_{(\dtp\tau-\dt\tau')} C +\hat{\delta}_{[\tau,\tau']^C} C
     + N[\tau,\tau']^C_0 . 
\end{align}
Once we sum \eqref{ddC0} with \eqref{dttC}, the terms involving $\dt\tau'$ clearly cancel.
In appendix \ref{AppAlgebraAction}, we evaluate the sum of the terms containing the variation $\hat\delta$. It reads
\begin{equation}
    \eqref{ddC02}+ \hat{\delta}_{[\tau,\tau']^C} C=\t0 \hat\delta_{\Dcal\tau'}C-\tp0 \hat\delta_{\Dcal\tau}C. \label{hatdTT}
\end{equation}
Finally, using the variation of $N$
\begin{equation}
    \dtp N= \pa_u(\dtp C)=\delta_{\dot \tau'}C+ \tp0 \dot N+2DN\tp1+3ND\tp1-6CN\tp2,
\end{equation}
and rewriting the spin $0$ commutator as
\begin{align}
[\tau,\tau']^C_0 &=\big(\t0 D\tp1+2\t1 D\tp0-3 C\t0\tp2\big) -\tau\leftrightarrow\tau' \cr &=\big(2\t1 D\tp0+\t0(\Dcal\tau')_0\big)-\tau\leftrightarrow\tau',
\end{align}
we obtain that
\begin{align}
    \eqref{ddC03}+N[\tau,\tau']^C_0 &=
     \Big(\big(\tp0\dt N+2\tp1 D(N\t0)+3 N\t0 D\tp1-6 CN\t0\tp2\big) -  \tau\leftrightarrow\tau'\Big) + N[\tau,\tau']^C_0 
    \nn \\ 
    &=  \t0\Big(-\dtp N +2 DN\tp1+3ND\tp1-6CN\tp2  + N(\Dcal\tau')_0\Big)  -\tau\leftrightarrow\tau'
    \nn\\
    &=  \t0\Big(-\delta_{\dot\tau'}C+ N(\Dcal\tau')_0  \Big) -\tau\leftrightarrow\tau'. \label{ddC5}
\end{align}
In the second equality we swapped $\tp0\dt N$ for $-\t0\dtp N$ and in the last equality we use that the term $-\t0\tp0 \dot{N}$ vanishes after anti-symmetrization. 
Adding \eqref{hatdTT} and \eqref{ddC5}, we finally obtain \eqref{algebroidRepinterm}:
\begin{equation}
     \big[\dt,\dtp\big]C+\delta_{\lbr \tau,\tau'\rbr}C =\tp0\delta_{(\dot\tau-\Dcal\tau)}C-
    \t0\delta_{(\dot\tau'-\Dcal\tau')}C.
\end{equation}

Note that the RHS vanishes when $\delta_{\dot\tau-\Dcal\tau}C  \in \Ccar{2,2}$ is proportional to $\t0$.  Since $\t0 \in \Ccar{-1,0}$ the proportionality coefficient is in $\Ccar{3,2}$. 
The only element of that weight that can be constructed in a local manner from $C$ is $\pa_u^2C$. This means that the RHS of 
\eqref{algebroidRepinterm} vanishes when 
$\delta_{(\dot\tau-\Dcal\tau)}C=\beta \t0 \pa_uN$, where $\beta$ is an arbitrary constant. 
In the following we choose the condition $\beta=0$ and therefore assume that
\be \delta_{(\dot\tau-\Dcal\tau)}C=0.
\ee
Since $\dt C$ involves only $\t0$, $\t1$ and $\t2$, this condition implies that the $\tau$'s satisfy the recursion relations\footnote{Here we precisely want to keep $\dt C\neq 0$ in general, so that we set the argument $\dot\tau-\Dcal\tau$ to 0, rather than restricting the map to its kernel.
}
\vspace{-0.2cm}
\be \label{trecursioninterm}
\dot\tau_s- (\Dcal\tau)_s= 0\qquad s=0,1,2.
\ee 
If we demand, in addition, that the condition \eqref{trecursioninterm} stays valid for all $\tau$'s in the image of the $\sfT$-bracket, i.e. parameters of the form $\tau= \lbr\tau',\tau''\rbr$, then this implies that
$\dot \tau_s=(\Dcal\tau)_s$ for all $s\geq 0$.
 
The proof goes as follows: since $\tau_s$ and $\tau'_s$ have to follow the differential equation \eqref{trecursioninterm} for $s=0,1,2$, the same has to hold for the bracket $\lbr\tau,\tau'\rbr_s$, $s=0,1,2$.
In particular, we find that in order to get
\begin{equation}
    \pa_u\lbr\tau,\tau'\rbr_1= D\lbr\tau,\tau'\rbr_2-4C\lbr\tau,\tau'\rbr_3, \label{tt1dot}
\end{equation}
we need to impose 
\begin{equation}
    \pa_u\t3=D\t4-6C\t5, \label{t3dot}
\end{equation}
and similarly for $\tau'$---which is nothing else than \eqref{trecursion} for $s=3$.
The general proof then goes on as such recursively, so that \eqref{trecursioninterm} does imply \eqref{trecursion}.
To give the gist of it, we report the demonstration of ``\eqref{tt1dot} implies \eqref{t3dot}'' in App.\,\ref{App:1implies3}.

This concludes the proof of this lemma and by extension the proof of the \hyperref[LemmaDualEOM]{lemma [Algebroid action and dual EOM]}.
The demonstration of the \hyperref[theoremAlgebroid]{theorem [$\A$-algebroid]} follows from combining the various lemmas.

\paragraph{Remark:}
The anchor map is a left anchor, which implies that it is  an anti-homomorphism\footnote{This is the same convention used for $\nkor$  in the \hyperref[ReminderAlgebroidbis]{reminder} ($\tau$ here plays the role of $\alpha$ there).}
of Lie algebras, i.e. satisfies\footnote{Since the notation shall never be confusing, we use $[\dt,\dtp]\equiv[\dt,\dtp]_{\mathrm{Lie}(J^\infty\bfm p)}$, cf. \eqref{actiondelta}.}
$[\dt,\dtp]\cdot=-\delta_{\lbr \tau,\tau'\rbr}\cdot\,$; the reason being that when acting on a functional $O$, the field space action $\dt O$ is realized as a differential operator $\LL_\tau O$ on $\scri$, namely $\dt O= \LL_\tau O$.
We purposefully chose the letter $\LL$ since $\LL_\tau$ can be viewed as a generalization of the spacetime Lie derivative.\footnote{After all, when $\tau$ is just a vector field $Y$, then $\LL_\tau$ \textit{is} the Lie derivative.}
Because the field space action $\dt$ commutes with the spacetime action $\LL_{\tau'}$ up to the derivative along the variation of $\tau'$,\footnote{In general we have $[\delta, \LL_V]= \LL_{\delta V}$ for an arbitrary vector field.} we have that $\dt\dtp O=\dt \LL_{\tau'} O=\LL_{\tau'}\dt O + \LL_{\delta_{\tau} \tau'}O=\LL_{\tau'} \LL_{\tau}O + \LL_{\dt\tau'}O$.
Therefore, $[\dt,\dtp]O=[\LL_{\tau'}, \LL_\tau]O+ \LL_{\dt\tau'}O- \LL_{\dtp \tau}O =\LL_{[\tau',\tau]}O+\LL_{\dt \tau'-\dtp\tau}O=\LL_{\lbr\tau',\tau\rbr}O =\delta_{\lbr\tau',\tau\rbr}O$, where we applied the usual commutation rule for our Lie derivative-like operator $\LL_\tau$.

\subsection{Initial condition \label{sec:InitialCondition}}

The $\A$-algebroid defined in the previous section only  uses the dual equation of motions $\E_s=0$ for $s\geq 0$. 
No condition on $\tau_{-1}$ was necessary. This is due to the fact that $ \tV_{-1}$ is central and acts trivially on the shear $C$.

This leaves free the possibility to impose additional conditions on $\tau_{-1}$ provided such conditions are compatible with the $\sfT$-bracket.
One option is simply to also demand that the dual equation of motion $\E_{-1}=0$ is satisfied at level $-1$.
Remarkably, there is another option, highly relevant for the canonical analysis, which is to impose the \textit{initial condition} denoted $\sI_\tau$ in the following and given by
\begin{equation}
    \boxed{\,D\t{-1}= C\t0.\,} \label{initialCondition}
\end{equation}  
Depending on which condition we impose for $\tau_{-1}$ we have two different time dependent parameter spaces.

\begin{tcolorbox}[colback=beige, colframe=argile]
\textbf{Definition [$\sfTh$-space \& $\sfTo$-space]}
\bs \label{defTbarThat}
\begin{align}
    \sfTh &:=\Big\{\,\tau\in\tV(\scri)~ \big|~\pa_u\t{s}=(\Dcal\tau)_s,~  s\geqslant -1\, \Big\}, \\
    \sfTo &:=\Big\{\,\tau\in\tV(\scri)~\big|~\pa_u\t{s}=(\Dcal\tau)_s,~  s\geqslant 0~\, \& ~ D\t{-1}= C\t0\,\Big\} . 
\end{align}
\es
\end{tcolorbox}

\ni Both $\sfTh$ and $\sfTo$ are subsets of $\sfT$.  $\sfTh$ and $\sfTo$ differ by the condition imposed on $\t{-1}$: the dual  equation of motion $\E_{-1}=0$ for $\sfTh$ or the initial condition \eqref{initialCondition} for $\sfTo$.
Quite remarkably, the $\sfT$-bracket is compatible with both spaces.

\begin{tcolorbox}[colback=beige, colframe=argile] \label{LemmaClosure2}
\textbf{Lemma [$\sfT$-bracket closure---degree $-1$]}\\
The $\sfT$-bracket closes on $\sfTh$ and on $\sfTo$, i.e. it satisfies
\begin{align}
    \tau,\tau' \in \sfTh \Rightarrow \lbr\tau,\tau'\rbr\in \sfTh  \quad \mathrm{and}\quad 
    \tau,\tau' \in \sfTo \Rightarrow \lbr\tau,\tau'\rbr\in \sfTo .
\end{align}    
\end{tcolorbox}

\paragraph{Proof:}
In practice this means that 
$D\lbr\tau,\tau'\rbr_{-1}=C\,\lbr\tau,\tau'\rbr_{0}$ 
when $\tau,\tau' \in \sfTo$ and that 
 $\pa_u\lbr\tau,\tau'\rbr_{-1}= \big(\Dcal\lbr\tau,\tau'\rbr\big)_{-1}$ when $\tau,\tau' \in \sfTh$.
The former is proven in appendix \ref{App:InitialConstraint} while the latter follows from the demonstration of the \hyperref[LemmaClosure]{lemma [$\sfT$-bracket closure]}, which was based on the computation of the Leibniz anomalies in App.\,\ref{App:dualEOM}, a result which holds at degree $-1$ too.\\

This lemma implies the following theorem. 
\begin{tcolorbox}[colback=beige, colframe=argile] \label{theoremAlgebroids}
\textbf{Theorem [$\Ah$- and $\Ao$-algebroid]}\\
Both spaces $\Ah\equiv\big(\sfTh,\lbr\cdot\,,\cdot\rbr,\delta\big)$  and 
$\Ao \equiv\big(\sfTo,\lbr\cdot\,,\cdot\rbr,\delta\big)$ are Lie algebroids 
equipped with the $\sfT$-bracket \eqref{Tbracket} and the anchor map $\delta: \A \to \X(\PS)  $ \eqref{defdelta} restricted to $\Ah$ and $\Ao$.
They are distinct sub-algebroids of $\A$.
Moreover, the intersection between these two algebroids defines an algebra\footnote{An algebra can be characterized as an algebroid with trivial anchor map.}  
\be 
\Wcal_C:=\Ah\cap \Ao. \label{WCIntersection}
\ee
which is a generalized wedge algebra on $\scri$.
\end{tcolorbox}

\paragraph{Proof:}
To prove the last part of the theorem, let us assume that  both the initial condition and the initial EOM are satisfied for $\tau$. This means that 
\begin{align}
    D \pa_u \tau_{-1} &= D^2 \tau_0 - 2 DC \tau_1 -2 C D\tau_1, \cr
    \pa_u D \tau_{-1} &= N\tau_0 + C D\tau_1 -3 C^2 \tau_2.
\end{align}
Therefore, taking the difference we get that 
\be \label{WCIntersection2}
\dt C=0, \quad \mathrm{when}\quad \tau \in \Wcal_C= \Ah\cap\Ao,
\ee 
When this condition holds\footnote{ $\Wcal_C$ is the time dependent analog of $\Wcal_\sigma$, the condition $\dt C=0$ replacing  $\hdT\sigma=0$. 
Note that the condition $\dt C=0$ seems to only imply $\pa_u\sI_\tau=-D\sE_{-1}(\tau)$.
However, using \eqref{AnomalyLeibniz-2}, we know that it actually implies $\sI_{[\tau,\tp{}]^C}=0$ as well, which then imposes $\sI_\tau=0$ by consistency and thus $D\sE_{-1}=0$.}
we have that the $C$-bracket $[\cdot\,,\cdot]^C$ satisfies the Jacobi identity, thanks to \eqref{JacCFC}, and thus,  there is  no need to use  field dependent transformation parameters and algebroid extension.\footnote{$\tau$ still  depends on $C$. This dependence is compatible with the condition $\dt\tau'=0$, since $\dt\tau'\propto \dt C=0$.}
In other words, the anchor map $\delta$ is identically $0$ on $\Wcal_C$, namely over its kernel, which means that $\Wcal_C$ is an algebra.

What needs to be established next is whether $\Wcal_C$ is non trivial.
The answer depends on the value of $C$: If $C$ is a non radiative shear, i.e. such that $\pa_u C=0$, then $\Wcal_C$ is isomorphic to  the celestial wedge algebra $ \Wcal_\sigma(S)$ for $\scri = \mathbb{R} \times S$ (see section \ref{sec:Wedge}). In our companion paper \cite{Cresto:2024fhd} we have shown that $ \Wcal_\sigma(S)$ is a non trivial algebra isomorphic to  $Lw_{1+\infty}$ when $S=S_2$.

In general when $N\neq 0$ but $\pa_u^{n+1} C=0$ for some $n>0$ we also expect $\Wcal_C\neq \{0\}$. 
For instance, we can find a two dimensional space of solutions such that $\tau_n=0$ for $n>0$. 
In this case, elements of $\Wcal_C$ are solutions of $D^2 \tau_0 - N \tau_0=0$. This is a second order differential equation on $S$ that admits a 2 dimensional set of solutions. 
If we denote by $ f := \frac{\tau_0}{\tau_0'}$ the ratio of these two solutions we have that 
\be 
-2 N=  \{ f, z\}:= D \left(\frac{D^2 f}{Df}\right)-\frac12 \left(\frac{D^2 f}{Df}\right)^2. 
\ee 
We recognize on the RHS  the  covariant Schwarzian derivative of $f$ along $z$. This shows that $\Wcal_C$ is non trivial.
Understanding the exact nature of $\Wcal_C$ is an interesting question which goes beyond the scope of this paper.

\paragraph{Remark:} As we are about to see in the next section,  the initial condition $D\t{-1}=C\t0$, which defines $\sfTo$,  is primordial to recover the Bondi mass aspect  as the canonical generator of supertranslations; i.e. when we show that  the algebroid constructed here can be represented, through Noether's theorem, in terms of  higher spin charges living on the gravitational phase space $\PS$.
It is quite remarkable that the algebroid $\A$ has ``room'' for this condition, in the sense that we are able to accommodate for such a constraint without changing anything to any of the previous discussions.

\paragraph{Remark:}\label{remarkH} Note that $\Ham$ is in $\sfTh$, but it is not in $\sfTo$. The time translation generator in $\sfTo$ is denoted $\hHam$ and given by $\hHam= (DG, 1, 0, 0, \ldots)$ where $G$ is the Goldstone, i.e. the  solution of $D^2 G = C$. $\hHam$ differs from $\Ham$ by an element of $\tV_{-1}$, which is central. 
Therefore the adjoint action of $\hHam$ coincides with the adjoint action of $\Ham$.
This means that  we could have used $\hHam$ everywhere instead of $\Ham$ in the previous section.  
We will see  in section \ref{SecNoether} that the shift from $\Ham$ to $\hHam$ corresponds to, at the level of charge aspect, shifting the mass generator from the covariant mass to the Bondi mass.

\section{Noether charge \label{SecNoether}}

The goal of this section is to prove that there exists a canonical representation of the higher spin algebroid bracket \eqref{Tbracket} on the Ashtekar-Streubel phase space. 
This phase space possesses a symplectic potential which depends on the shear $C(u,z,\bar{z})$ and the news $\bN\equiv \pa_u \bC$. 
We use the convention where 
$C:=\tfrac12C_{mm}= \frac{P^2}{2}C_{zz}$,\footnote{This is the Newman-Penrose's shear $\sigma_{\textsc{np}}$.} 
instead of the one of \cite{Freidel:2021ytz}. 
The holomorphic Ashtekar-Streubel symplectic potential \cite{ashtekar1981symplectic} is given by an integral over $\scri$,\footnote{$\delta$ here is the fields space exterior derivative, not to be confused with the anchor. 
By definition, both are related according to $I_{\dt}(\delta C)=\dt C$, for $I$ the fields space interior product.}
\be 
\Theta = \frac1{4\pi \GN} \int_{\scri}  \bN \delta C.
\ee 
It is related to the usual, real Ashtekar-Streubel symplectic potential by a \textit{complex} canonical transformation.
Indeed, 
\begin{equation}
    \Theta=\Theta^{\mathsf{AS}} +\frac{1}{8\pi\GN}\left(\delta\left(\int_\scri \bN C\right)-\int_\scri \pa_u(C\delta\bC) \right),
\end{equation}
where
\be 
\Theta^{\mathsf{AS}}=\frac1{8\pi \GN} \int_{\scri}\big(\bN \delta C +N \delta \bC\big).
\ee
This section aims to show that the \hyperref[theoremAlgebroids]{$\Ao$-algebroid} is represented canonically on the gravitational phase space.
This is the central result of our paper, which is valid provided we impose proper boundary conditions. 
In this work, we choose the fields $(C,\bC)$ to belong to the Schwartz space.\footnote{
The Schwartz space $\mathcal{S}$ is defined \cite{schwartz2008mathematics} as the set of continuous functions $f(u)$ that decay faster than any positive inverse power of $u$ as $|u| \rightarrow \infty$. Formally,
\be 
\mathcal{S} = \left\{f \in C^{\infty}\,|\, \forall \,\alpha, \beta \in \mathbb{N}, || f ||_{\alpha, \beta} < \infty \right\},
\ee
where $
||f||_{\alpha, \beta} = \underset{u \in \mathbb{R}}{\rm sup} \left|u^{\alpha} \pa_u^{\beta} f(u) \right|.$ 
\label{foot1}}

In quantum mechanics, the presence of symmetry associated with the Lie algebra $\g$ acting canonically on the phase space $\PS$, means that there exists a moment map\footnote{In the physics literature, it is customary to work with the co-moment map instead 
 $Q: \mathfrak{g} \to  \mathcal{C}(\PS)$, which is simply the dual map.} $Q^*: {\cal{P}} \to \mathfrak{g}^*$, \cite{da2001lectures, weinstein1981symplectic}. This is the essence of Noether's theorem \cite{noether1971invariant}.

In practice the Noether's theorem means that for any element $X\in \g$ we have a Noether charge $Q[X]$ which is a functional on phase space $\PS$ such that 
\be 
\poisson{Q[X], Q[Y]}= Q\big[[X,Y]_\g\big], \qquad 
\poisson{Q[X], O} =\delta_X O.
\ee 
where $O\in \mathcal{C}(\cP)$ is a phase space functional and  $\delta_X$ represents the action of $\g$ on $\cP$.
In our case, we have a symmetry algebroid and a generalization of the moment map theorem for the $\Ao$-algebroid.

\subsection{Master charge \label{secGR:MasterCharge}}

Let us start with the explicit construction of the algebroid Noether charge. 
For each $s\in \{-1,0,1,\cdots,\}$ we define a charge aspect $\tQ_s$ to be the spin $s$ charge aspect. 
This is an element of $\Ccar{3,s}$ and its evaluation at a point $(u,z,\bz)$ on $\scri$ is denoted $ \tQ_s(u,z,\bz)$. These charge aspects are defined recursively by imposing that 
\be 
\tQ_{-2}= \dot{\bN},\qquad \tQ_{-1}=  D\bN,\qquad \pa_u \tQ_{-1}= D\tQ_{-2},
\ee
and through the evolution equations
\be \label{Qevolve}
\pa_u \tQ_{s}= D\tQ_{s-1}+(s+1) C \tQ_{s-2}, \qquad s\geqslant 0.
\ee 
This set of evolution equations encodes the asymptotic evolution of the Weyl tensor \cite{Adamo:2009vu, Freidel:2021ytz, Geiller:2024bgf}.
Imposing the asymptotic condition that  $\tQ_s|_{u=+\infty}=0$ allows us to define $\tQ_s(u)$, after recursive $u$-integration, as a function of $(C,\bN)$ which is \emph{linear} in $\bN$ and polynomial of degree $\lfloor s/2\rfloor+1$ in $C$ and its derivatives\footnote{This condition is valid for $s\geqslant -2$, with $\lfloor n\rfloor$ the floor function.
To see it, just notice that the initial data $\tQ_{-2}$ and $\tQ_{-1}$ are of degree 0 in $C$ and afterwards $\deg[\tQ_s]=\deg[\tQ_{s-2}]+1$.} \cite{Freidel:2021ytz}.

At any given $u=\mathrm{const}$ cut of $\scri$ we  define, for $s\geqslant -1$, the smeared charges\footnote{In the following we shall write $\int_S:=\int_S\bfm{\epsilon}_S$.}
\begin{equation}
Q_s^u[\tau_s]:=
\frac{1}{4\pi \GN}\int_S(\tQ_s\t{s})(u,z,\bz), \label{Qstaus}
\end{equation}
where $\tau_s$ satisfy the dual evolution equations  $\E_s(\tau)=0$ for $s\geq 0$ \eqref{dualEOM}, where
\be 
 \E_s(\tau) := \dot \tau_s - D \t{s+1} + (s+3) C \t{s+2},\label{tausevolb}
\ee  
and are subject to the initial condition \eqref{initialCondition} $\sI_\tau=0$, where
\be  \label{tauintial}
\sI_\tau:=C\t0-D \t{-1}.
\ee 
When the context is clear, we henceforth use $\E_s$ instead of $\E_s(\tau)$.
The fact that $\t{s} \in \Ccar{-1,-s}$ while $\tQ \in \Ccar{3,s}$ implies that the product $\tQ_s \tau_s \in \Ccar{2,0}$ is a scalar density on the sphere and 
\eqref{Qstaus} actually defines, after integration on $S$, a pairing between $\Ao$ and its dual $\big(\Ao\big)^*$.

It is important to remember that $\cQ{s}{\tau_s}$ involves an integral over the sphere, so that we can freely integrate by parts sphere derivatives.
We then construct the time-dependent \emph{master charge} $\Qt^u$,
\begin{equation} \label{masterChargeGR}
\boxed{\Qt^u:= \sum_{s=-1}^{\infty} Q^u_s[\tau_s]}.
\end{equation}
The time evolution of the master charge is remarkably simple. 
One first uses the charge aspect evolution equation \eqref{Qevolve} to get 
\begin{align}
\pa_u Q^u_\tau  &= Q^u_{-1}[\dot\tau_{-1}]-Q^u_{-2}[D\t{-1}]+\sum_{s=0}^\infty \Big(Q^u_{s}[\dot \tau_s] - Q^u_{s-1}[D \t{s}]  + (s+1) Q^u_{s-2}[C \t{s}]\Big) \label{timeEvolQt}\\
& = Q^u_{-2}\big[-D \t{-1} +  C \t0\big] + Q^u_{-1}\big[\dot \tau_{-1} - D\t0 +2C\t1\big]+\sum_{s=0}^\infty Q^u_s\big[\dot\tau_s - D\t{s+1} +(s+3) C \t{s+2}\big]. \nn
\end{align}
Using the dual evolution equations \eqref{tausevolb} and initial condition \eqref{tauintial}, i.e. when $\tau \in \Ao$, only the second term survives.
The latter can be simplified using that $\tQ_{-1}=  D\bN$, so that 
\begin{align}
\dot{Q}^u_\tau &= - \bN \left[ D \dot\tau_{-1} - D^2 \t0 +  2 D(C \t1)\right] \nn \\
&= - \bN  \left[  \big(N\t0 + C(D \t1 - 3C \t2)\big) - D^2 \t0 + 2  D(C \t1)\right] \label{dotQtuinterm}\\
& =- {\bN} \left[ -\left( D^2-  N \right) \t0 + \big(2 DC+ 3 CD\big)\t1 - 3 C^2 \t2 \right]. \nn
\end{align}
Hence
\begin{equation}
\pa_u\Qt^u=- \frac{1}{4\pi \GN}\int_S \bN \dt C , \label{Qtdot}
\end{equation}
where we used the transformation \eqref{dtauC}
\begin{equation}\label{deltaC}
\dt C:= -D^2 \t0+ N\t0+ 2 DC\t1+ 3 CD \t1- 3 C^2\t2.
\end{equation}
On the RHS, all functions are evaluated at the same point $(u,z,\bz)$ where the shear $C$ is evaluated on the LHS. 
We have seen in sections \ref{secAlgebroid} and \ref{sec:InitialCondition} and equation \eqref{RepdeltaPS} that this transformation provides a representation of the $\Ao$-algebroid on phase space. 
Besides, notice that $\Qt^u$ is conserved in the absence of left-handed radiation, i.e. when $\bN\equiv 0$. 
From \eqref{Qtdot} and the final condition $\Qt^{+\infty}=0$,\footnote{Note that what matters for the definition of the charge is that $\lim_{u\to +\infty} \tQ_s\t{s}=0$, which is much less restrictive than imposing $\lim_{u\to +\infty} \tQ_s=0$ as it can be read as a condition on the transformation parameters $\t{s}$ instead. \label{foot2}}
we can express $\Qt^u$ as the integral 
\begin{equation}
\Qt^u = \int^{+\infty}_u  \bN[\dt C]. \label{defNoetherChargeinterm}
\end{equation}

\subsection{Covariant charge \label{Sec:CovCharges}}

It is interesting to note that the initial condition $D \t{-1} =  C \t0$  is crucial in order to obtain a series of charges which is conserved when $\bN=0$ and for which the total mass satisfies the Bondi  loss formula. 
We now show that the total mass is the charge associated to $\hHam$.

The initial condition ensures that $\t{-1}$ is not an independent parameter, so there is no independent spin $-1$ charge. Instead, it implies that the charge associated with $\tau_0$ is given by the combination,\footnote{Notice however that
\begin{equation}
    \poisson{\cQ{-1}{\t{-1}},C(u',z,\bz)}=- C\t0\,\delta(u-u').
\end{equation}
We thus see that the bracket with the $\tQ_{-1}$ charge contributes to a contact term which becomes 0 upon taking the limit $u\to -\infty$, cf. the definition of the Noether charge \eqref{defNoetherCharge}.}
\be
 Q^u_{-1}[\t{-1}] +Q^u_0[\t0]= \frac{1}{4\pi \GN}\int_S \left(  - \bN D\tau_{-1} + \tQ_0\t0 \right)=
 \frac{1}{4\pi \GN}\int_S \big(\tQ_0 - \bN C\big)\t0. \label{bondimass}
\ee 
$\tQ_0$ is the covariant mass aspect equal to the leading order in the $1/r$ expansion of the Weyl tensor component $\psi_2$. On the other hand the combination $\tQ_0 - \bN C$ is the Bondi mass aspect!\footnote{This was first noted by Moreschi in \cite{moreschi2004intrinsic}.} The total mass is the charge evaluated for $\tau=\hHam$ and is given by the Bondi flux formula
\be 
\pa_u Q^u_{\hHam}= -\frac{1}{4\pi \GN} \int_S \bN N \leq 0.
\ee

Note that a choice was made in \eqref{timeEvolQt} about the behavior of $\tau_{-1}$. 
If we only assume that the evolution equations $\E_s=0$ for $s\geq0$ are imposed, we obtain that 
\be 
\pa_u \mathds{Q}_{\tau}^u = Q^u_{-2}\big[-D \t{-1} +  C \t0\big] + Q^u_{-1}\big[\dot \tau_{-1} - D \t0 +  2 C \t1\big],
\ee
where the notation $\mathds{Q}_{\tau}$ refers to the fact that $\tau_{-1}$ is left unconstrained, i.e. that $\tau\in \sfT$.
From this expression we see that two natural possibilities open: either we impose the initial condition which gives a charge $\mathds{Q}^u_{\tau}\to Q^u_\tau$ conserved when $\bN=0$; or we impose the condition $\E_{-1}(\tau)=0$, namely
\be 
\dot \tau_{-1} = D \t0 -  2 C \t1. \label{dualEOM-1}
\ee 
In this case we denote the charge by $\mathds{Q}^u_{\tau}\to H^u_\tau$, which is conserved when $\dot\bN=0$. This is what we call the covariant charge in \cite{Freidel:2021qpz}. In order to understand the relationship between $H_\tau$ and $Q_\tau$, we reveal the important identity 
 \be
 \pa_u (D \t{-1} -  C \t0) &=
 D \dot \tau_{-1} -  N \t0 
 - C \dot \tau_0 \nn\\
 &= D (\dot\tau_{-1}-D \t0 +  2 C \t1 )
 +(D^2-N) \t0 -  2 DC \t1 - 2C D\t1  -C( D\t1 -3 C\t2) \nn\\
&= D (\dot\tau_{-1}-D \t0 +  2 C \t1)
 +(D^2-N) \t0 -  2 DC \t1 - 3 C D\t1  + 3 C^2\t2 \nn\\
&= D \E_{-1} - \dt C.
 \ee 
In other words,
\begin{equation}\label{-1id}
     \boxed{\dt C=D\E_{-1}+\pa_u\sI_\tau },
\end{equation}
with $\dt C$  given in \eqref{deltaC}. 
From this identity we conclude that if we impose the initial condition, then we come back to the definition of $\dt C$ and the original computation \eqref{dotQtuinterm} for which $\pa_u Q_\tau^u =- \bN[\dt C]$.
On the other hand, if we impose,   $\E_{-1}=0$, then we conclude  that  we can interpret
$\sI_\tau$ as the transformation of the potential $h:= \pa_u^{-1} C$.\footnote{$h=\pa_u^{-1}C$ appears as the twistor potential evaluated at $\scri$ \cite{Adamo:2021lrv,Donnay:2024qwq,Kmec:2024nmu}.} We denote this transformation by $\tdt$ and define 
\be
\boxed{\tdt h := \t0 \pa_u h  - D\t{-1}=\sI_\tau=-(\Dcal\tau)_{-2}}. \label{dtauh}
\ee
From the condition $\sE_{-1}=0$ we get that the charge associated to $\tilde\delta_\tau$ satisfies
\be \label{QH}
\pa_u H_{\tau}^u = Q^u_{-2}[C\t0-D \t{-1}] = \dot\bN\big[\tdt h\big].
\ee
This evolution equation shows that the charge is conserved when $\dot \bN=0$.

A priori the symmetry transformation $\tdt$ associated with the charge $\Ht$ is different than the one $\dt$ associated with
the charge $\Qt$ due to the different  conditions on $\t{-1}$.
However, the equation \eqref{-1id} means, under the condition
$\sE_{-1}=0$, that $\pa_u\tdt h = \dt C$. This signifies that although the charges $\Qt$ and $\Ht$ are different, their action on the shear $C$ coincides!
The difference only lies in the fact that the twistor potential $h$ is transformed by the $\Ht$ action since the $\Ht$ action of $\t{-1}$ is non trivial on $h$.

In \cite{Kmec:2024nmu} it was recently showed that the covariant charge $H_\tau$ can be naturally derived from the analysis of the twistorial formulation of self-dual gravity after a  gauge fixing that projects the twistor description onto spacetime.
Equation \eqref{QH} shows that the charge $H_\tau$ is naturally associated to the symplectic potential
\be 
\Theta_h = -\frac1{4\pi \GN} \int_{\scri}  \dot \bN \delta h.
\ee 
This potential is, after integration by part, equal to the holomorphic Ashtekar-Streubel symplectic potential provided we assume the boundary condition $\displaystyle{\lim_{u\to \pm \infty}  (\bN \delta h)(u,z,\bz) =0}$.
To avoid confusion, we also label the associated phase space with $h$, namely $\PS_h$.

In the appendix \ref{App:hE-1}, we show that we can readily adapt the $\Ah$-algebroid construction of Sec.\,\ref{sec:InitialCondition} to accommodate for this change of symplectic potential.
In particular, re-defining the algebroid bracket using $\tdt$ instead of $\dt$ for the anchor map, i.e.\footnote{$[\tau,\tau']^C_s$ is still the bracket \eqref{CFbrackettau}.}
\begin{equation}
    \lbr\tau,\tau'\rbr_s :=[\tau,\tau']^C_s+\tdtp\t{s}-\tdt\tp{s}, \label{hbracket}
\end{equation}
we prove that 
\begin{equation}
    [\tilde{\delta}_\tau, \tilde{\delta}_{\tau'}] h= 
- \tilde{\delta}_{\lbr \tau, \tau'\rbr } h. \label{algebroidReph}
\end{equation}
when $\tau,\tau'\in \sfTh$.

Moreover, if instead of requiring $\tau$ and $\tau'$ to satisfy the initial condition $\sI_\tau=0$ \eqref{tauintial}, we impose the dual EOM $\mathsf{E}_{-1}=0$ \eqref{dualEOM-1}, then the latter is preserved by $\lbr \tau, \tau'\rbr$:
\begin{equation}\label{bibi}
    \pa_u\lbr\tau,\tau'\rbr_{-1}
    =\big(\Dcal\lbr\tau,\tau'\rbr\big)_{-1}.
\end{equation}
The proof of \eqref{bibi} and that \eqref{hbracket} respects the EOM is the same as for the \hyperref[LemmaClosure]{lemma [$\sfT$-bracket closure]}.
In other words,
\begin{equation} \label{AnomalyLeibnizh}
    (\pa_u-\Dcal)\lbr\tau,\tau'\rbr= \big\lbr(\pa_u-\Dcal)\tau,\tau'\big\rbr +\big\lbr\tau,(\pa_u-\Dcal)\tau'\big\rbr+\tilde\delta_{(\pa_u-\Dcal)\tau}\tau'-\tilde\delta_{(\pa_u-\Dcal)\tau'}\tau,
\end{equation}
and the RHS vanishes for $\tau,\tau'\in\sfTh$.

We thus have two notions of $\A$-algebroid, the first one is $\Ao$, which is naturally adapted to the Ashtekar-Streubel symplectic potential; the other denoted $\Ah$ is adapted to the twistor symplectic potential.
Both share the same action on the shear.

We can then state the same theorem as in Sec.\,\ref{secAlgebroid}, but for the $\Ah$-algebroid.

\begin{tcolorbox}[colback=beige, colframe=argile] \label{theoremAlgebroidh}
\textbf{Theorem [$\Ah$-algebroid]}\\
The space $\Ah\equiv\big(\sfTh,\lbr\cdot\,,\cdot\rbr,\tilde\delta\big)$ equipped with the bracket \eqref{hbracket} and the anchor map,
\begin{align}
    \tilde\delta : \,&\Ah\to \X(\PS_h) \nn\\
     &\,\tau\mapsto\tilde\delta_\tau \label{defdeltah}
\end{align}
 forms a Lie algebroid over $\PS_h$.
\end{tcolorbox}

\ni Since the action $\tdt$ reduces to $\dt$ on $\PS$, we have that $\Ah$ here reduces to $\Ah$ from section \ref{theoremAlgebroids} if all the functionals $\tau$ depend only on $C$ and not $h$. 
To put it differently, $\big(\sfTh,\lbr\cdot\,,\cdot\rbr,\tilde\delta\big)$ is an extension of $\big(\sfTh,\lbr \cdot\,,\cdot\rbr,\delta\big)$ to $\PS_h$.

\subsection{Canonical action \label{Sec:CanAction}}

We now show that the action of $\A$ on the gravity phase space is canonical (here we use $\A$ to generically refer to $\Ao$ or $\Ah$).
So far, we have only described how $\A$ acts on $C$. We also need to describe its action on $\bN$.
For this, we use that
$\dt C$ is a functional of $C$ only. This action involves local terms $\pa_u^n D^m C$, but it also involves non local terms 
$\pa_u^{-n} D^m C$ where we choose\footnote{ The definition of $\pui$ is ambiguous and depends on a base point $\alpha$. We could also choose  $\pui O =\int_\alpha^u \rd u' O(u')\rd u'$ and more generally 
\begin{equation}
    \big(\pui[n]O\big)(u)=\int_\alpha^u \rd u_1\int_\alpha^{u_1}\rd u_2\ldots\int_\alpha^{u_{n-1}}\rd u_n O(u_n).
\end{equation} 
In this section we take $\alpha=+\infty$. Another choice will be used in the next section. The context makes it clear which inversion we use.}
$\pui O := \int_{+\infty}^u O(u') \rd u'$.
Moreover, if $\tau \in \A^{s+1}$ (cf.\,\eqref{defFiltration}), the non-locality is bounded since the most non-local term involves $\pui[s]C$ at most.\footnote{As we shall see explicit in \eqref{soltau43210}, $\dt C$ contains terms of the type $\pui(C\pui C)$, thence $\delta(\dt C)$ involves $\pui(\delta C\pui C)$ and $\pui\big(C\pui(\delta C)\big)$.
Using the generalized Leibniz rule \eqref{LeibnizRuleGeneral} for $\pui$, the first variation contributes to terms of the type $\big(\pui[(2+n)]C\big)(\pa_u^n\delta C)$ while the second variation involves terms like $\big(\pui[(1+n)]C\big)(\pa_u^{n-1}\delta C)$, for $n\in\N$.}
The general variation rule \eqref{actiondelta} then implies that we can write\footnote{Notice that by construction, the operator $\cD$ incorporates the variation of the field dependent symmetry parameters $\tau$.} 
\be 
\delta (\delta_\tau C)(y) =
\int_{\scri} \rd^3 x \, \cD_\tau(y,x)  \delta C(x),
\qquad
\cD_\tau(y,x) := \sum_{n=-s}^\infty \sum_{m=0}^{\infty}
\frac{ \delta (\delta_\tau C(y))}{\delta( \pa_u^nD^m C(x))}
\pa_u^nD^m,
\ee 
where $x=(u,z,\bz)$.
To write down the action of $\A$ on $\bN$ we need the dual operator $\cD_\tau^*(x,y)$, where the duality is defined by 
\be 
\int_{\scri} \rd^3 x \rd^3 y  \, \big(\cD_\tau^*(x,y) A(y)\big)   B(x):= \int_{\scri} \rd^3 x \rd^3 y \, A(x) \big(\cD_\tau(x,y) B(y)\big).
\ee 
In order to construct $\cD^*_\tau$ we therefore need to construct $(\pa_u^n D^m)^*$. For $D$, we have $D^*=-D$ since we are integrating over the sphere. 
Moreover, we also have $\pa_u^* =-\pa_u$ since we are assuming that the fields are Schwartzian, thence we can safely integrate by parts.
To construct $(\pa_u^{-1})^*$ we use that  that
\begin{align}
    \int_{-\infty}^\infty\rd u\, A(u)[\pui B](u) &=\int_{-\infty}^\infty\rd u\, \pa_u\left(\int_{-\infty}^u \rd u'A(u')\right)[\pui B](u) \nn\\
    &=\left.\left(\int_{-\infty}^u\rd u' A(u')[\pui B](u)\right)\right|_{u=-\infty}^{u=+\infty}-\int_{-\infty}^\infty\rd u\left[\int_{-\infty}^u\rd u' A(u')\right]B(u) \nn\\
    &=\int_{-\infty}^\infty\rd u\left[\int^{-\infty}_u\rd u' A(u')\right]B(u),
\end{align}
where the evaluation at the boundary drops since we picked $\pui=\int_{\infty}^u$.
Hence in that case, $(\pui)^*=\int_u^{-\infty}$.
We give the general expression of $(\pui)^*$ for arbitrary base point $\alpha$ in App.\,\ref{App:pui}.

The knowledge of $\big(D^*,\pa_u^*, (\pa_u^{-1})^* \big)$ defines $\cD_\tau^*$ and we find that the canonical action of $\A$ on $\bN$ is simply given by 
\be 
\delta_\tau \bN(x) = -\int_\scri \rd^3 y\, \cD^*_\tau (x,y) \bN(y).
\ee 
It is now straightforward to see that\footnote{We use that $[L_{\dt},\delta]=0$.}
\begin{align}
    4\pi \GN\big(L_{\dt}\Theta\big) &=L_{\dt} \int_\scri\bN \delta C =\int_\scri\Big(\dt\bN\delta C+\bN\delta(\dt C)\Big) \\
    &=\int_\scri \Big(-\big(\cD^*_\tau\bN\big)\delta C+\bN\big(\cD_\tau\delta C\big)\Big) =0. \nn
\end{align}
From there we infer that\footnote{Using $L_{\dt}=I_{\dt}\delta+\delta I_{\dt}$.} 
\begin{equation}
    \boxed{I_{\dt}\Omega=-\delta\Qt}, \label{canonicalRep}
\end{equation}
where $\Omega\equiv\delta\Theta$ is the symplectic form\footnote{The fact that the symmetry action is integrable is non-trivial and due to the fact that all the field dependency of $\tau_s$ is taken into account in the definition of the dual action operator $\cD^*_\tau$.} and 
\begin{equation}
    \boxed{\Qt:=\Qt^{-\infty}=I_{\dt}\Theta} \label{defNoetherCharge}
\end{equation}
is the Noether charge. This charge can be  written as an integral over $\scri$ as a direct consequence of \eqref{defNoetherChargeinterm}. Explicitly, this means that 
\be 
\boxed{\Qt =\frac{1}{4\pi \GN}\int_\scri \bN 
\Big( ( N-D^2 ) \t0+  2 DC\t1+ 3 CD \t1- 3 C^2\t2\Big)}. 
\ee 
Notice that the dependence on the higher spin $\tau_s$ is implicit through the time evolution $\E_s=0$ \eqref{tausevolb}.

The equations \eqref{canonicalRep} and \eqref{defNoetherCharge} prove that $\Qt$ is the Noether charge for the $\Ao$-algebroid action. 
It satisfies\footnote{We used that the morphism property $\big[\dtp,\dt\big]O =\delta_{\lbr\tau,\tau'\rbr}O$ of the symmetry action implies 
\begin{equation}
 \poisson{\Qtp,\{\Qt,O\}}-\poisson{\Qt,\{\Qtp,O\}}=\poisson{\{\Qtp,\Qt\},O} =\poisson{Q_{\lbr\tau,\tau'\rbr},O}.
\end{equation}
\vspace{-0.4cm}}
\be 
\boxed{\{Q_\tau, O\} =\delta_\tau O \qquad \textrm{and}\qquad
\poisson{Q_\tau, Q_{\tau'}}= -Q_{ \lbr \tau, \tau' \rbr }}, \label{NoetherChargeRep}
\ee
where $O$ is an arbitrary functional of $(C,\overbar{C})$.
Notice that we find an equivariant moment map,\footnote{Sign convention: $I_{\dtp}I_{\dt}\Omega=\Omega\big(\dt,\dtp\big)=\poisson{\Qt,\Qtp}=L_{\dt}\Qtp=\big(\mathrm{ad}^*_\tau (Q^*)\big)(\tau')=-Q(\mathrm{ad}_\tau\tau')=-Q\big(\lbr\tau,\tau'\rbr\big)\equiv-Q_{\lbr\tau,\tau'\rbr}$, with $\mathrm{ad}^*$ the coadjoint action \cite{Hall2013}.} namely a representation without any 2-cocycle.

\paragraph{Remark:}
We expect the discussion to generalize straightforwardly when working with the covariant news tensor $\hat N=N-\rho$, where $\rho=\rho_{AB}m^Am^B$ is the Geroch tensor \cite{Geroch1977, Compere:2018ylh, Dray:1984rfa}. 
The latter is necessary when considering a conformal compactification which is not the round sphere.
It also ensures that the BMS charge algebra is 2-cocycle free \cite{Rignon-Bret:2024wlu, Rignon-Bret:2024gcx}.
Besides, notice that the spin 1 charge matches with the Dray-Streubel charge \cite{Dray:1984rfa} and was proven to define a $\gbms$ moment-map in \cite{Barnich:2021dta, Freidel:2024jyf}.\\

The same analysis goes through if we consider the covariant charge $\Ht$ instead. 
In that case,
\begin{equation}
    \Omega_h=\delta\Theta_h,\qquad I_{\tdt}\Omega_h=-\delta\Ht,\qquad \Ht:=\Ht^{-\infty}=I_{\tdt}\Theta_h, \label{canonicalRepH}
\end{equation}
where the Noether charge is written as the following integral over $\scri$:
\be 
\boxed{\Ht =-\frac{1}{4\pi \GN}\int_\scri \dot{\bN}\big(C\t0-D\t{-1}\big)}. 
\ee 
$\Ht$ is the Noether charge for the $\Ah$-algebroid action. 
It satisfies
\be 
\boxed{\{\Ht, O\} =\tdt O \qquad \textrm{and}\qquad
\poisson{\Ht, H_{\tau'}}= -H_{ \lbr \tau, \tau' \rbr }}. \label{NoetherChargeRepH}
\ee

\subsection{Arbitrary cut \label{sec:ArbitraryCut}}

So far the charges we have constructed are associated with constant $u$ cuts. 
We need to be able to construct Noether charges associated with any cut  $S(U)=\{ u= U(z,\bz)\}\subset\scri$.
 where $U \in \Ccel{(-1,0)}$. 
Here, we generalize the Noether charge construction to accommodate any such cuts of $\scri$.
To do so, we promote the charge aspect to a 2-form on $S$. 
One defines\footnote{Recall that $\bfm\epsilon_S\equiv i\frac{\rd z \wedge \rd \bz}{P\bP}$.}
\begin{empheq}[box=\fbox]{align}
    \mathsf{Q}_s : = \tQ_s\,\bfm\epsilon_S + \tQ_{s-1} \,\rd u \wedge \frac{i\rd \bz}{\bP}, \qquad
 \mathsf{Q}_\tau := \sum_{s=-1}^\infty \tau_s \mathsf{Q}_s. \label{Qaspect2form}
\end{empheq}
Denoting $\rd^3 x:= \rd u \wedge \bfm{\epsilon}_S$ the volume element on $\scri$ and $\rd = \rd u \,\pa_u + \frac{\rd z}{P} D + \frac{\rd \bz}{\bP} \bD$ its differential, one evaluates
 \be
 \rd \mathsf{Q}_\tau = &
 \sum_{s=-1}^\infty
 \left(\tau_s \big(\pa_u \tQ_s
 - D\tQ_{s-1}\big) +\pa_u\tau_s \tQ_s - D\tau_s \tQ_{s-1}\right)
 \rd^3 x \cr
 = &
 \left( \sum_{s=0}^\infty (s+1) C \tau_s \tQ_{s-2} - D\tau_{-1} \tQ_{-2}
 + (\pa_u\tau_{-1}- D\t0) 
 \tQ_{-1}  + 
 \sum_{s=0}^\infty ( \pa_u\tau_s- D\tau_{s+1}) \tQ_s \right)
 \rd^3 x \cr 
 = &
 \left( \big(C\t0- D\t{-1}\big)\tQ_{-2}
 + \big(\pa_u\t{-1}- D\t0 + 2C\t1\big) 
 \tQ_{-1} \right)
 \rd^3 x,
\ee
where we used the EOM \eqref{Qevolve} in the second equality and the dual EOM \eqref{tausevolb} in the last one. 
Hence,
\begin{equation}
    \boxed{\rd \mathsf{Q}_\tau=\big(\sI_\tau \tQ_{-2} +\E_{-1}\tQ_{-1}\big)\rd^3 x}.
\end{equation}
The initial condition \eqref{tauintial} implies that the first term vanishes while we know that $D\E_{-1}=\dt C$, cf.\,\eqref{-1id}.
Therefore, since $\tQ_{-1}= D\bN$, we get that\footnote{$\mathsf{Q}_\tau=\mathsf{Q}_\tau'$ on a constant $u$ cut, as it should.}
\be \label{FluxBalanceLaw}
\rd \mathsf{Q}_\tau' 
=  - \big(\bN \dt C \big) \rd^3 x, \qquad
\mathsf{Q}_\tau'
:=
\mathsf{Q}_\tau + \E_{-1}(\tau) \bN \,\rd u \wedge \frac{i\rd \bz}{\bP}. 
\ee 
When the left-handed radiation vanishes, i.e. $\bN=0$, we have that 
$\mathsf{Q}_\tau'=\mathsf{Q}_\tau$ is covariantly conserved: $\rd \mathsf{Q}_\tau =0$.\footnote{Following the discussion in Sec.\,\ref{Sec:CovCharges}, if we impose $\E_{-1}=0$ rather than the initial constraint $\sI_\tau=0$, then $\rd\mathsf{H}_\tau=(\dot{\bN}\tdt h)\rd^3x$, or equivalently (using $\pa_u\sI_\tau=\dt C$)
\begin{equation} \label{Haspect2form}
    \rd\mathsf{H}'_\tau=-(\bN\dt C)\rd^3x,\qquad \mathsf{H}'_\tau:=\mathsf{H}_\tau-\sI_\tau\bN\bfm{\epsilon}_S.
\end{equation}}
One can use $\mathsf{Q}_\tau$ to define the Noether charge at any cut,
\be \label{QtArbitraryCut}
Q_\tau^{U} \equiv
\frac{1}{4\pi\GN}\int_{S(U)} \mathsf{Q}_\tau
= \frac{1}{4\pi\GN}\sum_{s=-1}^{\infty} \int_{S} \big(\tau_s \tQ_s + \tau_s \tQ_{s-1} DU \big)(U(z,\bz),z,\bz)\, \bfm\epsilon_S, 
\ee 
and similarly for  $\Qt^{'U}$.

\paragraph{Remark:}
The equation \eqref{FluxBalanceLaw} is the usual flux balance law satisfied by the Wald-Zoupas charge prescription \cite{Wald:1999wa}.
If there exists a spacetime realization in terms of bulk diffeomorphisms of the higher spin symmetries, it would be interesting to make contact with this prescription. 
See \cite{Rignon-Bret:2024wlu} and references therein for a recent review.

\section{Solution of the dual EOM \label{Sec:SolDualEOM}}

In this section, we study the solution to the dual EOM and give explicit solutions to all order in the $\GN$ expansion.
In particular, we introduce a Lie algebroid map $\bft$ which defines a unique $\bft(T)\in\Ao$ or $\bft(T)\in\Ah$ given an element $T\in\V(S)$.
As we will see in section \ref{sec:RenormCharge}, the knowledge of this explicit solution gives a geometrical and natural understanding of the charge renormalization procedure that was devised in \cite{Freidel:2021ytz, Geiller:2024bgf}.

We have proven that the symmetry algebroids $\Ao$ and $\Ah$, which are subalgebroids of $\A$ are canonically represented on the gravitational phase space.
This algebroid $\A$ is filtered and it is natural to understand the associated gradation. 
This is what we now describe. 

\subsection{Filtration and gradation \label{sec:filtration}}

There exists a natural filtration of $\A$,
\begin{equation}\label{defFiltration}
    \{0\} \subset\A^{-1}\subset\A^0\subset \A^1\subset \ldots\subset\A^s\subset \ldots\subset \A \quad\textrm{such that}\quad \A=\bigcup_{n+1\in \mathbb{N}}\A^n,
\end{equation}
where the $\A^s\subset\A$ are the subspaces for which $\t{n} = 0$ for $n>s$.
This filtration is compatible with the bracket in the sense that
 \begin{equation}
    \lbr\A^s,\A^{s'}\rbr\subseteq \A^{s+s'-1},\quad \mathrm{when}\quad s,s'>0. \label{filtrationCompatible}
 \end{equation}
 The spin $0$ is an exception since in general we only have that  $\lbr\A^s,\A^0\rbr\subseteq \A^s$.\footnote{Indeed, if $\tau\in\A^s$ and $\tau'\in\A^0$, then $\lbr\tau,\tau'\rbr_s=\dtp\t{s}$ which does not generically vanish.}
There exists an associated graded algebroid $\G{\A}$ to the filtered algebroid $\A$, defined as
\begin{equation}
\G{\A}=\bigoplus_{s=-1}^\infty\sfT_s,
\end{equation}
with 
\begin{equation}
    \sfT_{-1}=\A^{-1}\qquad \textrm{and}\qquad \sfT_s=\A^s/\A^{s-1}\quad\textrm{for }s\geqslant 0.  \label{defgradationTh}
\end{equation}
Each $\sfT_s$ is an equivalence class and we can write $\A^s= \bigoplus_{n=-1}^s\sfT_n$.
It is easy to show that we have the isomorphism 
\vspace{-0.2cm}
\be \label{IsoQ}
\sfT_s \simeq  \V_s.
\ee 
To prove this we just look at the evolution equations for the equivalence class $[\tau]\in \sfT_s$. 
Let us assume that $\tau\in\A^s$ for $s\geq 0$.
Since $\pa_u [\tau]_s=0$,
this means that $[\tau]_s$ is constant in time and therefore equal to its value at any cut of $\scri$. 
For definiteness we choose the cut to be at $u=0$, hence we have that  
\be 
[\tau]_s=\T{s},\quad  \mathrm{where}\quad  T_s:= \tau_s|_{u=0}\in \V_s.
\ee
The other values, $[\tau]_{s-n}\neq 0$ for $1\leq n\leq s$  are  determined recursively from $[\tau]_s$ by the equation of motion. 
Describing this construction explicitly is the purpose of the next subsections.

In the following, rather than working with equivalent classes, it is more convenient to pick the natural representative element $\tau(\T{s})\in\A^s,\,s\geq 0,$ which is fully determined by $\T{s}\in\V_s$.
To do so we use that 
the isomorphism \eqref{IsoQ} means that we have a projection map $p: \A^s \to \V_s$ which maps $\tau \to T_s = \tau_s|_{u=0}$. 
This map admits an ``inverse'', i.e. a section $\tau: \V_s \to \A^s$,
\begin{align}
    \tau :\,&\V_s\to \A^s \nn\\
     &\, T_s\mapsto \tau(T_s), \label{maptau}
\end{align} 
such that  $(p \circ \tau)(T_s) = T_s$.
This section denoted $\tau(T_s)(u,z,\bz) \in \A^s$ is the unique solution of the dual EOM with initial condition given by\footnote{By definition, we also have that $\tau_{s+n}(u,z,\bz)\equiv 0$ for all $n>0$ since  $\tau(T_s)\in \A^s$.}
\be 
\tau_s(T_s)\big|_{u=0}=T_s, \qquad \tau_{s-n}(T_s)\big|_{u=0}=0,\quad \mathrm{for} \quad 1 \leq n \leq s.
\ee 
These solutions are polynomial in $u$ and thus diverge at infinity.
This is however not an issue since the charge aspects $\tQ_s\in\cS$ are part of the Schwartz space, cf. footnotes \ref{foot1} and \ref{foot2}.
For simplicity, we study the cut $S\equiv S(0)$, i.e. $T$ is the value of $\tau$ at the cut $u=0$ of $\scri$. The construction done here can be adapted to any other choice of cut.
In particular, a constant finite cut $S(u_0),\,u_0\in\R$ is equivalent to $S$ by translation, i.e. by replacing $u\rightarrow u-u_0$ in the upcoming analysis, for instance in \eqref{soltaup}.

Since $\tau:\V_s\to \A^s$ is a linear map, it can be extended by linearity to a map $\bft: \Vs \to \A$  on the full space using that $\Vs =\bigoplus_{s=0}^\infty \V_s$:
\begin{align} \label{maptaubis}
    \bfm{\tau} :\,& \,
    \Vs \to \A \nn\\
     & \, T \mapsto \bft(T) := \sum_{s=0}^{\infty} \tau(\T{s}).
\end{align}

The filtration for $\A$ extends to a filtration for $\Ah$ and $\Ao$ which treats the modes $\t{-1}$ differently. 
On one hand, the construction of this subsection goes along the same lines with $\Ah$, where $\t{-1}$ is treated on par with the positive degrees.\footnote{We have in particular the map $\bft:\V(S)\to\Ah$.}
On the other hand, $\t{-1}$ is determined by $\t0$ through the initial condition when dealing with $\Ao$.\footnote{The only quantity that matters to define the Noether charge is $D\t{-1}$, which precisely is purely given in terms of $\t0$.
Furthermore, the transformation of the shear $\dt C$ does not involve $\t{-1}$ at all.
Therefore, it is sufficient to consider elements $\bft(T)$ with $T\in\Vs$.}
In this case, since $(\Ao)^{-1}$ is central,\footnote{Indeed, if $\tau\in(\Ao)^s$ and $\tau'\in(\Ao)^{-1}$, then $\lbr\tau,\tau'\rbr_n=0,\,n\geqslant 0$ and $\lbr\tau,\tau'\rbr_{-1}=-\dt\tp{-1}$. 
Since $D\tp{-1}=0$, $\tp{-1}$ is field independent and $\dt\tp{-1}=0$.
Therefore $\lbr(\Ao)^s,(\Ao)^{-1}\rbr=\{0\}$.}
we can consider the quotient Lie algebroid $\Ao/(\Ao)^{-1}$ for which $\G{\Ao/(\Ao)^{-1}} =\bigoplus_{s=0}^\infty\sfT_s$.
By construction, the bracket at degree $-1$ vanishes on this quotient space (and the other degrees are unchanged). 
In particular, the super-translations commute.
This is thus the right object to consider if we want to recover the generalized BMS algebra as a sub-case of the higher spin symmetries analysis.

\subsection{Explicit construction of the section for $s\leq 4$}

To illustrate the construction of the map $\bfm{\tau}$, we start with an explicit example. 
Let us assume that $\t{n}\equiv 0,\,n\geq 5$. 
Since $\pa_u\t{s}=D\t{s+1}-(s+3)C\t{s+2}$, we deduce that $\t4$ is constant in time.
By taking\footnote{In this section and the following one we anchor the inverse operator at the cut $0$ where the initial condition is set up. 
In the previous section this operator was anchored at $\infty$. We use the same notation $\pa_u^{-1}$ as the context is clear. } $\pui=\int_0^u$, we can then write $\t{s},\,s=0,1,2,3,4$ as follows:
\bs
\label{soltau43210}
\begin{align}
    \t4(T) &=\T4, \\
    \t3(T) &=\T3+uD\T4, \\
    \t2(T) &=\T2+uD\T3+\frac{u^2}{2}D^2\T4-5(\pui C)\T4, \\
    \t1(T) &=\T1 +uD\T2+\frac{u^2}{2}D^2\T3+\frac{u^3}{3!}D^3\T4-5\pui D\big((\pui C)\T4\big)-4\pui \big(C(\T3+uD\T4)\big), \\
    \t0(T) &=\sum_{k=0}^4\frac{u^k}{k!}D^k\T{k}-5(\pui D)^2\big((\pui C)\T4\big)-4(\pui D)\pui\big(C(\T3+uD\T4)\big) \nn\\
    &-3\pui\left(C\left(\T2+ uD\T3+\frac{u^2}{2}D^2\T4 -5(\pui C)\T4\right)\right).
\end{align}
\es
Now that the pattern is clear, we write a general solution for $\tau\in \A^s$.

\subsection{General solution}

When $\tau\in\A^s$, then $\dot\tau_s\equiv 0$ and we can solve the dual EOM \eqref{tausevolb} recursively.
The solution which satisfies $\t{p}(u=0,z,\bz)= 0$ for $p > s$ and
\begin{equation}
    \t{p}(u=0,z,\bz)=\T{p}(z,\bz), \quad 0\leqslant p\leqslant s,
    \label{intconstant}
\end{equation}
is given by $\t{s}=T_s$, $\t{s-1}= \T{s-1} + uD T_s$ and in general by
\begin{equation}
    \boxed{\t{s-p}(T)=\sum_{k=0}^{p}\frac{u^{p-k}}{(p-k)!}D^{p-k}T_{s-k}- \sum_{k=0}^{p-2}(s-k+1)\big(\pui D\big)^{p-k-2} \pui(C\t{s-k})\in\A_s\,}, \label{soltaup}
\end{equation}
for $2\leqslant p\leqslant s$.
These relations show that $\tau_{s-p}$ is defined recursively in terms of $\tau_s,\tau_{s-1}, \ldots, \tau_{s-p +2 }$ and $T_s, \ldots, T_{s-p}$.
It also shows that $\tau_{s-p}$ is a polynomial in the shear $C$ and its derivatives of degree $\lfloor p/2\rfloor$.
Solving \eqref{soltaup} can then be done recursively starting from the expression for  $\t{s}$, then $\t{s-1}$, $\t{s-2}$ and so on, as we did in the warm-up.
Doing so, \eqref{soltaup} is an explicit solution of \eqref{tausevolb}.
The reader can readily check that  this recursive definition satisfies the 
dual equation of motion
\be 
\dot\tau_{s-p}(T) &=\sum_{k=0}^{p-1}\frac{u^{p-k-1}}{(p-k-1)!}D^{p-k}\T{s-k}-\sum_{k=0}^{p-2}(s-k+1)\big(\pui D\big)^{p-k-2} (C\t{s-k})\cr
& =D\t{s-p+1}(T)-(s-p+3)C\t{s-p+2}(T).
\ee

Because when pairing $\t{s}$ with $\tQ_s$ the initial condition \eqref{initialCondition} amounts to a change of the spin-0 charge from $\tQ_0$ to $\tq_0:=\tQ_0-\bN C$---cf. \eqref{bondimass}---we do not need to solve for the parameter $\t{-1}$ when we restrict to $\tau(T)\in\Ao$. On the other hand, if we consider $\tau(T)\in\Ah$, then we just have to extend the range of $p$ to $2\leqslant p\leqslant s+1$ in the solution \eqref{soltaup}.

Next, let us choose a solution for which only $\T{s}\neq 0$.
We get that\footnote{Recall that $\bfm{\tau}(T)=\sum_s\tau(\T{s})$, where the sum runs over all the values of $s$ for which $\pi_s(T)\neq 0$. \label{foot:linearity}}
\begin{equation}
    \boxed{\t{s-p}(\T{s}) =\frac{u^{p}}{p!}D^{p}\T{s}-\sum_{k=0}^{p-2}(s-k+1)\big(\pui D\big)^{p-k-2}\pui (C\t{s-k})},\qquad p\geqslant 0. \label{soltaupTs}
\end{equation}
By construction, this $\tau(\T{s})$ is precisely a representative element of $\sfT_s$,\footnote{Notice however a subtlety in the fact that the gradation depends on the cut $S(U)$ of $\scri$, so that to be precise, we should have written $\mathcal{G}_{U}(\A)$ in section \ref{sec:filtration}.}
which makes explicit the aforementioned isomorphism with $\V_s$, since for a given shear $C$, $\tau(\T{s})$ is fully determined by $\T{s}\in\V_s$.

The form \eqref{soltaupTs} of the `polynomial class' of solutions of \eqref{tausevolb} is handy for section \ref{SecComparison}.
However, we also present a systematic approach in Sec.\,\ref{Sec:systApproach}, more convenient for Sec.\,\ref{sec:RenormCharge}.

\subsection{Systematic approach \label{Sec:systApproach}}

First of all, we introduce $\hs$ and $\hS$, respectively the spin and shift operators, acting on the series $\tau\in \tV(\scri)$ as\footnote{The same formulas are valid for any $\alpha_s\in \Ccar{\delta,-s}$ or $\alpha_s\in\Ccel{(\Delta,-s)}$.}
\begin{equation}
    \hs \tau_s =s\t{s}\qquad \textrm{and} \qquad 
    \hS\t{s}=\t{s+1}.
\end{equation}
Hence $\hs$ is an operator of degree $(0,0)$ and $\hS$ is an operator of degree $(0, 1)$. 
The usage of $\hS$ turns out to be convenient for any systematic approach.
Furthermore, $\hs$ and $\hS$ satisfy the following commutation relations with $D$ and $C$:\footnote{Notice that from this point of view, $C$ acts as an operator that shifts the degree by $-2$ and multiplies the graded vector by its ``eigenvalue'' $C$.}
\begin{equation}
    \hS D=D\hS,\quad\hS C=C\hS,\quad D\hs=(\hs+1)D,\quad\hs C=C(\hs-2),\quad\hs\hS=\hS(\hs+1).
\end{equation}
This formalism then allows us to write $\Dcal$ as follows:
\begin{equation}
    \Dcal \tau=\big(D-C (\hs+1)\big)\tau, \label{DcalOp}
\end{equation}
which in particular means that
\begin{align}
    (\Dcal \tau)_s
    =(D\tau)_s-(\hs+3 )(C\tau)_{s}=\big(D-C(\hs+1)\hS\big)\t{s+1}.
\end{align}
Hence $(\Dcal^n \tau)_s$ takes the compact form
\begin{equation}
    (\Dcal^n \tau)_s = \big(D-C(\hs+1)\hS\big)^n\t{s+n}.
\end{equation}

Next, we recast the dual evolution equation \eqref{tausevolb} as follows:
\begin{equation}
    \boxed{\tau=\pui\Dcal\tau +T}. \label{solEOMcompact}
\end{equation}
More concretely,
\begin{equation}
    \t{s}=\big(\pui D-\pui C(\hs+1)\hS\big)\t{s+1}+\T{s},
\end{equation}
where it is understood that $\pui$ acts on all products on its right.
Therefore, when looking for $\t{s-1}$, $\t{s-2},\ldots$ and so on, we see that formally, one does nothing more than taking an extra power of the operator $\pui\Dcal$ at each step.
Concretely, picking $\tau\in\A^s$, we can write
\begin{equation}
    \t{s-n}(T) =\sum_{k=0}^n
    \left(\pui D-\pui C(\hs+1)\hS
    \right)^k T_{s-n+k},
\end{equation}
or equivalently
\begin{equation}
    \t{n}(T)=\sum_{k=0}^{s-n}\left(\pui D-\pui C(\hs+1)\hS\right)^k\T{n+k}.
\end{equation}
At that stage, this solution is still very formal.
As we did in \eqref{soltaupTs}, the next step is to find a representative element of $\sfT_s$, i.e. we take only $\T{s}\neq 0$.
By inspection, we find that\footnote{The key to find the lower bound in the sum over $k$ is to realize that for each term in the sum, the shift operator acts between 0 and $k$ times.
Each term at a definite $k$ is thus a linear combination of $\T{n+k},\ldots,\T{n+2k}$.
$k$ needs to be sufficiently large, so that $n+2k\geqslant s$.
We therefore have to take the integer part of $\frac{s-n+1}{2}$.}
\begin{equation}
    \boxed{\t{n}(\T{s})=\sum_{k=\lfloor\frac{s+1-n}{2}\rfloor}^{s-n}\left(\pui D-\pui C(\hs+1)\hS\right)^k\T{n+k}}. \label{soltaunTs}
\end{equation}
Notice that we have to deal with an operator of the form $(A-B)^k$, where in our case $B$ is linear in the shear and in $\hS$.
A good way to organize the solution is thus to make patent this expansion in powers of $C$.
For this, we can use the following formula:
\begin{equation}
    (A-B)^k=\sum_{\l=0}^k(-1)^\l\sum_{P=k-\l}A^{p_0}BA^{p_1}BA^{p_2}\ldots BA^{p_\l},\qquad\textrm{with}\quad P=\sum_{i=0}^\l p_i.
\end{equation}
Therefore,
\begin{align}
    &\left(\pui D-\pui C(\hs+1)\hS\right)^k = \label{puiDcalk}\\
    &=\sum_{\l=0}^k(-1)^\l\sum_{P=k-\l}\pui[p_0]D^{p_0}\bigg(\pui\Big\{C(\hs+1)\pui[p_1]D^{p_1}\Big(\pui\Big\{C(\hs+1)\pui[p_2]D^{p_2}\Big(\ldots \nn\\
    &\qquad\qquad\qquad\ldots \pui\Big\{C(\hs+1)\pui[p_{\l-1}]D^{p_{\l-1}}\Big(\pui\Big\{C(\hs+1)\pui[p_\l]D^{p_\l}\hS^\l\Big\}\Big)\Big\}\ldots \Big)\Big\}\Big)\Big\}\bigg). \nn
\end{align}
For visual clarity, we used round and curly parentheses to distinguish (when necessary) between the action of $D$ versus $\pui$.

This way \eqref{soltaunTs} and \eqref{puiDcalk} of writing the solution of \eqref{tausevolb} is particularly convenient for the discussion about the renormalized charges in section \ref{sec:RenormCharge}.

This result is also an opportunity to show how algorithmic the construction is and how concise the results and the notation we have been using are.
Indeed, $\dt C$ is very simply written in terms of $\t0,\t1$ and $\t2$---cf. \eqref{dtauC}---but the amount of complexity hidden in the latter is tremendous.
Having proven that $\dt$ is a representation of the symmetry algebroid $\Ao$ on the phase space---and here we want to emphasize that $\tau$ contains all the symmetry parameters $\T{s}$ and arbitrarily high powers of $C$---is clearly a non-perturbative result.
To be more accurate, if we formally introduce $\gN$ and $\bgN$ as small dimensionless parameters, then by rescaling $C\rightarrow\gN C$ and $\bC\rightarrow\bgN\bC$, our result is non perturbative in $\gN$ and at leading order in $\bgN$.
We can use $\gN$ and $\bgN$ as a bookkeeping device.

We now turn to the next section, where we exploit the machinery developed so far in order to clarify the notion of renormalized charge introduced in \cite{Freidel:2021dfs, Freidel:2021ytz} and partially corrected in \cite{Geiller:2024bgf,Kmec:2024nmu}.

\section{Renormalized charge and its action on the shear \label{sec:RenormAction}}

In this section we show that the charge renormalization procedure devised in \cite{Freidel:2021ytz, Geiller:2024bgf}, amounts to evaluate the smearing variable at the cut $u=0$.
We then compute the action of the renormalized charges at quadratic order onto the shear to show that we recover the results of the aforementioned papers.

\subsection{Renormalized charge aspect \label{sec:RenormCharge}}

The goal of this subsection is to recast the Noether charge $\Qt$, when $\tau=\bft(\T{s})$, as the integral over the sphere of a certain charge aspect $\hq_s(z,\bz)$ smeared against the symmetry parameter $\T{s}$ that defines the $\tau(T_s)$.

As we did in section \ref{Sec:SolDualEOM}, let us start with an example for the lowest spin-weights.
Consider successively $\tau=\bft(\T{s})$ for $s=0,1,2,3,4$.
We denote
\begin{equation}
    \cq^u_{\T{s}}\equiv Q^u_{\bft(\T{s})}=\sum_{n=-1}^sQ^u_n\big[\t{n}(\T{s})\big].
\end{equation}
Using the explicit computation \eqref{soltau43210}, we deduce that
\bs
\begin{align}
    (4\pi \GN)\cq^u_{\T0} &=\int_S\big(\tQ_0-\bN C\big)\T0\equiv \int_S\tq_0\T0, \\
    (4\pi \GN)\cq^u_{\T1} &=\int_S\left(\tq_0 uD\T1+\tQ_1\T1\right)=\int_S\big(-uD\tq_0+\tQ_1\big)\T1, \\
    (4\pi \GN)\cq^u_{\T2} &=\int_S\left(\tq_0\left(\frac{u^2}{2}D^2\T2-3\T2\pui C\right)+\tQ_1 uD\T2+\tQ_2\T2\right) \nn\\
    &= \int_S\left(\frac{u^2}{2}D^2\tq_0-uD\tQ_1+\tQ_2-3\tq_0\pui C\right)\T2, \\
    (4\pi \GN)\cq^u_{\T3} &=\int_S\left(\tq_0\left(\frac{u^3}{3!}D^3\T3-4D\big(\T3\pui[2]C\big)-3D\T3\pui(uC)\right)\right. \nn\\
    &\qquad+\left.\tQ_1\left(\frac{u^2}{2}D^2\T3-4\T3\pui C\right)+\tQ_2 uD\T3+\tQ_3\T3\right) \nn\\
    &=\int_S\left(-\frac{u^3}{3!}D^3\tq_0+4D\tq_0\pui[2]C+3D\big(\tq_0\pui(uC)\big)\right. \\
    &\,\hspace{3cm}\left.+\frac{u^2}{2}D^2\tQ_1-4\tQ_1\pui C-uD\tQ_2+\tQ_3\right)\T3, \nn
\end{align}
\es
and similarly for $\cq^u_{\T4}$.
We can thus identify the charge aspects
\bs
\label{renormcharge}
\begin{align}
\tq_0 &\equiv q_0-\bN C, \\
\tq_1 &\equiv q_1+uD(\bN C), \\
\tq_2 &\equiv q_2-\frac{u^2}{2}D^2(\bN C)-3\tq_0\pui C, \\
\tq_3 &\equiv q_3+\frac{u^3}{3!}D^3(\bN C)+4D\tq_0\pui[2]C+3D\big(\tq_0\pui(uC)\big)-4\tQ_1\pui C, \\
\tq_4 &\equiv q_4-\frac{u^4}{4!}D^4(\bN C)-5D^2\tq_0\pui[3]C-4D\big(D\tq_0\pui[2](uC)\big)-\frac32 D^2\big(\tq_0\pui(u^2C)\big) \nn\\
& +5D\tQ_1\pui[2]C+ 4D\big(\tQ_1\pui(uC)\big)-5\tQ_2\pui C +15\tq_0\pui\big(C\pui C\big), \label{tq4}
\end{align}
\es
where 
\be 
q_s =\sum_{n=0}^s \frac{(-u)^n}{n!} D^n \tQ_{s-n}. \label{defqs}
\ee 
The reader can check that the $\tq_s$ defined by \eqref{renormcharge} are conserved when no radiation is present.
More precisely, $\pa_u\tq_s$ involves only terms that contain $\bN$---see App.\,\ref{App:tq4} for the explicit computation of $\pa_u\tq_4$.
We dub $\tq_s$ the \emph{renormalized charge} aspect of helicity $s$.\footnote{Such a renormalization procedure depends on the choice of cut $S(U)$.
For simplicity, as we did in section \ref{Sec:SolDualEOM}, we present the discussion associated with the cut $S\equiv S(0)$.}
Up to the term\footnote{This term is also just the consequence of swapping $\tQ_0$ for $\tq_0$ in the definition \eqref{defqs} of $q_s$.} $(-1)^{s+1}\frac{u^s}{s!}D^s(\bN C)$ and to the fact that the Bondi mass $\tq_0$ enters in our expressions rather than the covariant mass $\tQ_0$, the expressions for the spins $0,1,2$ were first given in \cite{Freidel:2021dfs} while the form of $\tq_3$ was first given in \cite{Geiller:2024bgf}.
The fact that the renormalized charges $\tq_s$ only involve the combination $\tQ_0-\bN C$, and never the charge $\tQ_0$ alone is perfectly natural in our construction, since this is a direct consequence of the initial constraint \eqref{initialCondition}, cf. \eqref{bondimass}.
The expression $\tq_4$ was also partially given  in \cite{Geiller:2024bgf}: the part linear in $(C,\tQ)$ is correctly reproduced by their equation (6.18); however the term of higher order, namely
\begin{equation}
    15\tq_0\pui\big(C\pui C\big), \label{cubicUs}
\end{equation}
is  incorrectly written as
\begin{equation}
    \frac{15}{2}\tQ_0(\pui C)(\pui C) \label{cubicMarc}
\end{equation}
in \cite{Geiller:2024bgf}.
Notice however that the time derivative of \eqref{cubicMarc}, namely
\begin{equation}
    \pa_u\left(\frac{15}{2}\tQ_0(\pui C)(\pui C)\right)= 15\tQ_0C\pui C+\frac{15}{2}\pa_u\tQ_0(\pui C)(\pui C),
\end{equation}
equals the time derivative of \eqref{cubicUs}, namely
\begin{equation}
    \pa_u\big(15\tq_0\pui(C\pui C)\big)=15\tq_0C\pui C+15\pa_u\tq_0\pui(C\pui C),
\end{equation}
up to terms that vanish when there is no radiation.
We understand that fact as an indication of how one can easily be misled in the construction of the renormalized charges.
Two terms can have the same behavior upon time derivation, but actually lead to a different action on the phase space.
Without a general procedure, it is incredibly cumbersome to work out the correct notion of $\tq_s$ for arbitrary $s$ and at arbitrary order in $\gN$.
The fact is, our construction precisely realizes this for free.
It guarantees that the charges form an algebra and is non-ambiguous when it comes to defining the renormalized charge.
Besides, as we mentioned already, our algorithm (understand the systematic usage of the dual EOM) is blind to the peculiar value of helicity one wishes to consider or to the order in $\gN$ one wishes to work at.

Summarizing the results so far, we showed that 
\begin{equation} \label{defcqTsu}
    \cq^u_{\T{s}}\equiv Q^u_{\bft(\T{s})}=\frac{1}{4\pi \GN}\sum_{n=-1}^s \int_S\tQ_n\t{n}(\T{s})=\frac{1}{4\pi \GN}\int_S\tq_s\T{s},\qquad s=0,\ldots,4.
\end{equation}
We are thus ready to state the following theorem.

\begin{tcolorbox}[colback=beige, colframe=argile]
\textbf{Theorem [Noether charge of spin $s$]}\\
The Noether charge $\cq_{\T{s}}$ associated to the symmetry parameter $\T{s}$ of helicity $s$ is written as the following corner integral:
\begin{equation}
    \cq_{\T{s}}\equiv \cq^{-\infty}_{\T{s}}=\frac{1}{4\pi \GN}\int_S\hq_s\T{s}, \qquad s\geqslant 0, \label{Noetherhatqs}
\end{equation}
with $ \cq_{\T{s}}^u$ given by \eqref{defcqTsu} for all $s\in\N$ and
\begin{equation}
    \hq_s(z,\bz)=\lim_{u\to -\infty}\tq_s(u,z,\bz).
\end{equation}
The renormalized charge aspect $\tq_s$ satisfies $\pa_u\tq_s=0$ in the absence of radiation $\bN=0$.
\end{tcolorbox}

\paragraph{Proof:}
We present the way to construct $\tq_s$ systematically in the appendix \ref{App:renormcharge}, cf. \eqref{deftqs}, where we also show that the renormalized charge $\th_s$ associated to the covariant charge $\Ht$ matches with the one recently proposed in \cite{Kmec:2024nmu}.
To show that $\pa_u\tq_s=0$ in the absence of left-handed radiation, simply notice on the one hand that, cf. \eqref{Qtdot},
\begin{equation}
    \pa_u \cq^u_{\T{s}} =-\frac{1}{4\pi \GN}\int_S\bN\delta_{\tau(\T{s})}C=0 \qquad\textrm{if}\qquad\bN=0,
\end{equation}
while on the other hand, we also have that
\begin{equation}
    \pa_u \cq^u_{\T{s}} =\frac{1}{4\pi \GN}\pa_u\left(\int_S\tq_s\T{s}\right)= \frac{1}{4\pi \GN}\int_S\pa_u\tq_s\T{s}.
\end{equation}
This concludes the proof.

\paragraph{Remark:}
In a non-radiative strip of $\scri$, where $N=0$ for $u\in[0,u_0]$, we get that
\begin{equation}\label{deftqs0rad}
    \boxed{\tq_s=\sum_{k=0}^s\frac{(-u)^k}{k!}\big(\Dcal^{\ast k}\tQcal\big)_s}.
\end{equation}
where $\tQcal_s\equiv\tQ_s$ for $s >0$; $\tQcal_0\equiv\tq_0=\tQ_0-\bN C$ and $\tQcal_s\equiv 0$ for $s< 0$; while $\big(\Dcal^*\tQ\big)_s=D\tQ_{s-1}+(s+1)C \tQ_{s-2}$. 
We present the proof of \eqref{deftqs0rad} at the end of App.\,\ref{App:renormcharge}.
We can easily check that this renormalized charge aspect is conserved in the strip if we also assume that $\bN=0$ since
\begin{equation}
    \pa_u\tq_s=-\sum_{k=1}^s\frac{(-u)^{k-1}}{(k-1)!}\big(\Dcal^{\ast k}\tQcal\big)_s +\sum_{k=0}^s\frac{(-u)^k}{k!}\big(\Dcal^{\ast k+1}\tQcal\big)_s=\frac{(-u)^s}{s!}\big(\Dcal^{\ast s+1}\tQcal\big)_s=0.
\end{equation}
In the last step, we used that $\big(\Dcal^{\ast s+1}\tQcal\big)_s$ only contains $\tQcal_s$ for $s<0$.

\subsection{Soft and Quadratic actions for arbitrary spin $s$ \label{SecComparison}} 

We now study the action $\dt C$ in details at quadratic order and show that we recover the result from \cite{Freidel:2021ytz} and \cite{Geiller:2024bgf}.

The Noether charge $\cq_{\T{s}}$, or equivalently the charge aspect $\hq_s$, depends linearly on $\bN$ and polynomially\footnote{In a generalized sense since the coefficients of the polynomial can be differential operators.} on $C$. 
More precisely we can decompose $\hq_s$ for $s\geqslant -2$  as
\be \label{hqsk}
\hq_s =\sum_{k=0}^{\lfloor s/2\rfloor+1} \q{s}{k},
\ee 
where $\q{s}{k}$ is homogeneous of degree $k$ in $C$ and linear in $\bN$.\footnote{As a reminder, this amounts to the expansion in terms of the coupling constant $\gN$ (while all charges are at leading order in $\bgN$).}
The charge aspect $\q{s}{0}\equiv\soft{\hq_s}$ is the soft charge while $\q{s}{1}\equiv\hard{\hq_s}$ is the hard charge and $\sum_{k=2}^{\lfloor s/2\rfloor+1} \q{s}{k}$ is the super-hard contribution.

In this section, we compute the soft and hard action for any spin. 
Actually \eqref{dtauC} allows us to predict much more than just the quadratic action, but for now we shall focus mostly on the latter and show how the action \eqref{dtauC} combined with the dual EOM \eqref{dualEOM} contain all the information previously obtained after tedious computations. 
Nevertheless, even if we let for further work the study of the super-hard action, we emphasize that the closure of the algebra generated by $\Qt$---in full generality, i.e. without relying on any sort of soft or hard truncation---is a huge achievement and a highly non-trivial consistency check that \eqref{dtauC} bears a fundamental status.

In order to classify the action as a function of the helicity, we define 
\begin{equation}
    \d{s}{\T{s}}C :=\big(\dt C\big)\big|_{\tau=\bft(\T{s})}= \poisson{\cq_{\T{s}},C}, \label{defdeltaTs}
\end{equation}
where $\bft(\T{s})$ is the solution \eqref{soltaupTs}.
We introduce a superscript $\d{s}{}$, which refers to the helicity of the charge the transformation is associated with.
We also use it to emphasize that $\d{s}{\T{s}}\neq\dT$ for $T=(0,\ldots,0,\T{s},0,\ldots)$.

\paragraph{Warm-up: Spin 0, 1 and 2:}

Before giving a general proof, we start with a warm-up and focus on the action of super-translations (spin 0), sphere diffeomorphisms (spin 1) and the helicity 2 charge. 
This was one of the results of \cite{Freidel:2021dfs}.
Using the solutions \eqref{soltau43210},
\begin{subequations}
\label{deltaC012}
\begin{align}
\d{0}{\T0} C&=-\big(D^2\T0\big)+\big(N\T0\big), \\
\d{1}{\T1} C&=-\big(uD^3\T1\big)+\big(uND\T1+3CD\T1+2DC\T1\big), \label{deltaYC}\\
\d{2}{\T2} C&= -\left(\frac{u^2}{2}D^4\T2\right)+ \left(\frac{u^2}{2}N D^2\T2+3D^2\big(\T2\pui C\big)+2 uDCD\T2+3uCD^2\T2\right) \nn\\
&\quad -\big(3N\T2\pui C +3 C^2 \T2\big),
\end{align}
\end{subequations}
where we have highlighted the soft, hard and super hard (in the spin 2 case) actions with the parentheses.
The reader can directly compare the 3 lines of \eqref{deltaC012} with the equations (59), (75) and (89) from \cite{Freidel:2021dfs}: they are identical upon changing $\T0\to T$, $\T1\to Y/2$, $\T2\to Z/3$ and $C\to C/2$.
These transformations were confirmed in \cite{Geiller:2024bgf}, cf. formula (6.14) upon changing $\T0\to T$, $\T1\to \overbar{\mathcal Y}$, $\T2\to Z$ and $C\to -C$.

In order to facilitate the comparison with the expressions of \cite{Freidel:2021ytz} this time, we rewrite \eqref{deltaC012} in this equivalent way:
\begin{subequations}
\label{deltaC012bis}
\begin{align}
\d{0}{\T0} C&=-\big(D^2\T0\big)+\big(\T0\big)\pa_u C, \\
\d{1}{\T1} C&=-\big(uD^3\T1\big)+\big(D\T1(u\pa_u+3)+2\T1 D\big)C, \\
\d{2}{\T2} C&= -\left(\frac{u^2}{2}D^4\T2\right)+ \left(D^2\T2\Big(\frac{u^2}{2}\pa_u^2+3u\pa_u+3\Big) +2 D\T2 D \big(u\pa_u+3\big)+3\T2 D^2\right)\pui C \nn\\
&\quad -\left(3\T2\pa_u(C\pui C)\right).
\end{align}
\end{subequations}
We emphasized that the hard action $\Hd{p}{\,\cdot}C$ can be viewed as a differential operator acting on $\pa_u^{1-p}C$. We will see that this property holds for any $p$ and not only for $p=0,1,2$ as we showed explicitly in this warp-up.

The transformation $\dt C$ \eqref{dtauC}, combined with the dual EOM \eqref{dualEOM}, thus reproduces the correct action of the charges of spin 0, 1 and 2 on the gravitational phase space. 

\paragraph{General case:}

We then focus on the soft and quadratic actions for arbitrary values of $s$, namely\footnote{$\soft{\cq_{\T{s}}} =\frac{1}{4\pi \GN}\int_S\soft{\hq_s}\T{s}$ and similarly for $\hard{\cq_{\T{s}}}$.}
\begin{equation}
    \boxed{\poisson{\soft{\cq_{\T{s}}},C} =\Sd{s}{\T{s}} C=-D^2\soft{\t0}(\T{s})} \label{deltatauSoft}
\end{equation}
and
\begin{equation}
    \boxed{\poisson{\hard{\cq_{\T{s}}},C}=  \Hd{s}{\T{s}}C= -D^2\hard{\t0}(\T{s})+N\soft{\t0}(\T{s})+ 2DC\soft{\t1}(\T{s})+3CD\soft{\t1}(\T{s})}. \label{deltatauHard}
\end{equation}
Before stating the lemmas, we define yet another piece of notation, namely 
\begin{equation}
    \td{s}{\T{s}}C\equiv\deg^u_0\big(\d{s}{\T{s}}C\big), \label{deftildedeltaTs}
\end{equation}
where by $\deg^u_0(O)$, we mean the coefficient of the term $u^0$ in the expression $O$ (when the latter is a polynomial in $u$).

\begin{tcolorbox}[colback=beige, colframe=argile]
\textbf{Lemma [Soft action]}\\
The soft action for arbitrary spin $s\in\N$ is given by
\begin{equation}
    \Sd{s}{\T{s}}C=-\frac{u^s} {s!} D^{s+2}\T{s}.
\end{equation}
\end{tcolorbox}

\ni Notice that $\Sd{s}{\T{s}}C=\frac{u^s}{s!}\Std{0}{D^s\T{s}}C$, where $\Std{0}{\T0}C=-D^2\T0$.
This means that the soft transformation of spin $s$, parametrized by the tensor $\T{s}$, is nothing more than the soft part of a super-translation parametrized by the tensor $D^s\T{s}$ (times the $u$ dependence $\frac{u^s}{s!}$).

\begin{tcolorbox}[colback=beige, colframe=argile]
\textbf{Lemma [Hard action][Part 1]}\\
The hard action for arbitrary spin $s\in\N$ takes the form
\begin{subequations}
\label{deltatildeHard}
\begin{equation}\label{deltatildeHardbb}
    \Hd{s}{\T{s}}C=\sum_{p=0}^s\frac{u^{s-p}}{(s-p)!}\Htd{p}{D^{s-p}\T{s}}C,
\end{equation}
where
\begin{equation}
    \Htd{p}{\T{p}}C=\sum_{k=0}^{\alpha_p}(-1)^k\frac{(p-2)_k}{k!}(p-k+1) D^{p-k}\big(D^k\T{p}\,\pa_u^{1-p}C\big), \label{deltatildeHardb}
\end{equation}
\end{subequations}
with $(x)_k=x(x-1)\ldots(x-k+1)$ the falling factorial, $\alpha_p=\max[\mathrm{mod}_2(p),p-2]$ where $p$ modulo 2 dominates if $p=0,1$.
\end{tcolorbox}

\ni Finally, the formula \eqref{deltatildeHardb} can also be written in another useful form:

\begin{tcolorbox}[colback=beige, colframe=argile]
\textbf{Lemma [Hard action][Part 2]}\\
\begin{equation}
    \Htd{p}{\T{p}}C=\sum_{k=0}^{\min[3,p]}\binom{3}{k}(p+1-k)D^k\T{p}\,D^{p-k}\pa_u^{1-p}C. \label{deltatildeHardbis}
\end{equation}
\end{tcolorbox}

\ni We present the proofs of these lemmas in Appendix \ref{AppSoftHardAction}.
The formulae \eqref{deltatildeHardbb} and \eqref{deltatildeHardbis} match with the equations (73) and (74) of \cite{Freidel:2021ytz}.

\paragraph{Remark:}
In a non-radiative strip of $\scri$, where $N=0$ for $u\in[0,u_0]$, we have that \eqref{soltaunTs} reduces to
\begin{equation}\label{deftn0rad}
    \t{n}(\T{s})=\sum_{k=\lfloor\frac{s+1-n}{2}\rfloor}^{s-n}\frac{u^k}{k!}\big(\Dcal^kT\big)_n,
\end{equation}
which implies that
\begin{equation}
    \d{s}{\T{s}}C=-\sum_{k=\lfloor\frac{s-1}{2}\rfloor}^s\frac{u^k}{k!}\big(\Dcal^{k+2}T\big)_{-2}.
\end{equation}
Since $\d{s}{\T{s}}C=\poisson{\cq_{\T{s}}, C}$, this expression parallels \eqref{deftqs0rad}.

\section{Algebroid section of the $\Wcal_\sigma$-algebra \label{Sec:algebroidTs}}

We now study the map $\bft$ in more details.
In particular, we prove that it is a Lie algebroid isomorphism.
Let us start with some preliminary calculations which follow from the previous sections.

Given $\tau,\tau'\in\sfTh$, we have seen in section \ref{sec{timeev}} that $\lbr\tau,\tau'\rbr\in\sfTh$ then satisfies the dual EOM.
It is therefore in the image of $\bft$ and we can write that
\begin{equation}
    \big\lbr\bfm{\tau}(T),\bfm{\tau} (T')\big\rbr\equiv\bfm{\tau}\big(\Tpp{}(T,T') \big)\in\sfTh. \label{tauppTTp}
\end{equation}
Indeed, since the LHS is fully determined by $T$ and $T'$, so should be the RHS.
We are thus looking for the symmetry parameter $T''$ determined in terms of $T$ and $T'$ such that $\bfm{\tau}(T'')$ matches with $\big\lbr\bfm{\tau}(T),\bfm{\tau} (T')\big\rbr\in\sfTh$.
By evaluating \eqref{tauppTTp} at $u=0$, we are then able to deduce this associated symmetry parameter $\Tpp{}$.
By construction of the map $\bft$ as a solution of the dual EOM, we have that $\bft(T)\big|_{u=0}=T$.
It is then clear that
\begin{equation}
    \big[\bfm{\tau}(T),\bfm{\tau} (T')\big]^C\Big|_{u=0}=[T,T']^\sigma,\qquad \sigma:=C\big|_{u=0}. \label{CsigmaBracket}
\end{equation}
Besides, using the concise form \eqref{solEOMcompact} plus the fact that $\pui=\int_0^u$ so that $\big(\pui(\ldots)\big)\big|_{u=0}=0$, we have that\footnote{Had we considered $\bft(T)\in\sfT$, then $\bft_{-1}(T)$ would not be defined (where $\bft_s\equiv\pi_s \circ\bft$).}
\begin{equation}
    \Big(\delta_{\bft(T)}\bft(T')\Big) \Big|_{u=0}=\Big(\pui\big( \delta_{\bft(T)}\big(\Dcal\bft(T')\big)\big)+\delta_{\bft(T)}T'\Big) \Big|_{u=0}=\dT T', \label{dtauTtauTp}
\end{equation}
with $\dT \sigma = \nu\T0  + \hdT \sigma$ and $\nu:=N|_{u=0}$,
so that $\dT\sigma$ is indeed $\big(\delta_{\bft(T)} C\big)\big|_{u=0}$.
\smallskip

We thus define the Lie algebroid bracket onto the sphere, denoted by $\lbr \cdot\,,\cdot \rbr^\sigma$:
\begin{equation}
    \boxed{~\big\lbr T,T'\big\rbr^\sigma= \big[T,T'\big]^\sigma + \dTp T-\dT T'~}. \label{TbracketSphere}
\end{equation}
The space $\big(\V(S),\lbr \cdot\,,\cdot\rbr^\sigma, \delta\big)$ is clearly an algebroid over $S$ since it results from the projection of the $\Ah$-algebroid at the cut $u=0$ of $\scri$.
Since we know that $\Ao$ is also an algebroid, we can similarly consider the projection coming from $\Ao$, for which it is guaranteed that $D\lbr T, T'\rbr^\sigma_{-1} =\sigma\lbr T, T' \rbr^\sigma_0$.\footnote{In that case, the map $\bft_{-1}(T)$ is implicitly defined via the constraint $D\bft_{-1}(T)=C\bft_{0}(T)$. We never need more than this relation.}
The previous results (\ref{tauppTTp}-\ref{dtauTtauTp}) show that the map $\bft$ is an isomorphism of Lie algebroids, i.e.
\be  \label{IsoAlgebroids}
\boxed{\big\lbr \bfm{\tau}(T), \bft(T') \big\rbr =  \bft\big( \lbr T,T'\rbr^\sigma  \big)}.
\ee 
Besides, $\bft$ maps the base space $S$ of the sphere algebroid, to the base space $\scri$ of $\Ah$ (or $\Ao$), via the inclusion map at a constant $u$ cut $\iota_S:S\to S(0)\subset\scri$.
We let for future work the careful discussion of an arbitrary cut $S(U)$; cf. also section \ref{sec:ArbitraryCut}.

\paragraph{Remark:} 
Why did we not name $\big(\V(S),\lbr \cdot\,,\cdot\rbr^\sigma, \delta\big)$ as $\A_\sigma$?
The subtlety comes from the transformation property of $\nu$.
Knowing about the $\A$-algebroid structure over $\scri$ (here we use $\A$ to refer irrespectively to $\Ah$ or $\Ao$), we have by construction that 
\begin{align}
    \dT\nu:=\big(\delta_{\bft(T)}N\big) \big|_{u=0} &=\Big(\delta_{\dot{\bft}(T)}C+ \bft_0(T) \dot N+2DN\bft_1(T)+3ND \bft_1(T)-6CN\bft_2(T)\Big)\Big|_{u=0} \nn\\
    &=\delta_{\Dcal T}\sigma+\Cc2\T0+2D\nu\T1+3\nu D\T1-6\sigma\nu\T2, \label{dTnu}
\end{align}
where $\Cc2:=\big(\pa_u N\big)\big|_{u=0}$ and we use the fact that $\pa_u\bft(T) =\Dcal\bft(T)$.
The important feature to notice here is that while the $\A$-algebroid is determined by only one parameter, namely the shear $C$, we expect its projection at a cut to depend on $\bfm{\sigma}:= \{\Cc{n}\}_{n=0}^\infty$, where $\Cc{n} \equiv \big(\pa_u^n C\big)\big|_{u=0}$.
We know that onto $\scri$ the action of the anchor map on $\pa_u^n C$ is given by $\pa_u^n(\dt C)$. 
This defines recursively $\dT\Cc{n}$  from the sphere viewpoint.
Repeating the computation \eqref{algebroidRepinterm} for the bracket $\lbr\cdot\,,\cdot\rbr^\sigma$, we find that $\dT\nu\equiv\dT\Cc1$ needs to be equal to \eqref{dTnu} in order for $[\dT,\dTp]\sigma+\delta_{\lbr T,T'\rbr^\sigma}\sigma=0$ to hold (where $\Cc0\equiv\sigma$).
Similarly we expect that $[\dT,\dTp]\Cc{n} +\delta_{\lbr T,T'\rbr^\sigma}\Cc{n}=0$ to follow from the transformation property of $\dT\Cc{n+1}$.
The systematic study (at the intrinsic corner level) of the potential $\A_{\bfCc}$-algebroid structure goes beyond the scope of the present manuscript and we let it for future investigation.
Notice however that $\A_{\bfCc}$ can still be defined as the restriction of $\A$ to an arbitrary cut.\\

A great simplification arises when the chosen cut at which $T=\tau|_{u=S{(U)}}$ is non-radiative. 
In this case $\nu=0$ so that $\dT\sigma\to\hdT\sigma$ and we can restrict $T$ to belong to the generalized wedge algebra $\Wcal_\sigma(S)$. 
The latter is characterized by $\hdT\sigma\equiv 0$ so that the algebroid bracket $\lbr\cdot\,,\cdot\rbr^\sigma$ reduces to the Lie algebra $\sigma$-bracket $[\cdot\,,\cdot]^\sigma$ that we studied in \cite{Cresto:2024fhd}.
We also know that $\hdT\sigma\equiv 0$ amounts to the condition $(\Dcal^{s+2}T)_{-2}=0$ on the graded vector $T$.\footnote{Note that $s=-1$ is the initial constraint $\sI_\tau\big|_{u=0}=0$.}
To consistently satisfy the non-radiation condition $\nu=0$, we impose that the transformation $\dT\nu$ vanishes as well.
Evaluating \eqref{dTnu} at $\nu=0$, we see that we have to take $\Cc2=0$.
Iterating this reasoning, we find that a non-radiative cut is characterized by $\Cc{n}=0,\,n>0$, so that $\Wcal_\sigma$ is indeed parametrized by the sole parameter $\sigma$.
In this case, we get that\footnote{We similarly get that $\dT\Cc{n}\big|_{\{\Cc{p}\}_{p=1}^n=0}=\hat \delta_{\Dcal^nT}\sigma=-(\Dcal^{n+2}T)_{-2}=0$ for $T\in\Wcal_\sigma$.}
\begin{equation}
    \dT\nu\big|_{\nu=0}=\hat\delta_{\Dcal T}\sigma=-(\Dcal^3T)_{-2}=0\quad\mathrm{for}\quad T\in\Wcal_\sigma.
\end{equation}

Then, as a consequence of equation \eqref{IsoAlgebroids}, we obtain the following theorem.

\begin{tcolorbox}[colback=beige, colframe=argile]
\textbf{Theorem [Representation of $\Wcal_\sigma$ on $\scri$]}\\
At any non-radiative cut of $\scri$, the Lie algebra $\Wcal_\sigma(S)$ admits a Lie algebroid section in $\Ao$, realized via the map $\bft$,
\begin{align}
    \bft :\,&\Wcal_\sigma(S) \to \Ao \nn\\
     &\, T\mapsto\bft(T),
\end{align}
which satisfies
\vspace{-0.3cm}
\begin{align}
    \bft\big([T,T']^\sigma\big)= \big\lbr\bft(T),\bft (T')\big\rbr. \label{tautoTbracket}
\end{align}
\end{tcolorbox}

When we study a gravitational system in between two non-radiative cuts $u_0$ and $u_0'$, with definite shears $\sigma,\sigma'$ and symmetry parameters $T,T'$ respectively, the covariant wedge conditions $\hdT\sigma=0=\hdTp\sigma'$ come from the requirement that the symmetry action leaves these shears unchanged.
In other words, the covariant wedge algebra is the Lie algebra that preserves the boundary conditions $C|_{u=u_0}=\sigma$ and $C|_{u=u'_0}=\sigma'$.
In the limit $u_0\to\infty$, this is the algebra that preserves the late time fall-off condition, cf. footnote \ref{foot1}.
Notice that imposing the normal wedge condition $D^{s+2}\T{s}=0$ is not sufficient since it only kills the inhomogeneous, i.e. soft, part of the shear transformation.

We let for further investigation the proper description of the transition in between $\Wcal_\sigma(S(u_0))$ and $\Wcal_{\sigma'}(S(u'_0))$. 
This is an interesting question that requires a careful treatment of radiation.

We conclude this paper by making contact with the twistor description of higher spin symmetries, especially the recent work \cite{Kmec:2024nmu}.

\section{Relation to twistor theory}\label{sec:twistor}

While finishing this project, the Oxford group published a remarkable work \cite{Kmec:2024nmu} which overlaps our work, although
their starting point is totally different from ours. 
They study self-dual GR in twistor space and show that via a gauge fixing adapted to the asymptotic twistor space, the residual gauge transformations form the $Lw_{1+\infty}$ algebra on twistor space.
From there, they build Noether charges on $\scri$ that realize the algebra to all order in $\GN$.
The construction of the Noether charges is non-perturbative because the renormalization scheme necessary to form a charge conserved in the absence of radiation to all order in $\GN$ only involves the knowledge of the dual EOM \eqref{tausevolb}. 
In their perspective, the latter come as a result of a gauge fixing, which is totally independent from our standpoint.

They also show that the charge conservation laws \eqref{Qevolve} can naturally be derived from a generalized Gauss's law in twistor space that follows from the twistor equation of motion on the Lagrange multiplier that imposes self-duality. 
Connecting these equations non perturbatively directly to the Bianchi identities for the Weyl tensor is still a challenge for us.
What is remarkable is that the twistor space construction is based on a twistor Poisson bracket which is independent of the field while we have seen that the canonical analysis done in this work involves a shear deformation of the original $\W$-bracket (cf. footnote \ref{footWbracket}).
In the next subsection, we show that the twistor  Poisson bracket equates our $C$-bracket on-shell of the dual EOM, cf. \eqref{TCbracket} and \eqref{TwistorPoissonBracket}.

\subsection{Trading the grading for an extra dimension}

The key element needed to understand the connection between our derivation of the $\Ah$-algebroid and the twistor derivation is the introduction of a spin $1$ variable $q \in \Ccar{0,1}$. This variable can be used to promote the series $\tau=(\tau_s)_{s+1\in \N}$ to a holomorphic function $\ft$ of $q$\footnote{We keep the Bondi coordinates dependence hidden. 
The full expression reads $\ft(q, u, z, \bz)=\sum_{s=-1}^{\infty} \tau_s( u, z, \bz)  q^{s+1}$. } 
valued onto vector fields on $\scri$,
\be
\ft(q)=\sum_{s=-1}^{\infty} \tau_s  q^{s+1} \in \tV_{-1}. \label{ftauDef}
\ee 
To avoid misunderstanding and shorten the notation, we denote this function by $\ft$.
The graded vector $\tau$ and the function $\ft$ can be viewed as two different representations of the same abstract vector in $\tV(\scri)$.
The covariant derivative operator is simply represented on these functions as the  functional
\be 
(\Dcal \ft)(q) = \sum_{s=-1}^{\infty} (\Dcal \tau)_s q^{s+1} =
 \pa_u \ft(q) - \sE_{\ft}(q), \label{Dcaltauq}
\ee 
where we introduced 
$
\sE_{\ft}(q):= \sum_{s=-1}^\infty \E_s(\tau) q^{s+1}.
$
Using this we can write our bracket in terms of the spin functionals. 
Quite remarkably we find that it is related to
the Poisson bracket in the $(q,u)$ plane, namely we introduce 
\be \label{bracketquPlane}
\{\ft,\ft'\}:= \pa_q \ft \pa_u \ft' - \pa_q \ft' \pa_u \ft.
\ee
Denoting $[\ft, \ft']^{C}(q) := \sum_s     [ \tau, \tau']_s^{C}q^{s+1}$, we find that 
\begin{align}
    \sum_{s=-1}^\infty [\tau, \tau']_s^{C}q^{s+1}&=\left(\sum_{s=-1}^\infty\sum_{n=0}^{s+1}(n+1)\t{n}q^n(\Dcal\tau')_{s-n}q^{s-n+1}\right)-\tau\leftrightarrow\tau'\cr
    &=\left(\sum_{n=0}^\infty\sum_{s=n-1}^\infty\pa_q(\t{n}q^{n+1})(\Dcal\tau')_{s-n}q^{s-n+1}\right)-\tau\leftrightarrow\tau'\\
    &=\pa_q\left(\sum_{n=0}^\infty \t{n}q^{n+1}\right)\left(\sum_{s=-1}^\infty(\Dcal\tau')_{s}q^{s+1}\right)-\tau\leftrightarrow\tau'. \nn
\end{align}
Using \eqref{Dcaltauq}, we conclude that
\begin{align}
\boxed{ [ \ft, \ft']^{C}=   
 \{ \ft, \ft'\} - \pa_q \ft \E_{\ft'} - \pa_q \ft' \E_{\ft} }.\label{TCbracket}
\end{align}
This shows that the $C$-bracket is given by the Poisson bracket, once we assume the dual equations of motion, i.e. for $\tau,\tau' \in \sfTh$.
The Poisson bracket is the canonical twistor Poisson bracket under the parametrization of the twistor fiber coordinates $\bar\mu^\alpha = u n^\alpha + q \lambda^\alpha$ where $\lambda^\alpha$ is the spinor parametrizing holomorphic homogeneous coordinates on the sphere and $n^\alpha$ is a spinor such that $\epsilon_{\alpha \beta}= \lambda_\alpha n_\beta-\lambda_\beta n_\alpha$.\footnote{It is customary to take 
$ n_\alpha = \frac{\hat{\lambda}_\alpha }{\lambda^\beta \hat\lambda_\beta}$ where $ \hat{\lambda}_\alpha=T_{\alpha \dot\alpha}\bar\lambda^{\dot\alpha} $ with $T$ a timelike vector.}${}^,$
This directly follows from $\pa_u = n^\alpha \frac{\pa}{\pa \bar\mu^\alpha}$ and $\pa_q = \lambda^\alpha \frac{\pa}{\pa \bar\mu^\alpha}$ so that 
\vspace{-0.1cm}
\be \label{TwistorPoissonBracket}
\{\ft,\ft'\} 
= (\lambda^\alpha n^\beta-\lambda^\beta  n^\alpha )\frac{\pa\ft}{\pa \bar\mu^\alpha}\frac{\pa\ft'}{\pa \bar\mu^\beta} = \epsilon^{\alpha\beta} \frac{\pa \ft }{\pa \bar\mu^\alpha} \frac{\pa \ft'}{\pa \bar\mu^\beta}.
\ee 
Finally we can establish that the transformation $\tdt h $ of the twistor potential given in \eqref{dtauh} is equivalent to a twistor gauge transformation.
One first introduces the covariant derivative 
\be 
\nabla := q\pa_u -D + C \pa_q,
\ee
and we obtain  that 
\begin{align}
    \nabla \ft(q) 
    &= \sum_{s=-1}^\infty \big(  \pa_u\tau_s - 
    D \tau_{s+1} 
     + (s+3) C \tau_{s+2}\big) q^{s+2}
    - D \tau_{-1} + C \tau_0 \cr
    &=  \sI_\tau + q \E_{\ft}(q) .\label{bNI}
\end{align}
The condition $\tau \in \sfTh$, which imposes $\E_{\ft}=0$, simply  reads, in twistor variables, $\pa_q \nabla \ft=0$.
This shows that when $\tau \in \Ah$, then $\nabla\ft$ is independent of $q$\footnote{This is natural since the asymptotic potential $h$ is only a function of $(u,z,\bz)$.} and
\be 
\tdt h= \nabla \ft = \sI_\tau.
\ee
We recognize \eqref{dtauh},
which establishes the equivalence between the twistor and phase space transformations for $\tau\in \Ah$.

Although we agree with most of the results of \cite{Kmec:2024nmu}, there seems to be a discrepancy with their equation (4.16) whose RHS misses the shear dependence of the term  $\sigma\paren{\cdot\,,\cdot}$ (cf.\,\eqref{DaliBracket}) in the $\sigma$-bracket \eqref{Wbracket}.
We suspect that the reason of the difference is that their bracket is evaluated at $\scri^+_-$, where according to the Schwartz boundary conditions, the shear vanishes.
This shear dependence of the $\sigma$-bracket when written on an arbitrary non-radiative cut of $\scri$ is the origin of the many subtleties of the present work.

\subsection{Good cut}

From \cite{Adamo:2021lrv,Kmec:2024nmu} we know that $q$ is a fiber coordinate associated with the projection $p:\mathbb{PT}\to \scri_{\mathbb{C}}$ given by $(\bar{\mu}^\alpha, \bar\lambda_{\dot\alpha}) \to (u =\bar{\mu}^\alpha \lambda_\alpha, \lambda_\alpha, \bar\lambda_{\dot\alpha})$.\footnote{The coordinate we call $(u,q)$ here are denoted $(\bar{u}, \bar{q})$ in \cite{Kmec:2024nmu}.} 
We now want to show that the spin coordinate $q$ possesses a natural geometrical interpretation from the point of view of  $\scri$ and its spacetime embedding.
First we know that $\scri$, being a null surface, is equipped with a Carrollian structure $(\ell^a, q_{ab})$ where $\ell$ is the vector tangent to the null congruence and the degeneracy vector for $q_{ab}$, i.e. $\ell^a q_{ab}=0$ \cite{Ashtekar:1981bq, Riello:2024uvs}. 
Choosing a complex structure on the metric amounts to the choice of a frame field $m_a$ such that $q_{ab}= m_a\bm_b +\bm_a m_b$. This frame is such that $\ell^a m_a=\ell^a \bm_a=0$. 
In order to equip $\scri$ with a connection we need to choose a ruling, that is a one form $k_a$ such that $k_a\ell^a=1$ \cite{Freidel:2022vjq,Freidel:2024emv}. 
In the Bondi analysis, one selects the ruling form to be exact and given by the differential of the Bondi time $k=\rd u$.
The carrollian vector is then simply $\ell=\pa_u$.
However, it is useful to allow more general choices such that $k$ carrying vorticity, i.e. $\rd k \neq 0$. 
For instance we can  choose $k_a$ to be associated to a null rigging structure \cite{Mars:1993mj}. 
This means that we can see the ruling form as deriving from a spacetime \emph{null} vector $k^a$ transverse to $\scri$. 
Such a transverse vector is understood as labeling a congruence of null geodesics transverse to $\scri$. 
Such general ruling vector, like the ones associated with geodesics null congruence, can be parametrized by a pair of spin variables $(q,\bar{q})$  and given by 
 \be
  k_{(q,\bar q)} = k + q \bm +\bar{q} m - q\bar{q} \ell.
 \ee 
The complex structure vectors normal to $k_q$ are then given by
 \be
 m_q= m- q\ell.
 \ee
The transformation $(k,m, \ell)\to( k_{(q,\bar q)},m_q,\ell_q)$, which fixes $\ell=\ell_q$ corresponds to a null boost with angle $q$.
This shows, as explained in the work of Adamo and Newman \cite{Adamo:2009vu, Adamo:2010ey}, that the value of $q$ at any point of $\scri$ is the stereographic angle which describes the null direction of each geodesic intersecting $\scri$.

The supertranslations $T=T(z,\bz)$ acts non trivially on the pair $(u,q)$\footnote{For the special case of a supertranslation, note that $\delta_{\tau(\T0)}\cdot=\d{0}{\T0}\cdot=\delta_{\T0}\cdot$, and here we just rename $\T0\to T$ since the context is clear.}
 \be 
 \delta_T u= T, \qquad \delta_T q = -D T.
 \ee 
The transformation for $u$ is the supertranslation definition. The transformation  for $q$ follows from the fact that while $m$ is a vector tangent to the cut $u=\mathrm{cst}$, the vector  $m_{q=-DT}=m + DT \ell$ is the vector tangent to the cut $u-T=\mathrm{cst}$.

In order to use the power of the null rotations we  consider a Goldstone field $G(u,z,\bz)$, which transforms linearly under supertranslation $\dT G= -T$.
This field defines a diffeomorphism of $\scri$ denoted  $\hG: \scri \to \scri$ and given by $\hG(u,z,\bz):= (G(u,z,\bz),z, \bz)$. 
Under this map the Bondi cuts $u=u_0$ are mapped onto supertranslated cuts $u= G(u_0,z,\bz)$. 
Moreover, the holomorphic derivative is shifted  by a spin $1$ connection $L$
\be \label{DefConnectionL}
D (F\circ \hG)= (D F + L\pa_u F)\circ \hG, \quad\mathrm{with}\quad L= DG\circ \hG^{-1}. 
\ee 
The Goldstone is defined to satisfy the good cut equation\footnote{For the equivalence, apply \eqref{DefConnectionL} to $ L\circ\hG $.}$^,$\footnote{Notice how a super-translation acts on $C\circ\hG$:
\vspace{-0.1cm}
\be
\dT\big(C\circ \hG\big)
 &= (\dT C)\circ \hG + \dT G (\pa_uC \circ \hG) \nn\\
&= (T\pa_u C - D^2 T )\circ \hG - T \pa_uC \circ \hG 
= -D^2T \circ \hG = -D^2T=\dT (D^2G).
\ee
\vspace{-0.4cm}
} \cite{newman_heaven_1976}
\be 
D^2 G = C \circ \hG\quad \Leftrightarrow\quad   DL + L \pa_u L=C, \label{GoodCutEq}
\ee 
which determines $G$ from the shear $C$ up to a 4 dimensional freedom interpreted as complex spacetime \cite{Adamo:2010ey, newman_heaven_1976}.

Given $\ft\in \sfT $,\footnote{Recall that the function $\ft$ is also a representation of the vector space $\sfT$.} we define the map $\mTG: \sfT \to \sfT $  with image $ \mTG[\ft]\in \sfT$ given by 
\be \label{TGdef} 
\boxed{\mTG[\ft](q):= \ft(q -L) \circ \hG}.
\ee
$\mTG$ is a map  which combines a shift of $q$ by the spin $1$ connection $L$ and a redefinition of time, both determined by the supertranslation Goldstone field. 
Explicitly, this means that  
\be 
\mTG[\ft](q,u,z, \bz)
= \ft\big(q - DG(u,z,\bz), G(u,z,\bz),z,\bz\big). \label{defmTGtwistor}
\ee
This transformation is such that
\begin{equation}
    \pa_u \mTG[\ft] =\dot G\mTG[\pa_u\ft]-D\dot G\mTG[\pa_q\ft].
\end{equation}
Similarly for $D$, we can write compactly\footnote{$D\dot G= {\dot G} (\pa_u L \circ\hG) $.}
\bs\label{DuTGq}
\begin{align}
\pa_u \mTG[\ft] &= \dot{G} \mTG\big[\pa_u \ft - \pa_u L \pa_q\ft\big], \\
  D \mTG[\ft] &= \mTG\big[D \ft + L \pa_u \ft - C\pa_q\ft \big].
\end{align}
\es
These relations can be inverted once  we introduce the frames 
\be
\ell^G:= \dot{G}^{-1}(\pa_u + D\dot{G} \pa_q),\qquad  
D^G:= D - DG \ell^G. 
\ee
We see that $DG$ and $D\dot{G}$ play the role of rotation coefficients respectively deforming the usual derivatives $(D, \pa_u)$ along $(\ell^G,\pa_q)$.
We recast \eqref{DuTGq} as\footnote{$\pa_q\mTG[\ft]=\mTG[\pa_q\ft]$.}
\begin{equation}\label{lGintertwine}
    \ell^G\mTG[\ft]=\mTG\big[\pa_u \ft\big], \qquad D^G\mTG[\ft]=\mTG\big[(D-C\pa_q)\ft\big].
\end{equation}
Notice also that
\begin{equation}
    \mTG[q\ft]=(q-DG)\mTG[\ft]\equiv q^G\mTG[\ft],
\end{equation}
where $q^G:=q-DG$ is the new $q$ variable after the change of frame.
The dual equation of motions for $\tau$ are then mapped into simpler equations where the shear has been removed.
Indeed, by combining the last two equations, we readily get that
\be \label{IntertT1}
\boxed{\big(q^G\ell^G-D^G\big)\mTG[\ft]= \mTG[\nabla\ft]}.
\ee 
For $\ft\in\sfTh$, we can use \eqref{bNI} to deduce the following form of the dual EOM:
\begin{equation}
    \pa_q \big(q^G\ell^G-D^G\big)\mTG[\ft]=0.
\end{equation}
It is interesting to formalize the previous findings. 
First we introduce the \emph{Newman space}\footnote{It is related to twistor space via the relations around \eqref{TwistorPoissonBracket}.} $\TN\equiv\C\times\scri_{\C} $ which is a one dimensional fibration of $\scri$ with fiber coordinate $q$ and we extend the map $\hG$ into a diffeomorphism $\hhG:\TN\to\TN$ given by 
\begin{align}
    \hhG:(q,u,z)\mapsto(q^G,G,z). \label{hhGdiff}
\end{align}
This map extends as an inverse pushforward to the vector fields $(\pa_q,\ell,m)$ according to\footnote{If we treat $\hhG_*$ as a holomorphic map, it leaves $\bm$ invariant.} 
\be 
\hhG_*(\pa_q,\ell,m) = \big(\pa_q,\ell^G,m^G\big).
\ee 
The map $\hhG$ is such that $\big(\pa_{q^G},\ell^G,m^G\big)(F \circ \hhG) = \big[(\pa_{q},\ell,m-C\pa_q) F\big]\! \circ \hhG $.
It is designed such that the shear variable $C$ is mapped onto 0!
Hence, the result \eqref{IntertT1} simply means that $\mTG$ implements the pullback $\hhG{}^*$ of $\ft$ from a radiative $\TN$ to a non radiative one.
In other words, the evolution operator $\nabla=(q\ell-m+C\pa_q)$ is mapped onto its non radiative version $\nabla^G:=(q^G\ell^G-m^G)$ such that
\begin{equation}
    \hhG{}^*(\nabla\ft)=\nabla^G\big(\hhG{}^*\ft\big).
\end{equation}

\subsection{Relation to the dressing map \label{sec:RelationDressingMap}}

We show that the map $\mTG$ defined here is the same as the dressing map $\mTG$ introduced in \cite{Cresto:2024fhd}.
In a way this follows from the previous calculations by noticing that\footnote{We can readily check that $(D-C\pa_q)\ft=-\sI_\tau+q\Dcal\ft$.}
\begin{equation}
    D^G\mTG[\ft]=\mTG \big[q\Dcal\ft\big]-\sI_\tau\circ\hG.
\end{equation}
Therefore $\mTG$ intertwines the action of $D^G$ and $q\Dcal$ when $\ft\in\sfTo$, which was the definition of the intertwining map in \cite{Cresto:2024fhd}.
It is also instructive to prove the isomorphism explicitly.
First notice that by definition \eqref{defmTGtwistor},\footnote{This equation defines the map $\mTG$ on the graded vector $\tau$.}
\begin{align}
    \mTG[\ft](q) =\sum_{s=-1}^\infty\mTG_s[\tau]q^{s+1} =\sum_{s=-1}^\infty\t{s}(G,z,\bz)\sum_{k=0}^{s+1}\binom{s+1}{k}q^{s+1-k}(-DG)^k.
\end{align}
Rearranging the summations we can write this expression in terms of the map 
$T^G$ introduced in \cite{Cresto:2024fhd}. Namely if we define
\begin{equation}
T^G_s[\tau]:=\sum_{k=0}^\infty \frac{(s+k+1)!}{k!(s+1)!}(-DG)^k\t{s+k}, \qquad s\geqslant -1,
\end{equation}
we have that 
\begin{align} \label{mTGdef}
    \mTG[\ft](q) =\sum_{k=0}^\infty \sum_{s=-1}^\infty\frac{(s+k+1)!}{k!(s+1)!}(-DG)^k q^{s+1}\big(\t{s+k}\circ\hG\big)= \sum_{s=-1}^\infty T^G_s \big[\tau\circ\hG\big]q^{s+1}.
\end{align}
This implies that while $T^G[\ft](q)= \ft(q-DG)$, we get that $\mTG$ is the composition of $T^G$ with the pullback of $\hG$:
\begin{equation}
   \mTG[\ft](q)=\ft (q-L)\circ \hG =(\ft\circ \hG) (q-DG) = T^G[\ft\circ \hG](q).
\end{equation}

Notice finally that in the particular case where $\hG$ is a supertranslation, i.e. when $G=G(z,\bz)$ is time independent, then
\begin{equation}
    \tau\circ\hG=\big(e^{G\pa_u}\tau\big) \big|_{u=0}=e^{G\Dcal}T.
\end{equation}
Therefore,
\begin{equation}
    \mTG[\tau]=T^G\big[\tau\circ\hG\big]= T^G\big[e^{G\Dcal}T\big]= \sum_{n=0}^\infty\frac{G^n}{n!}T^G\big[\Dcal^n T\big],
\end{equation}
as it was defined in \cite{Cresto:2024fhd}.
In our companion paper, we had denoted this map $\mTG[T]$ to emphasize its expression intrinsic to the sphere $S$.
Here we see explicitly that the peculiar dependence in $G$ and $\Dcal$ comes from $e^{G\Dcal}$ and reflects the evaluation of the Carrollian symmetry parameter $\tau$ at the super-translated cut defined by $G$.

\section{Conclusion \label{secGR:conclusion}}

In this paper, we constructed a symmetry algebroid $\Ao$ onto $\scri$, which is realized on the holomorphic Ashtekar-Streubel asymptotic gravitational phase space non linearly. 
We also built the algebroid $\Ah$, which admits a linear realization on a phase space that naturally appears from the reduction of twistor space to $\scri$. 
They differ on the way one deals with the symmetry transformation of degree $-1$. 
Both algebroids have the same action on the shear and both are realized canonically via Noether charges. 
Interestingly,  the charge associated with the time translation is the Bondi mass for $\Ao$ and the covariant mass \cite{Freidel:2021qpz} for $\Ah$.
These charges  satisfy a Hamiltonian flow equation valid for every spin and at all order in $\GN$ and are conserved in the absence of radiation.
The algebroid bracket which is represented on phase space is built out of the $C$-bracket \eqref{CFbrackettau}, which is a deformation of the celebrated $w_{1+\infty}$ bracket.
Deforming the $w_{1+\infty}$ algebra is essential to go beyond the wedge and realize the whole symmetry algebra canonically.

The key for our non-perturbative treatment is the introduction of time and field dependent symmetry parameters $\t{s}$ constrained by a set of evolution equations \eqref{dualEOM} dual to (the truncation of) the asymptotic Einstein's equations \eqref{Qevolve}.
The infinitesimal shear transformation under the algebroid is then parametrized entirely by $\t0$, $\t1$ and $\t2$ and given in \eqref{dtauC}.
The usual celestial symmetry parameters $\T{s}$ that generalize super-translations $\T0$ and sphere diffeomorphisms $\T1$, are understood as initial conditions for the dual EOM.
In other words, they correspond to the value of $\t{s}$ at a certain cut of $\scri$.
When evaluating $\t0,\t1,\t2$ on such a solution parametrized by $\T{s}$, the dual EOM encode all the non-linearity and non-locality of the shear variation.
Similarly, the Noether charge (defined at $\scri^+_-$) associated to the spin $s$ parameter $\T{s}$ is the result of a renormalization procedure fully determined by the dual EOM.

We also find that the symmetry transformations that preserve the shear at any non-radiative cut of $\scri$ define an algebra---called the covariant wedge algebra $\Wcal_\sigma(S)$---discussed more at length in \cite{Cresto:2024fhd}. 
This is the latter which represents the true symmetry transformations
preserving the late time fall-off condition of the shear. 
We explain how these algebras (one of for each shear) are embedded inside the algebroid. 
Our formalism guarantees that $\Wcal_\sigma(S)$ is canonically represented on $\scri$ through the $\Ao$-algebroid.

Moreover, we prove that the $C$-bracket amounts to the twistor Poisson bracket discussed in \cite{Kmec:2024nmu}.\footnote{While we agree with the twistor description \cite{Kmec:2024nmu}, which is linear, we disagree with their formula (4.16) which suggests that the $w_{1+\infty}$ algebra can be realized on any cut of $\scri$, without any deformation. As we show it is essential to include the shear deformation on the celestial sphere when the supertranslations are involved.}
The correspondence relies heavily on two essential features: 
The higher spin parameters can all be recast in terms of a generating functional which depends on a spin variable $q$. 
This variable was first introduced by Newman as the parameter that describes the choice of Carrollian ruling of $\scri$ \cite{Freidel:2024emv, Riello:2024uvs}. 
It also can be understood from the twistor perspective as the fiber coordinate of the twistor fibration over $\scri$. 
The second ingredient is the dual equations of motion that allow to recast the $C$-bracket defined on the celestial sphere, which explicitly depends on the shear, into a linear Poisson bracket. 
One therefore sees that the twistor description amounts to a linearization of the gravitational symmetry representation. 
\medskip

Our work opens up several questions. 
First the analysis of \cite{Kmec:2024nmu} clearly shows that the  asymptotic EOM \eqref{Qevolve} can be obtained as charge conservation in self-dual gravity. 
What is missing, from our perspective, is to understand clearly the relationship between these equations and the asymptotic Bianchi identities. 
It was conjectured in \cite{Geiller:2024bgf} that the conservation laws can be extracted from the Weyl Bianchi identities associated with self-dual gravity after a proper choice of asymptotic frame, but that still needs to be demonstrated.

On another front, while the symmetry algebras are identified as generalized wedge algebras, the algebroid, which goes beyond the wedge, has to be the appropriate algebraic entity to describe radiation.
Some questions are still to resolve, in particular it would be interesting to understand more deeply how the algebroid structure allows us to interpolate between two non radiative cuts. 
A better understanding of the algebra $\Wcal_C$ and of the algebroid structure $\A_{\{\sigma\}}$ will probably be necessary to address this question.

Also, while we constructed explicitly the renormalized symmetry charges parametrized in terms of $T_s$ defined at a cut $u=u_0$, we did not study directly the flux laws for these renormalized charges. 
This should allow us to define directly the charges in terms of data defined as limits  to $\scri^+_-$ and $\scri^+_+$, which would be interesting to construct.

Moreover, while we have mostly described the renormalized charges at constant $u=u_0$ cuts (see however section \eqref{sec:ArbitraryCut}), we can more generally define them at arbitrary cuts $u=U(z,\bar{z})$.
It would be desirable to understand better the covariance properties of the charges, and how does that relate to the notion of dressing map that enables to reabsorb the shear into a change of frame as described in the present work and in \cite{Cresto:2024fhd}. 

It would also be interesting to understand the (anti-)holomorphicity structure determined by the equation \eqref{dualEOM}, which does not involve the operator $\bD$. 
See \cite{Chen:2023rxf, Garner:2023zqn} for 3d generalization of the notion of chirality, relevant for extending the celestial modes expansion in $(z,\bz)$ to include the $u$ direction. 
Working solely with a cut of $\scri$ for instance does not capture the higher degree negative modes (i.e. the raviolo polynomials \cite{Garner:2023zqn}) which could be present when working globally on null infinity. 
The relevance, if any, of these modes in celestial holography seems worth investigating.

Besides, an important question concerns whether we could generalize the canonical analysis and relax the Schwartzian fall-off condition for the time dependency of the shear and the charges aspects. 
Recently the work \cite{Nagy:2024jua} investigated the possibility of constructing an extended phase space for Yang-Mills---generalizing the dressing field ideas of \cite{Donnelly:2016auv} to higher spin (see also \cite{Campiglia:2024uqq} for an application of the phase space extension to include GBMS)---in order to allow for a relaxation of the asymptotic boundary conditions. 
A similar construction should be available in gravity and it seems important to understand the connections between \cite{Nagy:2024jua} and the current work.

Of course, this suggests that the techniques developed here should also be applied to  non-abelian gauge theories and its higher spin symmetries.

Another important aspect left to understand is the link with OPEs and soft theorems. 
A natural question is that if we assume that the $\mathcal{S}$-matrix is invariant under the action of the Noether charge $\Qt$ (including the super-hard contributions), what does it imply for the scattering of soft gravitons? 
Is there a quantum anomaly in the charge algebra as suggested by \cite{Costello:2022upu, Bittleston:2022jeq} that can be re-derived from the canonical perspective? 
See \cite{Baulieu:2024oql, Baulieu:2025itt} for a BRST approach on that question for super-rotations.

Finally, from \cite{Kmec:2024nmu} it is clear that the higher spin symmetries described here are symmetries of self-dual gravity. 
However we also know that they are exact symmetry of full Einstein gravity up to spin $3$ and that spin $2$ generates the entire tower of higher spin charges. 
It therefore begs the question to understand what can be the use of the higher spin charges in full GR.

We can hope that the understanding of the quantization of the higher spin charges serves as a solid basis to describe asymptotic dressed states and develop a new type of interacting picture for full gravity. 
Whether or not part of the algebraic structure (or a deformation of it) will survive the inclusion of the corrections to the fundamental pattern \eqref{Qevolve} necessary to reproduce the full asymptotic Einstein's equations is a major open question.

\section{Glossary \label{secGR:glossary}}

\textbf{General notation, cf. section \ref{sec:preliminaries}:}
\begin{itemize}
    \item $S$: a 2d complex manifold.
    \item $S_n$: a 2d complex manifold with $n$ punctures.
    \item $(m,\bm)$: null dyad on $S$.
    \item $D=m^A D_A$: the covariant derivative on $S$ along $m$.
    \item $(u,r,z,\bz)$: Bondi coordinates.
    \item $C$ \& $N\equiv\pa_u C$: the shear and the news on $\scri$.
    \item $\sigma$: a deformation parameter of the $w_{1+\infty}$ algebra that plays the role of the shear at a cut $S$ of $\scri$.
    \item $\PS$ \& $\PS_h$: the Ashtekar-Streubel and the ``twistor/self-dual'' phase spaces.
    \item $\TN$: the Newman space \eqref{hhGdiff}.
\end{itemize}

\ni \textbf{Graded and filtered vector spaces:}
\begin{itemize}
    \item $\V(S)$: the space of celestial symmetry parameters $T$, cf.\,\eqref{Vsccel}.
    \item $\W$: the wedge subspace of $\V(S)$ \cite{Cresto:2024fhd}.
    \item $\W_\sigma$: the covariant wedge subspace of $\V(S)$, cf.\,\eqref{defCovWedgeSpaceGR}.
    \item $\tV(\scri)$: the space of time dependent symmetry parameters $\tau$,  cf.\,\eqref{VCcardef}.
    \item $\sfT$: a subspace of $\tV(\scri)$ where $\tau$ satisfies the dual EOM $\E_s=0$ for $s\in\N$, cf.\,\eqref{TspaceGR}.
    \item $\sfTh$: a subspace of $\tV(\scri)$ where $\tau$ satisfies $\E_s=0$ for $s+1\in\N$, cf.\,\eqref{defTbarThat}.
    \item $\sfTo$: a subspace of $\tV(\scri)$ where $\tau$ satisfies $\E_s=0$ for $s\in\N$ and $\sI_\tau=0$, cf.\,\eqref{defTbarThat}.
\end{itemize}

\ni \textbf{Graded vectors:}
\begin{itemize}
    \item $T$ \& $\tau$: graded vector in $\V(S)$ and $\tV(\scri)$ respectively. $\T{s}$ and $\t{s}$ are the degree $s$ elements,  cf.\,section \ref{secGR:tVspace}.
    \item $\ft$: a representation of a vector in $\tV$ in terms of a holomorphic function of a spin-1 variable $q$, cf.\,\eqref{ftauDef}.
\end{itemize}

\ni \textbf{Dual equations of motion (EOM):}
\begin{itemize}
    \item $\E_s(\tau)\equiv\E_s$: the dual EOM of degree $s$, \eqref{dualEOM} and \eqref{tausevolb}.
    \item $\sI_\tau$: the initial constraint \eqref{initialCondition} and \eqref{tauintial}.
    \item $\tau(T)$: a solution of $\E(\tau)=0$ for which $\tau(T)\big|_{u=0}=T$, cf.\,\eqref{maptaubis} and \eqref{solEOMcompact}.
\end{itemize}

\ni \textbf{Brackets:}
\begin{itemize}
    \item $[\cdot\,,\cdot]^\V$: the $\V$-bracket on $S$ \cite{Cresto:2024fhd}.
    \item $[\cdot\,,\cdot]^\W$: the $\W$-bracket \cite{Cresto:2024fhd}, a restriction of the $\V$-bracket to the wedge. It reduces to the $w_{1+\infty}$ bracket on $S_2$. Also the $\sigma$-bracket for $\sigma=0$.
    \item $[\cdot\,,\cdot]^\sigma$: the $\sigma$-bracket \eqref{Wbracket}, a deformation of $[\cdot\,,\cdot]^\W$, which is a Lie bracket over $\W_\sigma$.
    \item $[\cdot\,,\cdot]^{\tV}$: the $\tV$-bracket \eqref{tVBracket}. Same as $[\cdot\,,\cdot]^\V$ but over $\scri$.
    \item $\paren{\cdot\,,\cdot}$: the Dali-bracket \eqref{DaliBracket}.
    \item $[\cdot\,,\cdot]^C$: the $C$-bracket \eqref{CFbrackettau}. Same as $[\cdot\,,\cdot]^\sigma$ but over $\scri$.
    \item $\lbr\cdot\,,\cdot\rbr^\sigma$: the algebroid bracket \eqref{TbracketSphere} over $S$.
    \item $\lbr\cdot\,,\cdot\rbr$: the algebroid $\sfT$-bracket \eqref{Tbracket} over $\scri$.
\end{itemize}

\ni \textbf{Covariant derivative:}
\begin{itemize}
    \item $\Dcal T=\frac{1}{\Ham_0}[\Ham,T]^\sigma$: the adjoint action of the particular super-translation $\Ham$, \cite{Cresto:2024fhd} and \eqref{DcalC}.
    \item $\Dcal \tau=\frac{1}{\Ham_0}[\Ham,\tau]^C$: the adjoint action but for the $C$-bracket, cf.\,\eqref{DcalC}.
\end{itemize}

\ni \textbf{Algebra and algebroids:}
\begin{itemize}
    \item $\Wcal(S)\equiv\big(\W,[\cdot\,,\cdot]^\W\big)$: the wedge algebra \cite{Cresto:2024fhd}. See sections \ref{sec:Wedge} and \ref{Sec:algebroidTs}.
    \item $\Wcal_\sigma(S)\equiv\big(\W_\sigma,[\cdot\,,\cdot]^\sigma\big)$: the covariant wedge algebra \cite{Cresto:2024fhd}. See sections \ref{sec:Wedge} and \ref{Sec:algebroidTs}.
    \item $\A\equiv\big(\sfT,\lbr\cdot\, ,\cdot\rbr,\delta\big)$: the \hyperref[theoremAlgebroid]{$\A$-algebroid}.
    \item $\Ao\equiv\big(\sfTo,\lbr\cdot\, ,\cdot\rbr,\delta\big)$: the \hyperref[theoremAlgebroids]{$\Ao$-sub-algebroid}, that admits a representation on $\PS$.
    \item $\Ah\equiv\big(\sfTh,\lbr\cdot\, ,\cdot\rbr,\delta\big)$: the \hyperref[theoremAlgebroids]{$\Ah$-sub-algebroid}.
    \item $\Ah\equiv\big(\sfTh,\lbr\cdot\, ,\cdot\rbr,\tilde\delta\big)$: an extension of the \hyperref[theoremAlgebroidh]{$\Ah$-sub-algebroid} that admits a representation on $\PS_h$.
\end{itemize}

\ni \textbf{Anchor map and infinitesimal variation:}
\begin{itemize}
    \item $\delta$ \& $\tilde\delta$: anchor maps of the various algebroids.
    \item $\hat\delta$: `non-radiative' anchor map \eqref{hataction}.
    \item $\dt$: fields space vector field that generates an infinitesimal symmetry transformation on functional over $\scri$, \eqref{dtauC}.
    \item $\dT$: same as $\dt$ but on functional over $S$.
    \item $\d{s}{\T{s}}\cdot=\delta_{\tau(\T{s})}\cdot$: the infinitesimal variation of spin-weight $s$ along the parameter $\T{s}$, cf.\,\eqref{defdeltaTs}.
    \item $\td{s}{\T{s}}\cdot$: the coefficient of $u^0$ when $\d{s}{\T{s}}\cdot$ is a polynomial in $u$, cf.\,\eqref{deftildedeltaTs}.
\end{itemize}

\ni \textbf{Charges:}
\begin{itemize}
    \item $\tQ_s(u,z,\bz)$: charge aspect of helicity $s$ that satisfies the truncation of EE \eqref{Qevolve}.
    \item $\tq_s(u,z,\bz)$: renormalized charge aspect \eqref{defcqTsu}. Bondi version.
    \item $\th_s(u,z,\bz)$: renormalized charge aspect \eqref{RenormChargeCov}. Covariant version.
    \item $\hq_s(z,\bz)=\lim_{u\to-\infty}\tq_s$: `Bondi' renormalized charge aspect at $\scri^+_-$.
    \item $\hh_s(z,\bz)=\lim_{u\to-\infty}\th_s$: `Covariant' renormalized charge aspect at $\scri^+_-$.
    \item $\q{s}{k}$: the part of $\hq_s$ homogeneous of degree $k$ in $C$, cf.\,\eqref{hqsk}.
    \item $Q^u_s[\t{s}]\propto\int_S \tQ_s\t{s}$: $\tQ_s$ smeared over the sphere.
    \item $\mathds{Q}_{\tau}^u$: master charge, the sum over all spins of $Q^u_s[\t{s}]$. 
    \item $\Qt^u$: `Bondi' master charge \eqref{masterChargeGR}. Conserved when $\bN=0$.
    \item $\Ht^u$: `Covariant' master charge. Conserved when $\dot{\bN}=0$, cf.\,\eqref{QH}.
    \item $\Qt=\Qt^{-\infty}$: `Bondi' Noether charge \eqref{defNoetherCharge}.
    \item $\Ht=\Ht^{-\infty}$: `Covariant' Noether charge \eqref{canonicalRepH}.
    \item $\cq_{\T{s}}=Q_{\tau(\T{s})} \propto\int_S\hq_s\T{s}$: Noether charge of the helicity $s$ parameter, cf.\,\eqref{Noetherhatqs}.
    \item $\mathsf{Q}$: charge aspect 2-form \eqref{Qaspect2form}.
    \item $\mathsf{Q}_\tau$: master charge 2-form \eqref{Qaspect2form}. `Bondi' version.
    \item $\mathsf{H}_\tau$: master charge 2-form \eqref{Haspect2form}. `Covariant' version.
    \item $\Qt^U$: master charge on an arbitrary cut $S(U)$ of $\scri$, cf.\,\eqref{QtArbitraryCut}.
\end{itemize}

\ni \textbf{Goldstone:}
\begin{itemize}
    \item $G(z,\bz)$ \& $G(u,z,\bz)$: the Goldstone field \eqref{Goldstone} \& \eqref{GoodCutEq}.
    \item $\hG$: the Goldstone diffeomorphism on $\scri$ \eqref{GoodCutEq}.
    \item $\hhG$: the Goldstone diffeomorphism on $\TN$, \eqref{hhGdiff}.
    \item $\mTG$: the dressing map \cite{Cresto:2024fhd}, see also \ref{sec:RelationDressingMap}. 
\end{itemize}

\section*{Acknowledgments}
\addcontentsline{toc}{section}{Acknowledgments}

We would like to thank Lionel Mason, Romain Ruzziconi and Adam Kmec for the ten fruitful days of discussion while they were visiting the Perimeter Institute for the Celestial Summer School 2024, couple of weeks after the publication of their related work.
We would also like to thank Atul Sharma for an insightful comment on the relation to twistor space and Andrew Strominger for his remarks during the celestial school.
NC thanks Hank Chen for numerous hours of discussion and the comments that he brought along the development of this project.
Research at Perimeter Institute is supported by the Government of Canada through
the Department of Innovation, Science and Economic Development and by the Province of
Ontario through the Ministry of Colleges and Universities. This work was supported by the
Simons Collaboration on Celestial Holography.

\newpage
\appendix

\section{Proof of the Jacobi anomaly \label{AppJacobiDcal}}

Here we tackle the computation of the cyclic permutation of $\big[\tau,[\tau',\tau'']^C\big]^C$, making use of the Leibniz anomaly \eqref{AnomalyLeibniz1}, that we prove in App.\,\ref{App:dualEOM}.
We use that $\hdt C= -(\Dcal^2\tau)_{-2}$.
\begin{align}
    \big[\tau,[\tau',\tau'']^C\big]^C_s &=\sum_{n=0}^{s+1}(n+1)\Big(\t{n}\big(\Dcal[\tau',\tau'']^C\big)_{s-n}-[\tau',\tau'']^C_n(\Dcal\tau)_{s-n}\Big) \cr
    &=\sum_{n=0}^{s+1}(n+1)\bigg(\t{n}[\Dcal\tau',\tau'']^C_{s-n}+\t{n}[\tau',\Dcal\tau'']^C_{s-n} \cr
    &\hspace{3cm} +(s-n+3)\t{n}\Big(\tp{s-n+2}(\Dcal^2\tau'')_{-2}-\tpp{s-n+2}(\Dcal^2\tau')_{-2}\Big)\bigg) \cr
    &-\sum_{n=0}^{s+1}(n+1)[\tau',\tau'']^C_n(\Dcal\tau)_{s-n} \cr
    &=\left\{\sum_{n=0}^{s+1}\sum_{k=0}^{s-n+1}(n+1)(k+1)\Big(\t{n}(\Dcal\tau')_k(\Dcal\tau'')_{s-n-k}-\t{n}\tpp{k}(\Dcal^2\tau')_{s-n-k}\Big)\right. \cr
    &\quad -\sum_{n=0}^{s+1}\sum_{k=0}^{n+1}(n+1)(k+1)\tp{k}(\Dcal\tpp{})_{n-k}(\Dcal\tau)_{s-n} \label{jacobiDcalinterm}\\
    &\quad \left.+\sum_{n=0}^{s+1}(n+1)(s-n+3)\t{n}\tp{s-n+2}(\Dcal^2\tau'')_{-2}\right\}-(\tau'\leftrightarrow\tau''). \nn
\end{align}
The last equation then splits into three types of contributions, that we shall refer to as \circled{1}, \circled{2} and \circled{3}.
Let us start with the third line of \eqref{jacobiDcalinterm}, that we dub \circled{3}. 
Extracting the $n=0$ term and then taking appropriate cyclic permutations, we get
\begin{align}
    \circled{3}&\cyc (s+3)\t0\tp{s+2}(\Dcal\tau'')_{-2}-(s+3)\tp0\t{s+2}(\Dcal\tau'')_{-2} \\
    &+\sum_{n=1}^{s+1}(n+1)(s-n+3)\t{n}\tp{s-n+2}(\Dcal^2\tau'')_{-2}-\sum_{n=1}^{s+1}(n+1)(s-n+3)\tp{n}\t{s-n+2}(\Dcal^2\tau'')_{-2}. \nn
\end{align}
Changing $n\to s-n+2$ in the last sum, the two sums cancel and \circled{3} reduces to
\begin{equation} \label{jacobiDcalinterm3}
    \circled{3}\cyc (\Dcal\tau'')_{-2}\paren{\tau,\tau'}_s \cyc -\hdt C\paren{\tau',\tau''}_s, 
\end{equation}
which is precisely \eqref{JacCFC}. 
We thus have to show that all the other terms in \eqref{jacobiDcalinterm} cancel.

The trick first consists in isolating from the summation the boundary terms that contain degree $-1$ element such as $(\Dcal \tau)_{-1}$ or $(\Dcal^2 \tau)_{-1}$.
We denote all the latter by $\circled{1}$ and the rest by \circled{2}.
For instance, in the first line of \eqref{jacobiDcalinterm},
\begin{equation}
    \sum_{n=0}^{s+1}\sum_{k=0}^{s-n+1}= \sum_{n=0}^{s+1}(k=s-n+1)+\sum_{n=0}^s\sum_{k=0}^{s-n}
\end{equation}
and we collect the first  term in \circled{1} and the second  term in \circled{2}.
We proceed similarly with the second line of \eqref{jacobiDcalinterm} and obtain that
\begin{align}
    \circled{1} &=\left\{\sum_{n=0}^{s+1}(n+1)(s-n+2)\t{n}(\Dcal\tau')_{s-n+1}(\Dcal\tau'')_{-1}\right. \cr
    &\quad - \sum_{n=0}^{s+1}(n+1)(s-n+2)\t{n}\tpp{s-n+1}(\Dcal^2\tau')_{-1} \label{jacobiDcalinterm2}\\
    &\hspace{-1cm}\left.-\sum_{n=0}^s(n+1)(n+2)\tp{n+1}(\Dcal\tpp{})_{-1}(\Dcal\tau)_{s-n}-\sum_{k=0}^{s+2}(s+2)(k+1)\tp{k}(\Dcal\tau'')_{s+1-k}(\Dcal\tau)_{-1}\right\} -(\tau'\leftrightarrow\tau''). \nn
\end{align}
The second term of this equation evaluates  after skew-symmetrization to
\begin{equation}
    -(\Dcal^2\tau')_{-1} \sum_{n=0}^{s+1}(n+1)(s-n+2)\t{n}\tpp{s-n+1}+(\Dcal^2\tau'')_{-1} \sum_{n=0}^{s+1}(n+1)(s-n+2)\t{n}\tp{s-n+1}\cyc 0,
\end{equation}
where we used the cyclic permutation and the change of variable $n\to s-n+1$ in one of the sums. Hence, by extracting the terms $n=s$, $k=s+1$ and $k=s+2$ in the last line of \eqref{jacobiDcalinterm2} (and renaming $k\to n$), \circled{1} is equal to
\begin{align}
    \circled{1}
    &= \left\{\sum_{n=0}^{s+1}(n+1)(s-n+2)\t{n}(\Dcal\tau')_{s-n+1}(\Dcal\tau'')_{-1}-\sum_{n=0}^{s+2}(s+2)(n+1)\tp{n}(\Dcal\tau'')_{s+1-n}(\Dcal\tau)_{-1} \right. \nn\\
    &\quad \left. -\sum_{n=0}^{s}(n+1)(n+2)\tp{n+1}(\Dcal\tpp{})_{-1}(\Dcal\tau)_{s-n}\right\}-(\tau'\leftrightarrow\tau'') \cyc 0. 
\end{align}
These terms  vanish upon cyclic permutation, which becomes clear as soon as one puts $(\Dcal\tau)_{-1}$ as a prefactor of every term.

We still have to deal with \circled{2}, which can now very conveniently be written as
\begin{align}
    \circled{2}&=-\sum_{a+b+c=s}(a+1)(b+1)\Big(\t{a}\tpp{b}(\Dcal^2\tau')_c-\t{a}\tp{b}(\Dcal^2\tau'')_c\Big) \cr
    &\quad +\sum_{a+b+c=s}(a+1)(b+1)\Big(\t{a}(\Dcal\tau')_b(\Dcal\tau'')_c-\t{a}(\Dcal\tau'')_b(\Dcal\tau')_c\Big) \cr
    &\quad -\sum_{a+b+c=s}(a+b+1)(a+1)\Big(\tp{a}(\Dcal\tau'')_b(\Dcal\tau)_c-\tpp{a}(\Dcal\tau')_b(\Dcal\tau)_c\Big).
\end{align}
Notice that $a,b$ and $c$ run from 0 to $s$. 
Without extracting the degree $-1$ elements in \circled{1}, the ranges for $a,b$ and $c$ would have been different, preventing us from treating them on par. 
Using cyclic permutation, we have that
\begin{align}
    \circled{2} &\cyc -\sum_{a+b+c=s}(a+1)(b+1)\Big(\t{a}\tpp{b}(\Dcal^2\tau')_c-\tpp{a}\t{b}(\Dcal^2\tau')_c\Big) \cr
    &\quad +\sum_{a+b+c=s}(a+1)(b+1)\Big(\t{a}(\Dcal\tau')_b(\Dcal\tau'')_c-\t{a}(\Dcal\tau'')_b(\Dcal\tau')_c\Big) \cr
    &\quad -\sum_{a+b+c=s}(a+b+1)(a+1)\Big(\t{a}(\Dcal\tau')_b(\Dcal\tau'')_c-\t{a}(\Dcal\tau'')_b(\Dcal\tau')_c\Big) \\
    &\cyc -\sum_{a+b+c=s}a(a+1)\Big(\t{a}(\Dcal\tau')_b(\Dcal\tau'')_c-\t{a}(\Dcal\tau'')_b(\Dcal\tau')_c\Big) \cyc 0. \nn
\end{align}
Therefore $\circled{1}=0=\circled{2}$ and $\big[\tau,[\tau',\tau'']^C\big]^C_s=\circled{3}= \eqref{jacobiDcalinterm3}$.

\section{Proofs of closure for the $\sfT$-bracket \label{Apptppsevol}}

In this section we use that the variational derivative $\delta_\tau$ commute with the differentials 
\be 
[\pa_u, \delta_\tau]=0, \qquad 
[D, \delta_\tau]=0.
\ee 
This is a consequence of the fact that the variational Cartan calculus commutes with the differential Cartan calculus \cite{Freidel:2020xyx}.\footnote{In general we have $[\delta, \LL_V]= \LL_{\delta V}$ for an arbitrary vector field.} 

The action $\dt$ can be written explicitly as a vector field on the jet bundle  \cite{saunders1989geometry} $J^\infty\bfm p$ over $\scri$, where $\bfm p:\PS\to\scri$ is the line bundle of Carrollian fields of weight $(1,2)$:
\begin{align}
    \dt O   = \sum_{n,m=0}^\infty \int_{\scri}  \pa_u^nD^m\big(\dt C\big)
    \frac{\delta O}{\delta \big(\pa_u^nD^m C\big)}. \label{actiondelta}
\end{align}

\subsection{Leibniz anomalies \label{App:dualEOM}}

In this section we provide the proof of \eqref{AnomalyLeibniz}.
We start by computing explicitly $\big(\Dcal[\tau,\tau']^C\big)_{s-1}$, for $s\geq 0$, using the formula \eqref{CFbrackettauDcal} for the $C$-bracket. 
\begin{align}
    \big(\Dcal[\tau,\tau']^C\big)_{s-1} &=D[\tau,\tau']^C_s-(s+2)C [\tau,\tau']_{s+1}^C \nn\\
    &=\left\{\sum_{n=0}^{s+1}(n+1)\Big(D\t{n}(\Dcal \tau')_{s-n}+\t{n}D(\Dcal \tau')_{s-n}\Big)\right. \\
    & \left.-(s+2)C\sum_{n=0}^{s+2}(n+1)\t{n}(\Dcal \tau')_{s+1-n}\right\}-(\tau\leftrightarrow \tau'). \nn
\end{align}
We now add and subtract the necessary terms to transform the $D$ into $\Dcal$ in the first sum:
\begin{align}
    \big(\Dcal[\tau,\tau']^C\big)_{s-1}  &= \left\{\sum_{n=0}^{s+1}(n+1)\Big((\Dcal \tau)_{n-1}(\Dcal \tau')_{s-n}+\t{n}(\Dcal^2 \tau')_{s-n-1}\Big)\right. \nn\\
    &+\sum_{n=0}^{s+1}(n+1)(n+2)C\t{n+1}(\Dcal \tau')_{s-n}+\sum_{n=0}^{s+1}(n+1)(s-n+2)C\t{n}(\Dcal \tau')_{s-n+1} \nn\\
    &-\left.(s+2)C\sum_{n=0}^{s+1}(n+1)\t{n}(\Dcal \tau')_{s+1-n}-(s+2)(s+3)C\t{s+2}(\Dcal \tau')_{-1}\right\} -(\tau\leftrightarrow \tau') \nn\\
    &=\left\{\sum_{n=0}^{s+1}(n+1)\Big((\Dcal \tau)_{n-1}(\Dcal \tau')_{s-n}+\t{n}(\Dcal^2 \tau')_{s-n-1}\Big)\right. \nn\\
    &\quad +\sum_{n=1}^{s+2}n(n+1)C\t{n}(\Dcal \tau')_{s+1-n}-\sum_{n=1}^{s+1}n(n+1)C\t{n}(\Dcal \tau')_{s+1-n} \nn\\
    &\quad -(s+2)(s+3)C\t{s+2}(\Dcal \tau')_{-1}\Bigg\} -(\tau\leftrightarrow \tau') \\
    &=\left\{\sum_{n=0}^{s+1}(n+1)\Big((\Dcal \tau)_{n-1}(\Dcal \tau')_{s-n}+\t{n}(\Dcal^2 \tau')_{s-n-1}\Big)\right\} -(\tau\leftrightarrow \tau'). \nn
\end{align}
We now evaluate the second term in the anomaly,
\begin{align}
  & \big[\tau,\Dcal \tau'\big]^C_{s-1}+\big[\Dcal \tau,\tau'\big]^C_{s-1} \cr
 = &  
 \sum_{n=0}^{s}(n+1)\Big((\Dcal\tau)_{n}(\Dcal \tau')_{s-1-n}+ \t{n}(\Dcal^2 \tau')_{s-1-n}\Big) 
  -(\tau\leftrightarrow \tau').
  \cr
 = &  
 \left\{\sum_{n=1}^{s+1} n(\Dcal\tau)_{n-1}(\Dcal \tau')_{s-n}+  \sum_{n=0}^{s} (n+1)\t{n}(\Dcal^2 \tau')_{s-1-n} \right\}
  -(\tau\leftrightarrow \tau').
\end{align}
Therefore the difference gives 
\begin{align}
\cA_{s-1}\big([\tau,\tau']^C,\Dcal \big) &=
 \left\{\sum_{n=0}^{s+1} (\Dcal \tau)_{n-1}(\Dcal \tau')_{s-n}
 +(s+2) \t{s+1}(\Dcal^2 \tau')_{-2}\right\} -(\tau\leftrightarrow \tau') \cr
 &=(s+2) \left( \t{s+1}(\Dcal^2 \tau')_{-2} -\tp{s+1}(\Dcal^2 \tau)_{-2} \right).
\end{align}
In the last equality we use that the first term is symmetric in $\tau\leftrightarrow \tau'$ which follows from the relabelling $n\leftrightarrow s+1-n$.

Moreover, we can also consider the special case $\big(\Dcal[\tau,\tau']^C\big)_{-2}$ for which
\begin{align}
    \big(\Dcal[\tau,\tau']^C\big)_{-2} &= D[\tau,\tau']^C_{-1}-C[\tau,\tau']^C_0 \cr
    &=\Big(D\big(\t0(\Dcal\tau')_{-1}\big)-C\big(\t0(\Dcal\tau')_0+2\t1(\Dcal\tau')_{-1}\big)\Big)-\tau\leftrightarrow\tau' \cr
    &=\Big(D\t0(\Dcal\tau')_{-1}+\t0( \Dcal^2\tau')_{-2}-2C\t1(\Dcal\tau')_{-1}\Big)-\tau\leftrightarrow\tau' \cr
    &=\t0( \Dcal^2\tau')_{-2}-\tp0( \Dcal^2\tau)_{-2}. 
    \label{AnomalyLeibniz-2}
\end{align}
In the third line, we gathered $D(\Dcal\tau')_{-1}-C(\Dcal\tau')_0=(\Dcal^2\tau')_{-2}$ while in the last step, we expanded the rest of the terms and used the anti-symmetry between $\tau$ and $\tau'$.
Now notice that the $C$-bracket at degree $-2$ vanishes identically, $[\cdot\,,\cdot]^C_{-2}\equiv 0$. 
This is a natural extension of our definition for the $C$-bracket, that we discuss at length in our companion paper.
It implies that the Leibniz anomaly of $\Dcal$ on $[\cdot\,,\cdot]^C$ takes the general form
\begin{equation}
    \boxed{\cA_{s}\big([\tau,\tau']^C,\Dcal \big) =(s+3) \left( \t{s+2}(\Dcal^2 \tau')_{-2} -\tp{s+2}(\Dcal^2 \tau)_{-2} \right)},\qquad s\geqslant-2. \label{AnomalyLeibniz1bis}
\end{equation}
This establishes \eqref{AnomalyLeibniz1} since $(\Dcal^2 \tau)_{-2}=-\hdt C$.

We then consider the quantity
\begin{align}
    \big(\Dcal(\dtp\tau)\big)_s &=D(\dtp\t{s+1})-(s+3)C\dtp\t{s+2} \cr
    &=\dtp D\t{s+1}-(s+3)\dtp(C\t{s+2})+(s+3)\t{s+2}\dtp C \nn\\
    &=\dtp (\Dcal\tau)_s+ (s+3)\t{s+2}\dtp C.
\end{align}
This allows us to compute the anomaly to the Leibniz rule of $\Dcal$ onto the algebroid bracket,
\begin{align}
    \cA_s\big(\lbr\tau,\tau'\rbr,\Dcal \big) &= 
    \cA_s\big([\tau,\tau']^C,\Dcal \big) + \cA_s\big((\delta_{\tau'}\tau-\delta_\tau\tau'),\Dcal \big)\cr
    &= - \delta_{\Dcal\tau'}\tau + \delta_{\Dcal\tau}\tau' + (s+3) \left( \t{s+2}(\dtp C-\hdtp C)  -\tp{s+2}(\dt C-\hdt C)
    \right)\cr
    & =  - \delta_{\Dcal\tau'}\tau + \delta_{\Dcal\tau}\tau' + (s+3)N  \big( \t{s+2}\tp{0}  -\tp{s+2}\t{0}
    \big),
\end{align}
which corresponds to \eqref{AnomalyLeibniz2}.
Finally, we compute similarly the time derivative of $\lbr\cdot\,,\cdot\rbr$ and find that
\begin{align}
    \pa_u\lbr\tau,\tau'\rbr_s &=\Big([\tau,\pa_u\tau']_s^C-(s+3)N\t0\tp{s+2}+\dtp\pa_u\t{s}\Big) -(\tau\leftrightarrow\tau') \cr
    &=\Big(\lbr\tau,\pa_u\tau'\rbr_s -\delta_{\pa_u\tau'}\t{s}-(s+3)N\t0\tp{s+2}\Big) -(\tau\leftrightarrow\tau'),
\end{align}
which matches with \eqref{AnomalyLeibniz3}.

\subsection{Proof of \eqref{tt1dot} $\Rightarrow$ \eqref{t3dot} \label{App:1implies3}}

Here we assume only $\pa_u\t{s}=(\Dcal\tau)_s,\,s=0,1,2$ and compute $\pa_u \lbr\tau,\tau'\rbr_1-\big(\Dcal \lbr\tau,\tau'\rbr\big)_1$. 
We just need to use the Leibniz anomaly \eqref{AnomalyLeibniz}, which at degree 1 gives
\begin{align}
    \big((\pa_u-\Dcal)\lbr\tau,\tau'\rbr\big)_1 &= \big\lbr(\pa_u-\Dcal)\tau,\tau'\big\rbr_1 +\big\lbr\tau,(\pa_u-\Dcal)\tau'\big\rbr_1+\delta_{(\pa_u-\Dcal)\tau}\tp1-\delta_{(\pa_u-\Dcal)\tau'}\t1 \cr
    &=\big\lbr(\pa_u-\Dcal)\tau,\tau'\big\rbr_1 -\tau\leftrightarrow\tau',
\end{align}
where we use that $\delta_{(\pa_u-\Dcal)\tau}\tp1\propto \delta_{(\pa_u-\Dcal)\tau} C=0$ by hypothesis. 
Hence,
\begin{align}
     \big\lbr(\pa_u-\Dcal)\tau,\tau'\big\rbr_1 &=\sum_{n=0}^2(n+1)\Big(\cancel{\big(\pa_u\t{n}-(\Dcal\tau)_n\big)}(\Dcal\tau')_{1-n}-\tp{n}\big(\Dcal(\pa_u-\Dcal)\tau\big)_{1-n}\Big) \cr
     &=-\sum_{n=0}^2(n+1)\tp{n}D \cancel{\big(\pa_u\tau-\Dcal\tau\big)}_{2-n}+\sum_{n=0}^2(n+1)\tp{n}(4-n)C\big(\pa_u\tau-\Dcal\tau\big)_{3-n} \cr
     &=4C\tp0\big(\pa_u\t3-(\Dcal\tau)_3\big).
\end{align}
This means that  $\pa_u\lbr\tau,\tau'\rbr_1=\big(\Dcal\lbr\tau,\tau'\rbr\big)_1$ precisely if $\dot\tau_3-(\Dcal\tau)_3 = \t0 M_3$, where $M_3 \in \Ccar{1,-3}$. 
No such local functional can be constructed using only $C$ and its (holomorphic) derivatives and we therefore conclude that $\dot\tau_3=(\Dcal\tau)_3$ which is the dual equation of motion for $s=3$, namely \eqref{t3dot}.

\subsection{Initial condition \label{App:InitialConstraint}}

If we assume that $\tau,\tau'\in\sfTo$  
we show that the $\sfT$-bracket respects the initial condition \eqref{initialCondition}, i.e.
\begin{equation}
    \boxed{D\lbr\tau,\tau'\rbr_{-1}= C\lbr\tau,\tau'\rbr_0}.
\end{equation}
Indeed, notice that thanks to the computation \eqref{AnomalyLeibniz-2}, we have that
\begin{align}
    \big(\Dcal\lbr\tau,\tau'\rbr\big)_{-2}&= \Big(\big(\Dcal[\tau,\tau']^C\big)_{-2}+\big(\Dcal(\dtp\tau)\big)_{-2}\Big)-\tau\leftrightarrow\tau' \cr
    &=\Big(\t0(\Dcal^2\tau')_{-2} +\dtp (D\t{-1})-C\dtp\t0\Big)-\tau\leftrightarrow\tau' \cr
    &=\Big(\t0(\Dcal^2\tau')_{-2} +\t0\dtp C\Big)-\tau\leftrightarrow\tau \\
    &=0, \nn
\end{align}
since $\dtp C=-(\Dcal^2\tau')_{-2}+N\tp0$ and we leveraged the anti-symmetry in $\tau,\tau'$.

\section{Proof of formula \eqref{hatdTT} \label{AppAlgebraAction}}

We compute the following quantity:
\begin{align}
    \Big(\Big\{2\tp1 D\big(\hdt C\big)+3D\tp1\big(\hdt C\big)-6C\tp2\big(\hdt C\big)\Big\}-\tau\leftrightarrow\tau'\Big)+\hat\delta_{[\tau,\tau']^C}C, \label{morphismInterm}
\end{align}
using $\hdt C=-(\Dcal^2\tau)_{-2}$ and the Leibniz anomaly \eqref{AnomalyLeibniz1bis}.
This is a complementary demonstration to the one already performed in App.\,E of \cite{Cresto:2024fhd}.
Indeed,
\begin{align}
    \hat\delta_{[\tau,\tau']^C}C &=-\big(\Dcal^2[\tau,\tau']^C\big)_{-2} =
    -\Big(\Dcal[\Dcal\tau,\tau']^C+ \Dcal[\tau,\Dcal\tau']^C+\Dcal\cA\big([\tau,\tau']^C,\Dcal\big)\Big)_{-2} \cr
    &=-\big[\Dcal^2\tau,\tau' \big]^C_{-2}-2 \big[\Dcal\tau, \Dcal\tau'\big]^C_{-2} -\big[\tau,\Dcal^2\tau'\big]^C_{-2} \\
    &\quad \,-\cA_{-2}\big([\Dcal\tau,\tau']^C, \Dcal\big)-\cA_{-2}\big([\tau, \Dcal\tau']^C, \Dcal\big)-D\cA_{-1}\big([\tau,\tau']^C, \Dcal\big)+C \cA_0\big([\tau,\tau']^C, \Dcal\big). \nn
\end{align}
We can now exploit the fact that the $C$-bracket at degree $-2$ vanishes identically, $[\cdot\,,\cdot]^C_{-2}\equiv 0$. 
Therefore, we just have to evaluate the various anomalies using \eqref{AnomalyLeibniz1bis}. 
We get
\begin{align}
    \hat\delta_{[\tau,\tau']^C}C &=\Big((\Dcal\tau)_0\hdtp C-\tp0\hat\delta_{\Dcal\tau}C+D \big(2\t1\hdtp C\big)-3C\t2\hdtp C\Big)-\tau\leftrightarrow\tau'.
\end{align}
Hence,
\begin{align}
    \eqref{morphismInterm} &=\Big(D\tp1\big( \hdt C\big)-3C\tp2\hdt C -(\Dcal\tau')_0\hdt C+\t0\hat \delta_{\Dcal\tau'}C\Big)-\tau\leftrightarrow\tau' \cr
    &=\t0\hat \delta_{\Dcal\tau'}C-\tp0\hat \delta_{\Dcal\tau}C,
\end{align}
which concludes the proof.

\section{Symmetry action on $h$ \label{App:hE-1}}

On the one hand,
\begin{equation}
    \big[\tdt,\tdtp\big]h=\tdt\big(C\tp0-D\tp{-1}\big)-\tau\leftrightarrow\tau'=\big(\tp0\tdt C+\tilde\delta_{\tdt\tau'}h\big)-\tau\leftrightarrow\tau'.
\end{equation}
On the other hand (using \eqref{AnomalyLeibniz-2}),
\begin{align}
    \tilde\delta_{\lbr\tau,\tau'\rbr}h=-\big(\Dcal[\tau,\tau']^C\big)_{-2}+ \tilde\delta_{\tdtp\tau-\tdt\tau'}h &=\big(\tp0(\Dcal^2\tau)_{-2}-\tilde\delta_{\tdt\tau'}h\big)-\tau\leftrightarrow\tau' \cr
    &=-\big(\tp0\dt C+\tilde\delta_{\tdt\tau'}h\big)-\tau\leftrightarrow\tau',
\end{align}
where in the last step, we used that $(\Dcal^2\tau)_{-2}=-\hdt C$ and we then replaced $\hdt C$ by $\dt C$ since the two differ by a term that vanishes upon anti-symmetrization in $\tau,\tau'$.
Finally, using the fact that $\tdt C=\pa_u(\tdt h)=\dt C$, we readily see that $\big[\tdt,\tdtp\big]h+\tilde\delta_{\lbr\tau,\tau'\rbr}h=0$.

\section{General expression for $(\pui)^*$ \label{App:pui}}

If we keep $\alpha$ arbitrary in the definition of $\pui$, namely $\pui=\int_\alpha^u$, then
\begin{align}
    \int_{-\infty}^\infty\rd u\,A(u)[\pui B](u) &=\int_{-\infty}^\infty\rd u\, A(u)\left[\int_\infty^u \rd u'B(u')\right]+\int_{-\infty}^\infty\rd u\,A(u)\left[\int_\alpha^\infty \rd u'B(u')\right] \nn\\
    &=\int_{-\infty}^\infty\!\!\rd u\left[\int^{-\infty}_u\!\!\!\rd u' A(u')\right]\!B(u)+\int_{-\infty}^\infty\!\!\rd u\left[\int_{-\infty}^\infty\!\!\rd u'\,\theta(u-\alpha)A(u')\right]\!B(u)\nn\\
    &\equiv\int_{-\infty}^\infty\rd u\big[(\pui)^*A\big](u)B(u).
\end{align}
We thus get that in general, 
\begin{equation}
    (\pui O)(u)=\int_\alpha^u O(u')\rd u',\quad\big((\pui)^*O\big)(u)=\int_u^{-\infty}\!\!O(u')\rd u'+\theta(u-\alpha)\int_{-\infty}^\infty\!O(u')\rd u'.
\end{equation}
Equivalently,
\begin{equation}
    \big((\pui)^*O\big)(u)=\left\{
    \begin{array}{lr}
        \int_u^{-\infty}\rd u'O(u') & \rm{if}~u<\alpha, 
        \vspace{0.2cm}\\
        \int_u^{\infty}\rd u'O(u') & \rm{if}~u\geqslant\alpha.
    \end{array}\right.
\end{equation}
Notice that if we require $\big((\pui)^*O\big)(u)$ to be continuous, then 
\begin{equation}
    \int_\alpha^\infty\rd u'O(u')=\int_\alpha^{-\infty}\rd u'O(u')\qquad\Leftrightarrow\qquad\int_{-\infty}^\infty\rd u'O(u')=0,
\end{equation}
so that any memory effect between $\scri^+_-$ and $\scri^+_+$ is excluded, which is expected from imposing Schwartz falloffs.

\section{Time derivative of $\tq_4$ \label{App:tq4}}

We compute the time derivative of 
\begin{align}
    \tq_4 &\equiv q_4-\frac{u^4}{4!}D^4(\bN C)-5D^2\tq_0\pui[3]C-4D\big(D\tq_0\pui[2](uC)\big)-\frac32 D^2\big(\tq_0\pui(u^2C)\big) \nn\\
    & +5D\tQ_1\pui[2]C+ 4D\big(\tQ_1\pui(uC)\big)-5\tQ_2\pui C +15\tq_0\pui\big(C\pui C\big).
\end{align}
We get
\begin{align}
    \pa_u\tq_4 &=-\sum_{n=1}^4\frac{(-u)^{n-1}}{(n-1)!}D^n\tQ_{4-n}+\sum_{n=0}^4\frac{(-u)^n}{n!}D^n\big(D\tQ_{3-n}+(5-n)C\tQ_{2-n}\big) \nn\\
    &-\frac{u^3}{3!}D^4(\bN C)-\frac{u^4}{4!}D^4\pa_u(\bN C)-5D^2\pa_u\tq_0\pui[3]C-5D^2\tq_0\pui[2]C-4D\big(D\pa_u\tq_0\pui[2](uC)\big) \nn\\
    &-4D\big(D\tq_0\pui(uC)\big)-\frac32 D^2\big(\pa_u\tq_0\pui(u^2C)\big)-\frac32 D^2\big(\tq_0 u^2C\big)+5D\tQ_1\pui C \\
    &+5D\big(D\tQ_0+2C\tQ_{-1}\big) \pui[2]C + 4D\big((D\tQ_0+2C\tQ_{-1}) \pui(uC)\big) + 4D\big(\tQ_1 uC\big) \nn\\
    &-5\big(D\tQ_1+3C\tQ_0\big)\pui C-5\tQ_2 C+15\pa_u\tq_0\pui\big(C\pui C\big)+15\tq_0\big(C\pui C\big). \nn
\end{align}
Let us reorganize the terms to emphasize the simplifications:
\begin{align}
    \pa_u\tq_4 &=-\sum_{n=0}^3\frac{(-u)^n}{n!}D^{n+1}\tQ_{3-n}+\sum_{n=0}^3\frac{(-u)^n}{n!}D^n\big(D\tQ_{3-n}\big) +\frac{u^4}{4!}D^4\pa_u\tQ_0-\frac{u^4}{4!}D^4\pa_u(\bN C) \nn\\
    &+\sum_{n=0}^3\frac{(-u)^n}{n!}D^n\big((5-n)C\tQ_{2-n}\big)-\frac{u^3}{3!}D^4(\bN C)-\frac32 D^2\big(\tq_0 u^2C\big)+ 4D\big(\tQ_1 uC\big) -5\tQ_2 C\nn\\
    &-5D^2\tq_0\pui[2]C -4D\big(D\tq_0\pui(uC)\big)+5D^2\tQ_0 \pui[2]C + 4D\big(D\tQ_0 \pui(uC)\big)\nn\\
    &+5D\tQ_1\pui C-5D\tQ_1\pui C-15C\tQ_0\pui C+15\tq_0\big(C\pui C\big) \\
    &-5D^2\pa_u\tq_0\pui[3]C-4D\big(D\pa_u\tq_0\pui[2](uC)\big)-\frac32 D^2\big(\pa_u\tq_0\pui(u^2C)\big)+15\pa_u\tq_0\pui\big(C\pui C\big) \nn\\
    &+10D\big(C\tQ_{-1}\big) \pui[2]C+ 8D\big(C\tQ_{-1} \pui(uC)\big). \nn
\end{align}
The first line reduces to
\bs 
\label{leftover}
\begin{equation}
    \frac{u^4}{4!}D^4\pa_u\tq_0,
\end{equation}
the second line to
\begin{equation}
    -\frac{u^3}{3!}D^4(\bN C)+\frac{3u^2}{2}D^2(\bN C^2)-\frac{2u^3}{3!}D^3\big(C\tQ_{-1}\big),
\end{equation}
where we used that $\tQ_{-1} = D \bN$.
The third line reduces to
\begin{equation}
    5D^2(\bN C)\pui[2]C +4D\big(D(\bN C)\pui(uC)\big),
\end{equation}
and the fourth to
\begin{equation}
    -15\bN C^2\pui C,
\end{equation}
\es
while the last and penultimate lines always involve $\pa_u\tq_0$ or $\tQ_{-1}= D\bN$, which are 0 if $\bN\equiv 0$.
Similarly, the leftover \eqref{leftover} annihilates when $\bN\equiv 0$, so that $\pa_u\tq_4=0$ when no radiation is present.

\section{Proof of Theorem \hyperref[Noetherhatqs]{[Noether charge at spin $s$]} \label{App:renormcharge}}

We construct $\tq_s$ for a general $s\geqslant 0$.
For this we have to leverage the computation \eqref{soltaunTs} and \eqref{puiDcalk}.
We define $\shs$ and $\shS$ the dual spin and shift\footnote{$\shS$ and $\hS$ are actually equal.} operators according to
\begin{equation}
    \shs=-\hs,\qquad \shs\tQ_s=s\tQ_s\qquad\textrm{and}\qquad \shS\tQ_s=\tQ_{s-1},
\end{equation}
with commutation relations given by
\begin{equation}
    \shS D=D\shS,\quad\shS C=C\shS,\quad D\shs=(\shs-1)D,\quad\shs C=C(\shs+2),\quad\shs\shS=\shS(\shs-1).
\end{equation}
By definition, and using \eqref{soltaunTs},
\begin{equation}
    4\pi G\,\cq^u_{\T{s}}=\int_S\tQ_{-1} \t{-1}+ \sum_{n=0}^s \sum_{k=\lfloor\frac{s+1-n}{2}\rfloor}^{s-n}\int_S\tQ_n\left(\pui D-\pui C(\hs+1)\hS\right)^k\T{n+k}.
\end{equation}
Then, since
\begin{equation}
    \sum_{n=0}^s\sum_{k=\lfloor\frac{s+1-n}{2}\rfloor}^{s-n}=\sum_{k=0}^s\sum_{0\leqslant n=s-2k}^{s-k}, \label{changesum}
\end{equation}
we need to compute, using \eqref{puiDcalk},\footnote{Recall that $P=\sum_{i=0}^\l p_i$.}
\begin{align}
    &\sum_{n=s-2k}^{s-k}\int_S\tQ_n \left(\pui D-\pui C(\hs+1)\hS\right)^k\T{n+k} = \nn\\
    =&\sum_{n=s-2k}^{s-k}\int_S\tQ_n\Bigg( \sum_{\l=0}^k(-1)^\l\sum_{P=k-\l}\pui[p_0]D^{p_0}\bigg(\pui\Big\{C(\hs+1)\pui[p_1]D^{p_1}\Big(\ldots \label{intermrenorm}\\
    &\hspace{7cm}\ldots\pui\Big\{C(\hs+1) \pui[p_\l]D^{p_\l}\hS^\l\Big\}\ldots \Big)\Big\}\bigg)\Bigg)\T{n+k}. \nn
\end{align}
Recall that we study $\tau\in\sfT_s$, so that only $\T{s}\neq 0$.
Hence, for each term in the sum over $n$, there is only one term in the sum over $\l$ that survives, namely the one for which $n+k+\l=s$.
We thus get\footnote{We use that $\pui[p] T_s = \frac{u^p}{p!} T_s.$}
\begin{align}
    \eqref{intermrenorm} &=\!\int_S\bigg(\tQ_{s-k}(-1)^0\frac{u^k}{k!}D^k\T{s}+\tQ_{s-k-1}(-1)^1\!\!\!\sum_{P=k-1}\!\!\pui[p_0]D^{p_0}\Big(\pui\Big\{C(\hs+1)D^{p_1}\T{s}\pui[p_1](1)\Big\}\Big) \nn\\
    &+\tQ_{s-k-2}(-1)^2\!\!\!\sum_{P=k-2} \!\!\pui[p_0]D^{p_0}\Big(\pui\Big\{C(\hs+1)\pui[p_1]D^{p_1}\Big(\pui\Big\{C(\hs+1)D^{p_2}\T{s}\pui[p_2](1)\Big\}\Big)\Big\}\Big) \nn\\
    &+\ldots+\tQ_{s-2k}(-1)^k\pui\Big\{C(\hs+1)\pui\Big\{C(\hs+1)\ldots \pui\Big\{C(\hs+1)\T{s}\Big\}\ldots\Big\}\Big\}\bigg) \label{intermrenorm2}\\
    &=\int_S\bigg((-1)^k\frac{u^k}{k!}D^k\tQ_{s-k}\T{s}+\!\!\sum_{P=k-1}\!(-1)^{p_0+1}D^{p_0}\tQ_{s-k-1}\pui[p_0-1]\Big(C(\hs+1)\pui[p_1](1)\Big)D^{p_1}\T{s} \nn\\
    &+\!\!\sum_{P=k-2}\!(-1)^{p_0+2}D^{p_0}\tQ_{s-k-2}\pui[p_0-1]\Big\{C(\hs+1)\pui[p_1]D^{p_1}\Big( \pui\Big\{C(\hs+1)D^{p_2}\T{s}\pui[p_2](1)\Big\}\Big)\Big\} \nn\\
    &+\ldots+(-1)^k\Big((\shs+3)\tQ_{s-2k}\Big)\pui\Big\{C\pui\Big\{C(\hs+1)\ldots \pui\Big\{C(\hs+1)\T{s}\Big\}\ldots\Big\}\Big\}\bigg), \nn
\end{align}
where in the last line we trade the action of $\hs$ for its dual.
Indeed, $(\hs+1)$ acts on the object $C^{k-1}\T{s}$ (the $\pui$ are irrelevant for counting the spin), so that
\begin{equation}
    (\hs+1)(C^{k-1}\T{s})=(s-2k+3)C^{k-1}\T{s}.
\end{equation}
On the other hand,
\begin{equation}
    (\shs+3)\tQ_{s-2k}=(s-2k+3)\tQ_{s-2k},
\end{equation}
so that the prefactors do match.
We proceed further,\footnote{When trading $\hs$ for $\shs$ in the first line for instance, we use that $(\hs+1)D^{p_1}\T{s}=(s-p_1+1)D^{p_1}\T{s}=(s+p_0-k+2)D^{p_1}\T{s}$, while $(\shs+3)D^{p_0}\tQ_{s-k-1}=(s+p_0-k+2)D^{p_0}\tQ_{s-k-1}$.}
\begin{align}
    \eqref{intermrenorm2} &=\int_S\bigg((-1)^k\frac{u^k}{k!}D^k\tQ_{s-k}\T{s}+(-1)^k\!\!\sum_{P=k-1}\!\!D^{p_1}\Big((\shs+3)D^{p_0}\tQ_{s-k-1}\pui[p_0-1]\Big\{C\pui[p_1](1)\Big\}\Big)\T{s} \nn\\
    &+\!\!\!\sum_{P=k-2}\!(-1)^{2+p_0+p_1}D^{p_1}\Big((\shs+3)D^{p_0}\tQ_{s-k-2}\pui[p_0-1]\Big\{C\Big)\pui[p_1-1]\Big\{C(\hs+1)D^{p_2}\T{s}\pui[p_2](1)\Big\}\Big\} \nn\\
    &+\ldots+(-1)^k(\shs+3)\Big((\shs+3)\tQ_{s-2k}\pui\Big\{C\Big)\pui\Big\{C\ldots\pui\Big\{C(\hs+1)\T{s}\Big\}\ldots\Big\}\Big\}\Bigg) \nn\\
    &=\int_S(-1)^k\bigg(\frac{u^k}{k!}D^k\tQ_{s-k}+\!\!\sum_{P=k-1}\!\!D^{p_1}\Big((\shs+3)D^{p_0}\tQ_{s-k-1}\pui[p_0-1]\Big\{C\Big)\pui[p_1](1)\Big\} \\
    &+\!\!\!\sum_{P=k-2}\!D^{p_2}\Big((\shs+3)D^{p_1}\Big((\shs+3)D^{p_0}\tQ_{s-k-2}\pui[p_0-1]\Big\{C\Big)\pui[p_1-1]\Big\{C\Big)\pui[p_2](1)\Big\}\Big\} +\ldots\nn\\
    &+(\shs+3)\Big(\ldots\Big((\shs+3) \Big((\shs+3) \tQ_{s-2k}\pui\Big\{C\Big)\pui\Big\{C\Big)\ldots C\Big)\pui\Big\{C\Big\}\ldots\Big\}\Big\}\Bigg)\T{s} \nn\\
    &=\int_S\bigg((-1)^k \sum_{\l=0}^k\sum_{P=k-\l}D^{p_0}\Big((\shs+3)D^{p_1}\Big((\shs+3)D^{p_2}\Big(\ldots \Big((\shs+3)\hS^{*\l}D^{p_\l}\tQ_{s-k}\cdot \nn\\
    &\hspace{4cm}\cdot\pui[p_\l-1]\Big\{C\Big)\ldots\Big)\pui[p_2-1]\Big\{C\Big)\pui[p_1-1]\Big\{C\Big)\pui[p_0](1)\Big\}\Big\}\ldots\Big\}\bigg)\T{s}. \nn
\end{align}
Most of the work is now done.
To relate this expression to $\cq^u_{\T{s}}$, we need two slight modifications.
Firstly, as we discussed already in several occasions, adding the term $Q_{-1}^u[\t{-1}]$ amounts to replacing every appearance of $\tQ_0$ by $\tQ_0-\bN C$.
For that purpose, we introduce $\tQcal_s\equiv \tQ_s$, for $s>0$ and $\tQcal_0=\tQ_0-\bN C$.
Secondly, according to \eqref{changesum}, the variable $n$ had to be positive.
The easiest way to implement this restriction is to take $\tQcal_s\equiv 0$, for $s<0$.
Finally, taking the sum $\sum_{k=0}^s$, we get the expression for $\tq_s$, namely
\begin{empheq}[box=\fbox]{align}
    \tq_s=\sum_{k=0}^s(-1)^k\sum_{\l=0}^k &\sum_{P=k-\l} D^{p_0}\Big((\shs+3)D^{p_1}\Big((\shs+3)D^{p_2}\Big(\ldots\Big((\shs+3)\hS^{*\l} D^{p_\l}\widetilde\Qcal_{s-k}\cdot \nn\\
    &\cdot\pui[p_\l-1]\Big\{C\Big)\ldots\Big)\pui[p_2-1]\Big\{C\Big)\pui[p_1-1]\Big\{C\Big)\pui[p_0](1)\Big\}\Big\}\ldots\Big\}. \label{deftqs}
\end{empheq}

This proof is straightforwardly amended if we wish to work with the covariant charge $\Ht$ instead.
Indeed, since the element $\T{-1}$ now behaves on par with $\T0,\T1,\ldots$, we just have to consider
\begin{equation}
    \cH^{\,u}_{\T{s}}= \sum_{n=-1}^s\sum_{k=\lfloor\frac{s+1-n}{2}\rfloor}^{s-n}\int_S\tQ_n\left(\pui D-\pui C(\hs+1)\hS\right)^k\T{n+k},
\end{equation}
and then notice that
\begin{equation}
    \sum_{n=-1}^s\sum_{k=\lfloor\frac{s+1-n}{2}\rfloor}^{s-n}=\sum_{k=0}^{s+1}\sum_{-1\leqslant n=s-2k}^{s-k}. \label{changesumH}
\end{equation}
We thus find that 
\begin{equation} \label{RenormChargeCov}
    \cH^{\,u}_{\T{s}}=\frac{1}{4\pi G}\int_S\th_s\T{s},\qquad s\geqslant -1,
\end{equation}
for $\th_s$ given by the formula \eqref{deftqs} where $\tQcal_s$ is now defined as $\tQcal_s\equiv\tQ_s, \,s\geqslant -1$ and $\tQcal_s=0,s<-1$.
As an example,
\begin{align}
    \th_{-1} &=\tQ_{-1},\nn\\
    \th_0 &=q_0-uD\tQ_{-1}, \nn\\
    \th_1 &=q_1+\frac{u^2}{2}D^2\tQ_{-1}-2\tQ_{-1}\pui C, \\
    \th_2 &=q_2-\frac{u^3}{3!}D^3\tQ_{-1}+3D\tQ_{-1}\pui[2]C+2D\big(\tQ_{-1}\pui(uC)\big)-3\tQ_0\pui C, \nn
\end{align}
where $q_s$ is given in \eqref{defqs}.
The reader can check that $\pa_u\th_s=0$ when $\dot\bN=0$, as expected.
Of course, the associated Noether charge is defined as 
\begin{equation}
    \cH_{\T{s}}=\frac{1}{4\pi G}\int_S\hh_s\T{s},\qquad s\geqslant -1,
\end{equation}
with $\hh_s= \lim_{u\to -\infty}\th_s$.
\medskip

Furthermore, it is interesting to consider the special case of \eqref{deftqs} where $N=0$ in a strip $u\in[0,u_0]$, namely a non-radiative strip.
The renormalized charge aspect then takes a very compact form.
Indeed, first notice that in this strip, $C=C\big|_{u=0}=\sigma$ and the shear goes through the inverse $u$ derivatives\footnote{Where $\pui=\int_0^u$, with $u\in[0,u_0]$.} of \eqref{deftqs}.
The latter then reduces to 
\begin{equation}
    \tq_s=\!\sum_{k=0}^s\frac{(-u)^k}{k!}\sum_{\l=0}^k \sum_{P=k-\l} D^{p_0}\Big(\sigma(\shs+3)D^{p_1}\Big(\sigma(\shs+3)D^{p_2}\Big(\!\ldots\!\Big(\sigma(\shs+3)\hS^{*\l} D^{p_\l}\widetilde\Qcal_{s-k}\Big)\ldots\Big)\!\Big)\!\Big),
\end{equation}
where we used that
\begin{equation}
    \pui[p_\l-1]\cdots\pui[p_1-1]\pui[p_0](1)=\pui[(P+\l)](1)=\pui[k](1)=\frac{u^k}{k!}.
\end{equation}
Defining the dual operator $\Dcal^*$ acting on $\tQ$ as
\begin{equation}
    \big(\Dcal^*\tQ\big)_s=D\tQ_{s-1}+(s+1)C\tQ_{s-2}=\big(D+C(\shs+3)\shS\big)\tQ_{s-1},
\end{equation}
we conclude using formula \eqref{puiDcalk} that when $N=0$ in a strip of $\scri$, then the renormalized charge aspect inside this interval is given by \eqref{deftqs0rad}
\begin{equation}
    \boxed{\tq_s=\sum_{k=0}^s\frac{(-u)^k}{k!}\big(\Dcal^{\ast k}\widetilde\Qcal\big)_s}.
\end{equation}

\section{Proofs of Lemmas \hyperref[deltatildeHard]{[Soft and Hard actions]} \label{AppSoftHardAction}}

We start with few preliminary computations. 
From \eqref{soltaupTs}, written again here,
\begin{equation}
    \t{p}(\T{s}) =\frac{u^{s-p}}{(s-p)!}D^{s-p}\T{s}-\sum_{n=p+2}^s(n+1)\big(\pui D\big)^{n-p-2}\pui (C\t{n}),
\end{equation}
we find in particular that
\begin{equation}
    \soft{\t{p}}(\T{s})= \frac{u^{s-p}}{(s-p)!}D^{s-p}\T{s},\qquad 0\leqslant p\leqslant s.
\end{equation}
Hence
\begin{equation}
    \soft{\t0}(\T{s})=\frac{u^s}{s!}D^s\T{s}\qquad\textrm{and}\qquad \soft{\t1}(\T{s})=\frac{u^{s-1}}{(s-1)!}D^{s-1}\T{s}. \label{soltau01Soft}
\end{equation}
Moreover
\begin{equation}
    \hard{\t0}(\T{s})=-\sum_{n=2}^s(n+1)\big(\pui D\big)^{n-2}\pui \left(C\soft{\t{n}}\right),\qquad s\geqslant 2.
\end{equation}
Using the generalized Leibniz rule for pseudo-differential calculus \cite{PseudoDiffBakas},
\begin{equation}
    \pui[\alpha](fg)=\sum_{n=0}^\infty\frac{(-\alpha)_n}{n!}(\pa_u^nf)\pui[(n+\alpha)]g, \qquad \alpha\in\R, \label{LeibnizRuleGeneral}
\end{equation}
which implies in particular that
\begin{equation}
    \pui[(n-1)]\left(\frac{u^{s-n}}{(s-n)!}C\right)=\sum_{k=0}^{s-n}(-1)^k\binom{n+k-2}{k} \frac{u^{s-n-k}}{(s-n-k)!}\pui[(k+n-1)]C,
\end{equation}
we obtain
\begin{align}
     \hard{\t0}(\T{s}) &=-\sum_{n=2}^s\sum_{k=0}^{s-n}(-1)^k\binom{n+k-2}{k}(n+1)\frac{u^{s-n-k}}{(s-n-k)!}D^{n-2}\Big(D^{s-n}\T{s}\,\pui[(k+n-1)]C\Big) \nn\\
     &=-\sum_{p=2}^s\sum_{k=0}^{p-2}(-1)^k\binom{p-2}{k}(p-k+1)\frac{u^{s-p}}{(s-p)!}D^{p-k-2}\Big(D^{s-p+k}\T{s} \,\pa_u^{1-p}C\Big), \label{tau0HardTs}
\end{align}
where we changed variable $p=n+k$ in the last step.

\subsection{Proof of Lemma [Soft action]}

Using \eqref{deltatauSoft}, combined with \eqref{soltau01Soft} and the definition \eqref{defdeltaTs}, we infer that
\begin{equation}
    \Sd{s}{\T{s}}C=-D^2\soft{\t0}(\T{s})=-\frac{u^s}{s!}D^{s+2}\T{s}
\end{equation}
and in particular
\begin{equation}
    \Sd{0}{\T0}C=-D^2\T0\equiv\Std{0}{\T0}C.
\end{equation}
Therefore
\begin{equation}
    \boxed{\Sd{s}{\T{s}}C=\frac{u^s}{s!}\Std{0}{D^s\T{s}}C}.
\end{equation} 

\subsection{Proof of Lemma [Hard action][Part 1]}

The simplest way to prove \eqref{deltatildeHard} is to expand the result and check that it matches with \eqref{deltatauHard} (combined with \eqref{defdeltaTs}).
For $\alpha=\max[\mathrm{mod}_2(p),p-2]$ we have that\footnote{Recall that the falling factorial satisfies $(x)_0=1$.}
\begin{align}
    \Hd{s}{\T{s}}C &=\sum_{p=0}^s\frac{u^{s-p}}{(s-p)!}\sum_{k=0}^{\alpha}(-1)^k\frac{(p-2)_k}{k!}(p-k+1) D^{p-k}\big(D^{s-p+k}\T{s}\,\pa_u^{1-p}C\big) \nn\\
    &=\sum_{p=2}^s\frac{u^{s-p}}{(s-p)!}\sum_{k=0}^{p-2}(-1)^k\binom{p-2}{k}(p-k+1) D^{p-k}\big(D^{s-p+k}\T{s}\,\pa_u^{1-p}C\big) \nn\\
    &+\frac{u^s}{s!}D^s\T{s}\,\pa_u C+\frac{u^{s-1}}{(s-1)!}\Big(2D\big(D^{s-1}\T{s}C\big)-(-1)_1D^s\T{s}C\Big) \\
    &=-D^2\hard{\t0}(\T{s})+N\soft{\t0}(\T{s}) +2DC\frac{u^{s-1}}{(s-1)!} D^{s-1}\T{s} +3C\frac{u^{s-1}}{(s-1)!}D^s\T{s}  \nn\\
    &=-D^2\hard{\t0}(\T{s})+N\soft{\t0}(\T{s})+2DC\soft{\t1}(\T{s})+3CD\soft{\t1}(\T{s}), \nn
\end{align}
as desired.

\subsection{Proof of Lemma [Hard action][Part 2]}

We have to distinguish the special cases $s=0$ and 1.
\medskip

$\bullet$ \textbf{Spin 0:} 
\begin{equation}
    \boxed{\Hd{0}{\T0}C=\T0\pa_u C\equiv\Htd{0}{\T0}C}. \label{deltatildeHard0}
\end{equation}

$\bullet$ \textbf{Spin 1:} 
\begin{equation}
    \Hd{1}{\T1}C= D\T1\big(u\pa_u+3\big)C+2\T1DC\equiv u\,\Htd{0}{D\T1}C+\Htd{1}{\T1}C,
\end{equation}
where we identify
\begin{equation}
    \boxed{\Htd{1}{\T1}C=2\T1 DC+3CD\T1}. \label{deltatildeHard1}
\end{equation}
The equations \eqref{deltatildeHard0} and \eqref{deltatildeHard1} indeed agree with the general formula \eqref{deltatildeHardbis}.
\medskip

$\bullet$ \textbf{Spin $p\geqslant 2$:} 
\medskip

Next let us massage \eqref{deltatildeHardb} for $p\geqslant 2$:
\begin{align}
    \Htd{p}{\T{p}}C &=\sum_{k=0}^{p-2}(-1)^k\frac{(p-2)_k}{k!}(p-k+1) D^{p-k}\big(D^k\T{p}\,\pa_u^{1-p}C\big) \nn\\
    &=\sum_{k=0}^{p-2}\sum_{n=0}^{p-k}(-1)^k\frac{(p-2)_k}{k!}\frac{(p-k)_n}{n!}(p-k+1) D^{k+n}\T{p}\,D^{p-k-n}\pa_u^{1-p}C. \label{intermdeltatilde}
\end{align}
We can change variable $k+n\to m$ such that the sums schematically become
\begin{equation}
    \sum_{k=0}^{p-2}\sum_{n=0}^{p-k}= \sum_{m=2}^p\sum_{n=2}^m+\sum_{m=1}^{p-2}\sum_{n=0}^1+\big[k=0;n=0\big] +\big[k=p-2;n=1\big].
\end{equation}
Explicitly, \eqref{intermdeltatilde} turns into
\begin{align}
    \Htd{p}{\T{p}}C &= \sum_{m=2}^p D^m\T{p}\,D^{p-m}\pa_u^{1-p}C\sum_{n=2}^m (-1)^{m-n}\frac{(p-2)_{m-n}}{(m-n)!}\frac{(p+n-m)_n}{n!}(p+n-m+1) \nn\\
    &+ \sum_{m=1}^{p-2}D^m\T{p}\,D^{p-m}\pa_u^{1-p}C\sum_{n=0}^1 (-1)^{m-n}\frac{(p-2)_{m-n}}{(m-n)!}\frac{(p+n-m)_n}{n!}(p+n-m+1) \nn\\
    &+(p+1)\T{p}\,D^p\pa_u^{1-p}C+6(-1)^p D^{p-1}\T{p}\,D\pa_u^{1-p}C.
\end{align}
We then split the first line into three parts: $[m=p-1]+[m=p]+\sum_{m=2}^{p-2}$. 
Similarly we split the second line into two terms: $[m=1]+\sum_{m=2}^{p-2}$. 
Therefore,
\begin{align}
    \Htd{p}{\T{p}}C &=\sum_{m=2}^{p-2} D^m\T{p}\,D^{p-m}\pa_u^{1-p}C\sum_{n=0}^m (-1)^{m-n}\frac{(p-2)_{m-n}}{(m-n)!}\frac{(p+n-m)_n}{n!}(p+n-m+1) \nn\\
    &-D^{p-1}\T{p}\,D\pa_u^{1-p}C\sum_{n=2}^{p-1}(-1)^{p-n}\frac{(p-2)_{p-1-n}}{(p-1-n)!}(n+1)(n+2) \nn\\
    &+D^p\T{p}\,\pa_u^{1-p}C\sum_{n=2}^p (-1)^{p-n}\frac{(p-2)_{p-n}}{(p-n)!}(n+1) \label{intermdeltatilde2}\\
    &-D\T{p}\,D^{p-1}\pa_u^{1-p}C\sum_{n=0}^1(-1)^n(p-2)_{1-n}(p+n-1)_n(p+n) \nn\\
    &+(p+1)\T{p}\,D^p\pa_u^{1-p}C+6(-1)^p D^{p-1}\T{p}\,D\pa_u^{1-p}C, \nn
\end{align}
where for now we only simplified trivial factorial factors.
In the first line, we have to deal with the expression
\begin{align}
    &\sum_{n=0}^m (-1)^{m-n}\frac{(p-2)_{m-n}}{(m-n)!}\frac{(p+n-m)_n}{n!}(p+n-m+1) \nn\\
    =&\sum_{n=0}^m (-1)^n\binom{m}{n}\frac{(p-2)!(p-n+1)!}{m!(p-2-n)!(p-m)!} \nn\\
    =& ~(p+1)\binom{p}{m}\sum_{n=0}^m (-1)^n\binom{m}{n}\frac{(p-2)_n}{(p+1)_n} \nn\\
    =& ~(p+1)\binom{p}{m} {}_2F_1(-m,2-p;-(p+1);1) \nn\\
    =& ~(p+1)\binom{p}{m}\frac{(3)_m}{(p+1)_m} \nn\\
    =& ~(p+1-m)\frac{(3)_m}{m!} =\left\{
    \begin{array}{lc}
       0 & \textrm{if } m>3, \\
       (p+1-m)\binom{3}{m}  & \textrm{if } m\leqslant 3,
    \end{array}\right.
\end{align}
where we used the fact that the hyper-geometric function (written in falling factorial convention using $(n)^{(k)}=(-1)^k(-n)_k$) simplifies to---cf. formula 7.3.5.4 in \cite{HyperGmathbook}---
\begin{equation}
    {}_2F_1(-m,-b;-c;1)=\frac{(c-b)_m}{(c)_m}.
\end{equation}
Then notice that in the penultimate line of \eqref{intermdeltatilde2},
\begin{equation}
    -\sum_{n=0}^1(-1)^n(p-2)_{1-n}(p+n-1)_n(p+n)=3p.
\end{equation}
Moreover, the second line of \eqref{intermdeltatilde2} recombines with the term proportional to 6 in the last line of the same equation:
\begin{align}
    -\sum_{n=2}^{p-1}(-1)^{p-n}\frac{(p-2)_{p-1-n}}{(p-1-n)!}(n+1)(n+2)+6(-1)^p &=\sum_{n=0}^{p-2} (-1)^{p-n}\binom{p-2}{n}(n+2)(n+3) \nn\\
    &=6(-1)^p{}_2F_1\big(-(p-2),4;2;1\big) \nn\\
    &=6(-1)^p\frac{(2)_{p-2}}{(-2)_{p-2}} \nn\\
    &=6\frac{(2)_{p-2}}{(p-1)!}=\left\{
    \begin{array}{ll}
       6 & \textrm{if ~} p=2,3, \\
       2 & \textrm{if ~} p=4, \\
       0 & \textrm{if ~} p\geqslant 5.
    \end{array}\right.
\end{align}
Concerning the third line of \eqref{intermdeltatilde2},
\begin{align}
    \sum_{n=2}^p (-1)^{p-n}\frac{(p-2)_{p-n}}{(p-n)!}(n+1) &=\sum_{n=0}^{p-2}(-1)^{p-n}\binom{p-2}{n}(n+3) \nn\\
    &=3(-1)^p{}_2F_1\big(-(p-2),4;3;1\big) \nn\\
    &=3(-1)^p\frac{(1)_{p-2}}{(-3)_{p-2}} \nn\\
    &=6\frac{(1)_{p-2}}{p!}=\left\{
    \begin{array}{ll}
       3 & \textrm{if ~} p=2, \\
       1 & \textrm{if ~} p=3, \\
       0 & \textrm{if ~} p\geqslant 4.
    \end{array}\right.
\end{align}
Gathering those simplifications, \eqref{intermdeltatilde2} reduces to\footnote{$\delta_{p,n}$ is just the Kronecker symbol.}
\begin{align}
    \Htd{p}{\T{p}}C &=\sum_{m=2}^{\min[3,p-2]} \binom{3}{m}(p+1-m)D^m\T{p}\,D^{p-m}\pa_u^{1-p}C \nn\\*
    &+3p\,D\T{p}\,D^{p-1}\pa_u^{1-p}C+(p+1)\T{p}\,D^p\pa_u^{1-p}C +\delta_{p,4}\big(2D^3\T4 D\pui[3]C\big)\nn\\*
    &+\delta_{p,2}\big(6D\T2 D\pui C+3D^2\T2\pui C\big)+\delta_{p,3}\big(6D^2\T3 D\pui[2]C+D^3\T3\pui[2]C\big) \nn\\
    &= \sum_{m=0}^{\min[3,p]} \binom{3}{m}(p+1-m)D^m\T{p}\,D^{p-m}\pa_u^{1-p}C,
\end{align}
which is indeed \eqref{deltatildeHardbis}.

\section*{Statements and Declarations}
\addcontentsline{toc}{section}{Statements and Declarations}

\paragraph{Funding and competing interests:}
The authors declare they have no financial or conflict of interests.
Research at Perimeter Institute is supported by the Government of Canada through the Department of Innovation, Science and Economic Development and by the Province of Ontario through the Ministry of Colleges and Universities. This work was supported by the Simons Collaboration on Celestial Holography.

\paragraph{Data Availability Statement:}

No datasets were generated or analyzed during the current study.


\newpage
\addcontentsline{toc}{section}{References}

\bibliographystyle{JHEP}
\bibliography{sourcesNoether}
\end{document}